\documentclass[prd,showpacs,altaffilletter,twocolumn,
superscriptaddress,floatfix,nofootinbib]{revtex4} 

\usepackage{amsfonts,amsmath,graphicx,bm,siunitx}
\usepackage[pdftex,
       colorlinks=true,
       pdftitle={Lattice QCD calculation of form factors describing B meson
decays to charmless vector mesons},
       pdfauthor={R R Horgan, Z Liu, S Meinel, M Wingate},
       pdfpagemode=None,
       bookmarksopen=true]{hyperref}

\usepackage{ifpdf}

\newcommand{\cambridge}{Department of Applied Mathematics and
Theoretical Physics, University of Cambridge, Cambridge CB3 0WA, UK}

\newcommand{\massit}{Center for Theoretical Physics, Massachusetts Institute
  of Technology, Cambridge, MA 02139, USA}

\newcommand{\ihepbeijing}{Institute of High Energy Physics and
  Theoretical Physics Center for Science Facilities, Chinese Academy
  of Sciences, Beijing 100049, China}

\newcommand{\VEC}[1]{{\bf \bm{#1}}} 
\newcommand{\subrm}[1]{{\scriptscriptstyle\mathrm{#1}}}
\def\today{\number\day\space\ifcase\month\or
January\or February\or March\or April\or May\or June\or
July\or August\or September\or October\or November\or December\fi
\space\number\year}
\newcount\mins \newcount\hours
\def\now{\hours=\time \mins=\time
	\divide\hours by60 \multiply\hours by60 \advance\mins by-\hours
	\divide\hours by60 
	\number\hours:\ifnum\mins<10 0\fi\number\mins }

\begin{document}

\title{Lattice QCD calculation of form factors describing the rare decays
$\bm{B\to K^*\ell^+\ell^-}$ and $\bm{B_s\to \phi\,\ell^+\ell^-}$}

\author{Ronald~R.~\surname{Horgan}} 
\affiliation{\cambridge}
\author{Zhaofeng~\surname{Liu}}
\affiliation{\ihepbeijing}
\author{Stefan~\surname{Meinel}}
\affiliation{\massit}
\author{Matthew~\surname{Wingate}\,}
\affiliation{\cambridge}

\preprint{DAMTP-2013-45}

\begin{abstract} 
The rare decays $B^0 \to K^{*0} \mu^+\mu^-$ and $B_s \to \phi
\mu^+\mu^-$ are now being observed with enough precision to test
Standard Model predictions.  A full understanding of these decays
requires accurate determinations of the corresponding hadronic form
factors.  Here we present results of lattice QCD calculations of the
$B\to K^*$ and $B_s\to \phi$ form factors.  We also determine the form
factors relevant for the decays $B_s\to K^* \ell \nu$ and $B_s\to
\bar{K}^{*0} \ell^+\ell^-$.  We use full-QCD configurations including $2+1$
flavors of sea quarks using an improved staggered action, and we
employ lattice non-relativistic QCD to describe the bottom quark.
\end{abstract}

\maketitle


\section{Introduction}
\label{sec:intro}

The weak decay of one flavor of quark to another of the same charge is
relatively rare.  It is much more likely for a bottom quark to decay
to a charm or an up quark than to a strange or a down quark.  For
example, the decay $\bar{B}^0 \to \bar{K}^{*0} \mu^+ \mu^-$ is 100
times rarer than $\bar{B}^0 \to \rho^+ \mu^- \bar{\nu}_\mu$
\cite{Amhis:2012bh,Beringer:2012zz}. In the context of the Standard
Model, this is understood by the absence of flavor-changing neutral
currents (FCNCs) in the Lagrangian; $b\to s$ decays occur only at the
one-loop level.  If the Standard Model is viewed as only the
lowest order of a low-energy effective field theory, an
approximation to a more complete theory ``beyond the Standard Model''
(BSM), then one would expect FCNCs to appear as higher-dimension
operators in the effective Lagrangian.  It is natural to hope that the
loop-suppression of FCNCs in the Standard Model will provide an
opportunity to discover and probe effects due to BSM physics.

The study of bottom quarks decaying to strange quarks is now
experimentally possible and is becoming more precise.  In particular,
the quantity and quality of experimental measurements of exclusive
$b\to s$ decays have increased greatly and will continue to do so as
the LHC experiments analyze their current data and then begin to take
more in the next run.

This paper describes lattice QCD calculations of the form factors
parametrizing hadronic matrix elements governing exclusive
semileptonic and radiative decays of the $B$ and $B_{s}$ mesons to
light vector mesons.  Due to the formulation we use, our results are
most accurate when the final-state meson recoils softly, the so-called
low-recoil or large $q^2$ regime.  Corresponding experimental
measurements have been reported over the past few years, mostly
studying $B\to K^*\ell^+\ell^-$
\cite{Wei:2009zv,Aaltonen:2011cn,Aaij:2012cq,Lees:2012tva,Aaij:2013hha,Aaij:2013iag,ATLAS:2013ola,Aaij:2013qta,Chatrchyan:2013cda,Gorelov:2013kja},
but also $B_s \to \phi\mu^+\mu^-$ \cite{Aaij:2013aln}.  The data are
presently being combined with theoretical and phenomenological
calculations in order to test the Standard Model and to constrain
classes of BSM models
\cite{Bobeth:2010wg,Altmannshofer:2011gn,Straub:2012jb,Altmannshofer:2012he,Hambrock:2012dg}.
The constraints on coefficients in effective Hamiltonians depend on
the certainty with which we know $B \to K^*$ (and related) form
factors.

The full decay $B \to K^*(\to K \pi) \mu^+ \mu^-$ is useful
phenomenologically since a full angular analysis is described by up to
24 observables
\cite{Kruger:1999xa,Faessler:2002ut,Kruger:2005ep,Bobeth:2010wg,Altmannshofer:2011gn,DescotesGenon:2012zf,Descotes-Genon:2013vna}.
Recently, some authors have found significant discrepancies, or
``anomalies'', compared to the Standard Model
\cite{Descotes-Genon:2013wba,Altmannshofer:2013foa}, while others
conclude that the Standard Model is still a good fit to global data
\cite{Bobeth:2012vn,vanDyk:2013uaa,Hambrock:2013zya,Beaujean:2013soa}.
The improvement made here in determining the form factors may aid
future analyses.

The same short-distance physics underlies the decays $B\to K\ell^+\ell^-$
\cite{Wei:2009zv,Aaij:2012cq,Lees:2012tva} and $\Lambda_b \to
\Lambda\ell^+\ell^-$ \cite{Aaltonen:2011qs,AAij:2013hna}.  These are not the
subject of the present calculation, but unquenched LQCD results for the
relevant form factors have recently appeared
\cite{Bouchard:2013mia,Bouchard:2013eph,Detmold:2012vy}.  Comprehensive
analysis of observables in each of these decays may be necessary to obtain a
full picture of BSM contributions.

The $B \to V$ form factors have been computed using lattice QCD,
but only in the quenched approximation
\cite{Bowler:1993rz,Bernard:1993yt,Burford:1995fc,Abada:1995fa,Becirevic:2006nm,Abada:2002ie,Bowler:2004zb}.
The calculation we present here removes this approximation by using
``full QCD'' gauge-field ensembles; the effects of up, down, and strange
sea quarks are included using an improved staggered quark action.
These ensembles were generated and made public by the MILC Collaboration
\cite{Bazavov:2009bb}.  In addition we improve upon previous work by
computing a large statistical sample of correlation functions and by
using nonrelativistic QCD (NRQCD) to treat the $b$ quarks.

In Sec.~\ref{sec:strategy} we review the construction of the $b\to
s$ effective Hamiltonian in order to set the notation and put in context
the present lattice QCD calculation.  Section~\ref{sec:details}
contains the computational details: a description of the correlation
functions from which we determine the form factors, a brief summary of
lattice actions and parameter values, and an overview of the analysis
methods used.  We describe in Sec.~\ref{sec:shape} our fits to the
shape of the form factors taking into account lattice spacing and
quark mass effects, to the extent that these can be seen given the
statistical uncertainties.  Our final results are given in
Sec.~\ref{sec:results} along with discussion of systematic
uncertainties.  Conclusions are given in Sec.~\ref{sec:concl}.  While
the main motivation for this work is the study of $b\to s$ decays, the
form factors describing the $b\to u$ decay $B_s \to K^* \ell\nu$ and
the $b\to d$ decay $B_s \to \bar{K}^{*0} \ell^+\ell^-$ are also
computed, with the results given in the Appendix.

Preliminary form factor results have appeared in several conference
proceedings as we tested formulations and methods for improving the
precision of the numerical data \cite{Liu:2009dj,Liu:2011ra}.  In
another paper we investigate the phenomenological consequences
of the improved form factor determinations for ${B}^0 \to {K}^{*0} \mu^+\mu^-$ 
and ${B}_s \to \phi \mu^+\mu^-$ observables \cite{Horgan:2013pva}.

\section{Theoretical framework}
\label{sec:strategy}

Since the form factors calculated in this paper will be most useful
in studies of $b\to s$ decays, we briefly review the theoretical
framework for describing them. 
At hadronic energies of a few GeV, $b\to s$ decays are governed
by the effective Hamiltonian \cite{Grinstein:1987vj,Grinstein:1990tj,Buras:1993xp,Ciuchini:1993ks,Ciuchini:1993fk,Ciuchini:1994xa}
\begin{equation}
\mathcal{H}_{\mathrm{eff}}^{b\to s} ~=~ -\frac{4 G_F}{\sqrt{2}}
V_{ts}^* V_{tb} \sum_i C_i O_i \,.
\label{eq:Heff}
\end{equation}
In principle, over 20 local operators $O_i$ could appear in the sum in
(\ref{eq:Heff}).  The Wilson coefficients $C_i$ depend on the details
of the high-energy electroweak theory and must be computed within that
theory or determined experimentally.  In the Standard Model,
presently our best candidate theory of weak interactions, the Wilson
coefficients have been calculated to very good accuracy
\cite{Buras:2002tp,Gambino:2003zm,Altmannshofer:2008dz}.  The
operators which dominate short-distance effects in $b\to s \ell\ell$
decays are
\begin{eqnarray}
O_9 &=& \frac{e^2}{16\pi^2}\bar{s} \gamma^\mu P_L b\, \bar\ell \gamma_\mu \ell
\nonumber \\
O_{10} &=& \frac{e^2}{16\pi^2}\bar{s} \gamma^\mu P_L b\, \bar\ell \gamma_\mu 
\gamma^5 \ell
\end{eqnarray}
and the electromagnetic dipole operator
\begin{equation}
O_7 \;=\; \frac{m_b e}{16\pi^2} \bar{s}\sigma^{\mu\nu} P_R b \, F_{\mu\nu} 
\end{equation}
where $P_{L/R} = \tfrac{1}{2}(1 \mp \gamma^5)$ and $\sigma^{\mu\nu} =
\tfrac{i}{2}[\gamma^\mu,\gamma^\nu]$.  Long-distance effects arise from
multiple sources, one of the most important being the
production of charmonium resonances via current-current operators
\begin{eqnarray}
O_1 &=&  \bar{s}^\alpha\gamma^\mu P_L c^\beta \; \bar{c}^\beta\gamma_\mu P_L b^\alpha \nonumber \\
O_2 &=&   \bar{s}^\alpha\gamma^\mu P_L c^\alpha \; \bar{c}^\beta\gamma_\mu P_L b^\beta
\end{eqnarray}
where $\alpha$ and $\beta$ are color indices.  Theoretical and
phenomenological work has been done which suggests long distance
effects could be small if the momentum transferred to the dilepton
pair $\sqrt{q^2}$ is significantly less than
\cite{Khodjamirian:1997tg,Khodjamirian:2010vf} or larger than
\cite{Grinstein:2004vb,Beylich:2011aq} the $J/\psi$ or $\psi'$ masses.
Recently, however, the charmonium resonance $\psi(4160)$ has been seen
in the decay $B^+\to K^+ \mu^+ \mu^-$ with a branching fraction enhanced
by interference effects \cite{Aaij:2013pta}.  The extent to which
resonances above open-charm threshold inhibit studies of short-distance
physics is an open issue requiring further investigation.

The separation between low- and high-energy in (\ref{eq:Heff}) depends
on an energy scale.  The perturbative matching between the effective
Hamiltonian and the Standard Model (or any BSM extension) is done at
$\mu_{\mathrm{match}} = m_W$, then the renormalization group equations
are used to determine the Wilson coefficients at the scale $\mu = m_b$
relevant for the matrix elements of the operators $O_i$
\cite{Altmannshofer:2008dz}.

Traditionally form factors governing the decays of a pseudoscalar
meson to a vector meson (via $b \to q$ currents) are defined through
the following expressions (with momentum transfer $q = p-k$)
\begin{widetext}
\begin{eqnarray}
\langle V(k,\varepsilon) | \bar{q}\gamma^\mu b| B(p)\rangle
&=& \frac{2 i  V(q^2)}{m_B + m_V}
\epsilon^{\mu\nu\rho\sigma} \varepsilon^*_\nu k_\rho p_\sigma \\[4mm]
\langle V(k,\varepsilon)|\bar{q}\gamma^\mu \gamma^5 b| B(p)\rangle
&=&  2 m_V A_0(q^2) \frac{\varepsilon^* \cdot q}{q^2} q^\mu 
 + (m_B + m_V)A_1(q^2)\left(
\varepsilon^{*\mu} - \frac{\varepsilon^* \cdot q}{q^2} q^\mu \right)\nonumber \\ 
&&- \; A_2(q^2) \frac{\varepsilon^*\cdot q}{m_B + m_V}\left[
(p+k)^\mu - \frac{m_B^2 - m_V^2}{q^2} q^\mu\right] \\[4mm]
q^\nu\langle V(k,\varepsilon) | \bar{q}\sigma_{\mu\nu}b| B(p)\rangle
&=&  2 T_1(q^2) \epsilon_{\mu\rho\tau\sigma} \varepsilon^{*\rho}p^\tau k^\sigma 
\\[4mm]
q^\nu\langle V(k,\varepsilon) | 
\bar{q}\sigma_{\mu\nu} \gamma^5 b| B(p)\rangle
& = &  i  T_2(q^2)
[(\varepsilon^* \cdot q)(p+k)_\mu - \varepsilon^*_\mu (m_B^2-m_V^2) 
] \nonumber \\
&&+\; i T_3(q^2)(\varepsilon^*
\cdot q)\left[ \frac{q^2}{m_B^2
-m_V^2}(p+k)_\mu - q_\mu \right]\,.
\end{eqnarray}
Above, $\varepsilon(k,s)$ denotes the polarization vector of the
final-state meson with momentum $k$ and spin polarization $s$. We
compute correlation functions which do not project out definite
polarizations of the final-state vector meson.  The amplitude
we obtain from correlator fits is of the form (with $j = 1,2,3$)
\begin{equation}
\sum_s \varepsilon_j(k,s) \langle V(k,\varepsilon(k,s)) | 
\bar{q} \Gamma b| B(p)\rangle \,.
\end{equation}
As a consequence we find it difficult to directly isolate $A_2$ and $T_3$
form factors.  Instead we obtain results for the form factors
\begin{align}
A_{12}(q^2) \;=\; & \frac{(m_B + m_V)^2(m_B^2 - m_V^2 - q^2) A_1(q^2) -
\lambda A_2(q^2)}{16 m_B m_V^2 (m_B + m_V)} \\
T_{23}(q^2) \;=\; & \frac{m_B+m_V}{8m_B m_V^2}
\left[\left(m_B^2 + 3m_V^2 - q^2\right)T_2(q^2)
- \frac{\lambda T_3(q^2)}{m_B^2 - m_V^2} \right] 
\end{align}
where we have introduced the conventional kinematic variable 
$\lambda = (t_+ - t)(t_- - t)$, with $t_\pm = (m_B \pm m_V)^2$ and $t=q^2$.
Therefore the main results of this paper are determinations of the seven
linearly independent form factors $V$, $A_0$, $A_1$, $A_{12}$, $T_1$,
$T_2$, $T_{23}$.  In addition, we also quote the following linear
combinations, which, together with $A_0$, $A_{12}$, $T_{23}$, form the
helicity basis:
\begin{align}
V_\pm(q^2) \;=\;& \frac12\left[\left(1 + \frac{m_V}{m_B}\right)A_1(q^2)
\mp \frac{\sqrt{\lambda}}{m_B(m_B+m_V)}\,V(q^2)\right] \\
T_\pm(q^2) \;=\;& \frac{1}{2m_B^2}\left[(m_B^2 - m_V^2)T_2(q^2) \,\mp\,
\sqrt{\lambda}T_1(q^2)\right] \,.
\label{eq:helicity_vpm_tpm}
\end{align}
The benefits of using the helicity basis in constraining Standard
Model physics and searching for new physics have been discussed
recently \cite{Feldmann:2012sus,Jaeger:2012uw}.
\end{widetext}

\section{Details of the calculation}
\label{sec:details}

\subsection{Correlation functions}
\label{ssec:corrfn}

We use local interpolating operators $\Phi_B \sim \bar\psi_{q'}
\gamma_5 \Psi_b$ and $\Phi_V \sim \bar\psi_{q'}\gamma_j \psi_q$ 
to annihilate ${B}$ and $V$ mesons, respectively.  
At leading order in $\Lambda_{\subrm{QCD}}/m_b$ in the lattice-to-continuum
matching (see details in Sec.~\ref{ssec:matching}), the renormalized
$b\to q$ currents are
\begin{equation}
\mathcal{J}^A \;=\; Z_{\Gamma^A} \bar\psi_q \Gamma^A \Psi_b
\end{equation}
where $\Gamma^A$ is a $4\times 4$ Dirac matrix.  For later convenience
we use the abbreviated index $A = 0$, $k$, $05$, $k5$, $[0\ell]$, $[k\ell]$,
$[0\ell]5$, $[k\ell]5$ to correspond to $\Gamma^A = \gamma^0$, $\gamma^k$,
$\gamma^0\gamma^5$, $\gamma^k\gamma^5$, $\sigma^{0\ell}$, $\sigma^{k\ell}$,
$\sigma^{0\ell}\gamma^5$, $\sigma^{k\ell}\gamma^5$, respectively, 
where $k,\ell \in [1,2,3]$.  Sometimes we will refer to pairs of
terms using, e.g., $\mu \in [0,1,2,3]$.

With these operators, we compute several correlation functions which
project onto hadrons with specific momenta.  We ultimately extract
form factors from three-point functions of the form
\begin{eqnarray}
C_\mathcal{J}(\VEC{p}, \VEC{k}, \tau, T) &=& \sum_{\VEC{y},\VEC{z}}
\langle \Phi_V(0) \mathcal{J}(y) \Phi_{{B}}^\dagger(z)\rangle \nonumber \\
&&\times \;e^{i \VEC{k}\cdot \VEC{y}- i \VEC{p}\cdot(\VEC{y}-\VEC{z})}
\label{eq:3ptcorrfn}
\end{eqnarray}
where $\tau = |y_0|$ and $T=|z_0|$.  (We suppress Lorentz indices here and
later in this subsection to avoid cluttered expressions.  Generally
there is an index associated with the component of vector meson spin
and one or two more indices due to the vector or tensor operator in the
three-point function.  For simplicity, expressions here and below define 
the origin to coincide with an interpolating operator; in the
computation we place the source location randomly
within a specific time slice.)  We also need the $B$ and $V$ two-point
correlation functions,
\begin{eqnarray}
C_{BB}(\VEC{p}, \tau) &=&  \sum_{\VEC{y}}
\langle \Phi_{{B}}(0) \Phi_{{B}}^\dagger(y)\rangle e^{i \VEC{p}\cdot\VEC{y}}
 \nonumber \\
C_{VV}(\VEC{k}, \tau) &=&  \sum_{\VEC{y}}
\langle \Phi_V(0) \Phi_V^\dagger(y)\rangle e^{i \VEC{k}\cdot\VEC{y}}
\label{eq:2ptcorrfn}
\end{eqnarray}
in order to divide the three-point functions (\ref{eq:3ptcorrfn}) by factors
associated with the interpolating operators.  In the limit of large
Euclidean-time separations between the meson interpolating operators
and the current insertion, only the lowest-energy states contribute to
the correlation functions,
\begin{eqnarray}
C_\mathcal{J}(\VEC{p}, \VEC{k}, \tau, T) &\to & A^{(\mathcal{J})}e^{-E_V\tau}
e^{-E_B^{\mathrm{sim}}(T-\tau)} 
\nonumber \\
C_{BB}(\VEC{p}, \tau) &\to & A^{(BB)} e^{-E_B^{\mathrm{sim}}\tau} \nonumber \\
C_{VV}(\VEC{k}, \tau) &\to & A^{(VV)} e^{-E_V\tau} \,.
\label{eq:BVcorr}
\end{eqnarray}
Since we use NRQCD for the heavy quark, the energy appearing in the
heavy-meson correlation functions, $E_B^{\mathrm{sim}}$, contains an 
energy shift, as we explain further in Sec.~\ref{ssec:configs}.

From the ground-state amplitude we obtain the matrix elements necessary for
computing the form factors,
\begin{equation}
A^{(\mathcal{J})}_j ~=~ \frac{\sqrt{\Xi_V \Xi_B}}{4E_V E_B}
\sum_s \varepsilon_j(k,s)\langle V(k,\varepsilon)|\mathcal{J}|{B}(p)\rangle \,.
\end{equation}
The coefficients are extracted from the ground-state amplitude of
two-point correlation functions $A^{(BB)} = \Xi_B/(2E_B)$ and 
$A^{(VV)}_j = \Xi_V/(2E_V)\sum_s\varepsilon^*_j(k,s)\varepsilon_j(k,s)$.
For convenience later, let us denote the matrix elements we extract
by
\begin{equation}
\mathcal{M}_j^A \;=\; \sum_s\varepsilon_j(k,s)\langle V(k,\varepsilon) | 
 \mathcal{J}^A|{B}(p)\rangle \,.
\label{eq:matrixels}
\end{equation}

In the $B$ rest frame the kinematic variable $\lambda$ is equal to $4
m_B^2 |\VEC{k}|^2$ and the longitudinal polarization of the current
is given by $\varepsilon_0(q^2) = (|\VEC{q}|, q^0
\VEC{q}/|\VEC{q}|)/\sqrt{q^2}$.  We obtain the form factors
from the matrix elements (\ref{eq:matrixels}) using the following
relations ($j$ is not summed over in the formulae below, although we
do average the data over all equivalent directions):
\begin{align}
V &\;=\; \frac{ i (m_B+m_V)}{2m_B}\, (\epsilon^{0\mu j \rho} k_\rho)^{-1} 
\mathcal{M}_j^{\mu} \;\mbox{(no $\mu$ sum)}
\label{eq:Vformula} \\
A_0 &\;=\; -\frac{m_V}{2 k_j m_B E_V} \, q_\mu  \mathcal{M}_j^{\mu 5} \\
A_1 &\;=\; -\frac{1}{m_B+m_V} \, \mathcal{M}_j^{j 5} ~~\mbox{for}~k_j=0  \\
A_{12} &\;=\; -\frac{\sqrt{q^2}}{8m_B} \frac{|\VEC{k}|}{k_j E_V}\,
\varepsilon^*_{0,\mu}(q) \, \mathcal{M}_j^{\mu 5} \\
T_1 &\;=\; -\frac{1}{2m_B \epsilon^{0\mu j \rho} k_\rho} \, q_\nu 
\mathcal{M}_j^{[\mu\nu]} ~\mbox{(no $\mu$ sum)}\\
T_2 &\;=\; -\frac{i}{m_B^2 - m_V^2} \, q_\nu \mathcal{M}_j^{[j\nu]5} \\
T_{23} & \;=\; -\frac{i m_V(m_B+m_V)}{4 E_V k_j m_B }\,
\varepsilon^*_{0,\mu}(q) \,q_\nu \mathcal{M}_j^{[\mu\nu]5}  \,.
\label{eq:T23formula}
\end{align}

\subsection{Lattice actions}
\label{ssec:configs}

\begin{table}
  \caption{\label{tab:gaugeparam}Parameters of the MILC 2+1 AsqTad gauge
    field configurations used in this work. $r_1/a$ values come from 
    Ref.\ \cite{Bazavov:2009bb}. We take $r_1 = 0.3133(23)$ fm from 
    Ref.~\cite{Davies:2009tsa}.}
\begin{center}
\begin{tabular}{cccccc}
  \hline \hline
  Ens. & \# & $N_x^3\times N_t$ &
  $u_P a m^{\mathrm{sea}}_\ell/u_P a m^{\mathrm{sea}}_s$ & $r_1/a$ & $a^{-1}$(GeV)  \\ \hline
  c007 & 2109 & $20^3\times 64$ & $0.007/0.05$ & $2.625(3)$ & $1.660(12)$ \\
  c02 & 2052 & $20^3\times 64$ & $0.02/0.05$ & $2.644(3)$ & $1.665(12)$ \\
  f0062 & 1910 & $28^3\times 96$ & $0.0062/0.031$ & $3.699(3)$ & $2.330(17)$ \\ 
  \hline \hline
\end{tabular}
\end{center}
\end{table}

We used a subset of the MILC collaboration gauge-field configurations 
\cite{Aubin:2004wf,Bazavov:2009bb}.
These lattice ensembles were generated using the Symanzik-improved 
gauge action (with coefficients determined through $O(\alpha_s)$)
\cite{Luscher:1985zq,Luscher:1984xn}. Effects due to $2+1$ flavors 
of dynamical fermions were included using the $O(a^2)$ tadpole-improved 
(AsqTad) staggered quark action 
\cite{Naik:1989bn,Orginos:1998ue,Lepage:1998vj,Orginos:1999cr,Bernard:1999xx}.
The fourth-root procedure was used to account for the multiple
tastes present in staggered fermion formulations 
(e.g.\ see \cite{Sharpe:2006re,Kronfeld:2007ek}).

\begin{table}
\caption{\label{tab:valparams}Valence quark parameters, including
the fourth-root of the plaquette, $u_P$, and the mean Landau-gauge link
$u_L$.}
\begin{center}
\begin{tabular}{ccccccc}
\hline \hline
Ensemble & \# & $u_P a m^{\mathrm{val}}_\ell /u_P am^{\mathrm{val}}_s$ & $u_P$ 
& $a m_b$ & $n$ & $u_L$ \\ \hline
c007 & 16872 &  $0.007/0.04$ & $0.8678$ & $2.8$ & $2$ & $0.836$ \\
c02 & 16416 &  $0.02/0.04$ & $0.8678$ & $2.8$ & $2$ & $0.837$ \\ 
f0062 & 15280 & $0.0062/0.031$ & $0.8782$ & $1.95$ & $2$ & $0.8541$ \\
\hline \hline
\end{tabular}
\end{center}
\end{table}

We chose the subset listed in Table~\ref{tab:gaugeparam} in order to
vary both the up or down sea quark mass $m_\ell^{\mathrm{sea}}$ and the
lattice spacing $a$.  We chose two ensembles (c007 and c02) with a
common, coarse lattice spacing on which to test quark mass dependence
and one ensemble (f0062) with a fine lattice spacing which has
approximately the same Goldstone pion mass as on the c007 ensemble.  A
calculation of $B\to\pi\ell\nu$ form factors on a similar subset of
MILC lattices \cite{Gulez:2006dt} found very mild quark mass
dependence and no statistically significant dependence on the lattice
spacing.  Since the signal-to-noise ratio is much worse for
correlation functions involving vector mesons in place of pseudoscalar
mesons, we chose to invest our computational effort in obtaining a large
statistical sample on these three ensembles rather than including more
ensembles.  As will be shown in Sec.~\ref{sec:results}, this set of
configurations is sufficient given the other sources of uncertainties.

We use the same action (AsqTad) for the light and strange valence
quarks as was used in the configuration generation.  After inverting
the staggered Dirac operator, we convert the staggered fields to
four-component ``naive'' fields for use in the interpolating operators
and currents \cite{Wingate:2002fh}. On each configuration we computed
eight light and strange quark propagators yielding more than 15000
measurements on each ensemble.  Precise figures are given in
Table~\ref{tab:valparams}.  (In fact, we compute correlation functions
forward and backward in Euclidean time and average the results
together. Counting these as independent would double the number of
measurements quoted.)  The eight point sources are evenly distributed on four
time slices with a random offset for the locations on each configuration
in order to reduce correlations.

For the heavy quark, we use lattice NRQCD \cite{Lepage:1992tx}.  The specific
form of the action is the same $O(v^4)$ action as was used in earlier work by
the HPQCD collaboration (e.g.\ \cite{Gulez:2006dt}).  Because we make use of
an effective field theory to treat the $b$ quark, the net energy of a $B$
meson is obtained by adding a contribution associated with the $b$ quark mass
to the energy of the $B$ meson in the Monte Carlo calculation
$E_{\mathrm{sim}}$.  For a $B$ meson with spatial momentum $\VEC{p}$ relative
to the lattice rest frame,
\begin{equation}
aE(\VEC{p}) \;=\; aE_{\mathrm{sim}}(\VEC{p}) + C_v \,.
\end{equation}
The additional term is renormalized by interactions:
\begin{equation}
C_v \;=\; Z_m  \,am_b + aE_0\,.
\end{equation}
(At tree level, $Z_m=1$ and $E_0=0$.)
The multiplicative and additive renormalization constants have been
computed perturbatively \cite{Gulez:2003uf}; however, we can determine
them nonperturbatively from Monte Carlo calculations of hadron dispersion 
relations using \cite{Foley:2004rp}
\begin{equation}
C_v \;=\; \frac{a^2\VEC{p}^2 - a^2[E_{\mathrm{sim}}^2(\VEC{p}) -
    E_{\mathrm{sim}}^2(0)]}{2n_Q a [E_{\mathrm{sim}}(\VEC{p}) -
    E_{\mathrm{sim}}(0)]}
\label{eq:CvNP}
\end{equation}
where $n_Q$ is the number of heavy quarks in the hadron.  We
spin-average $C_v$ over $\eta_b(1S)$ and $\Upsilon(1S)$ states with
momentum $|\VEC{p}| = 2\pi/(a N_x)$.  We find consistent results if we
use $|\VEC{p}| = 4\pi/(a N_x)$, and both agree with the perturbative
determination.  Within the $0.15\%$ statistical uncertainties, we find
no dependence on the sea quark mass.  Central values for the coarse
and fine lattices are given in Table~\ref{tab:renormparam}.

In Table~\ref{tab:valparams} we also give the
tadpole improvement parameters $u_P$, determined from the fourth root
of the mean plaquette, and $u_L$, determined from the Landau-gauge
mean link; these values are used in the AsqTad and NRQCD
actions, respectively.  In the table, $n$ denotes the NRQCD stability
parameter.

In this work we consider only correlation functions with the $B$ meson
at rest in the lattice frame.  We investigated the use of moving NRQCD
to extend and improve the kinematic range of the calculation
\cite{Horgan:2009ti}, but we concluded that it was more expedient to
concentrate on a high-statistics study with $\VEC{p}=0$. We also
investigated the use of stochastic sources to improve the precision of
correlation functions \cite{Davies:2007vb}.  For vector meson final
states we found it would be more efficient to use many local sources,
which could be used for any final state momentum, instead of using
many stochastic sources, each of which would correspond to a distinct
$\VEC{k}$ \cite{Liu:2009dj}.  In order to improve the statistical
signal for the $B$ meson two-point function we perform a $2 \times 2$
matrix fit to correlators obtained with both local and smeared sources
and sinks.

\subsection{Operator matching}
\label{ssec:matching}

\begin{table}
  \caption{\label{tab:renormparam}Heavy quark and heavy-light current 
    renormalization constants (for the parameters as in 
    Table~\ref{tab:valparams}) \cite{Gulez:2003uf,Gulez:2006dt,Muller:2010kb}. 
  For the tensor current matching, the matching scale is taken to be $m_b$.}
\begin{center}
\begin{tabular}{crr}
\hline\hline
 & \multicolumn{1}{c}{Coarse}  & \multicolumn{1}{c}{Fine} 
\\ \hline
$C_v$ & $2.825$ & $1.996$ \\
$\rho^{(0)}$ & $0.043$ & $-0.058$ \\
$\zeta_{10}^{(0)}$ & $-0.166$ & $-0.218$ \\
$\rho^{(k)}$ & $0.270$ & $0.332$  \\
$\zeta_{10}^{(k)}$ & $0.055$ & $0.073$ \\
$\rho^{([0\ell])}$  & $0.076$  & $0.320$ \\
$\zeta_{10}^{([0\ell])}$ & $-0.055$ & $-0.073$ \\
$\rho^{([k\ell])}$ & $0.076$ & $0.320$ \\
$\zeta_{10}^{([k\ell])}$ & $-0.055$ & $-0.073$ \\
\hline \hline
\end{tabular}
\end{center}
\end{table}

We must match the currents involving NRQCD $b$ quarks and
naive/staggered light quarks to the continuum currents of
interest. The matching of the leading-order currents is such that
\begin{equation}
(\bar{q} \Gamma^A b)|_{\mathrm{cont}} ~\doteq~ \mathcal{J}^A \;=\;
Z_{\Gamma^A} (\bar{\psi}_q \Gamma^A \Psi_b)|_{\mathrm{latt}}
\label{eq:LOmatch}
\end{equation}
where the $\doteq$ symbol means that the operators on either side of the
relation have the same matrix elements up to the stated accuracy.  For the
temporal ($\mu = 0$) and spatial ($\mu = k$) components of the vector $\Gamma^A
= \gamma^\mu$ and axial vector currents $\Gamma^A = \gamma^\mu\gamma^5$, we write
\begin{equation}
Z_{\gamma^\mu} \;=\; Z_{\gamma^\mu\gamma^5} \;=\; 1 + \alpha_s \rho^{(\mu)} \,,
\end{equation}
where $\rho^{(0)} \ne \rho^{(k)}$ because we use the NRQCD action.
(The remnant chiral symmetry of staggered fermions assures the first
equality.) The tensor matching coefficients, i.e.\ for $\Gamma^A =
\sigma^{\mu\nu}$ and $\Gamma^A = \sigma^{\mu\nu}\gamma^5$, are defined through
\begin{equation}
Z_{\sigma^{\mu\nu}} \;=\; Z_{\sigma^{\mu\nu}\gamma^5} \;=\; 
1 + \alpha_s \rho^{([\mu\nu])} \,.
\end{equation}
The tensor current is not conserved; it runs logarithmically with a scale
$\mu$.  This scale dependence is implicitly included in the coefficient
$\rho^{([\mu\nu])}$ \cite{Muller:2010kb}.  

Higher dimension operators must be included at next-to-leading order
in the heavy-quark expansion.  Denoting the leading-order currents by
$J_0^A = (\bar{\psi}_q\Gamma^A \Psi_b)|_{\mathrm{latt}}$, we also
compute matrix elements of the dimension-4 operators $J_1^A =
-\frac{1}{2m_b}(\bar{\psi}_q\Gamma^A \bm{\gamma}\cdot \nabla
\Psi_b)|_{\mathrm{latt}}$.  The NLO matching reads
\begin{equation}
\mathcal{J}^A \;=\; 
Z_{\Gamma^A} J_0^A + J_1^A - \alpha_s \zeta_{10}^{(A)} J_0^A
\label{eq:NLOmatch}
\end{equation}
The last term in (\ref{eq:NLOmatch}) accounts for the fact that matrix
elements of $J_1^A$ include not only the nonperturbative NLO
corrections of order $\Lambda_{\subrm{QCD}}/m_b$ but also a
perturbative mixing-down with $J_0^A$ of order $1/am_b$.  The
matching (\ref{eq:NLOmatch}) neglects corrections of order $\alpha_s
\Lambda_{\subrm{QCD}}/m_b$ and of order $\alpha_s^2$.  Results for
$\rho^{(0)}$ \cite{Gulez:2003uf}, $\rho^{(k)}$ \cite{Gulez:2006dt},
$\rho^{([\mu\nu])}$ \cite{Muller:2010kb} are reproduced in
Table~\ref{tab:renormparam}, as are the mixing coefficients
$\zeta_{10}^{(A)}$, provided by private communication from
E.~M\"{u}ller.

When we determine the currents to leading order in
$O(\Lambda_{\subrm{QCD}}/m_b)$ (\ref{eq:LOmatch}) or
next-to-leading order (\ref{eq:NLOmatch}), we perform the matching at a scale
$q^* = 2/a$ with the motivation that the truncated terms can be
minimized by such a choice \cite{Lepage:1992xa}.  Taking
$\alpha_V^{(3)}(\mathrm{7.5~GeV}) = 0.21$ \cite{Mason:2005zx} and
running to the lower scale $2/a$ gives $\alpha_V$ of 0.30 and 0.24 on
the coarse and fine lattices, respectively -- these are the values we
used in the matching.  Instead if we had chosen $q^* = 3/a$, we would have
used $\alpha_V = 0.24$ and $0.22$ for coarse and fine lattices.
Using the coefficients in  Table~\ref{tab:renormparam},
the variation in the matching due to $q^*$ uncertainty is largest for
the spatial components of vector and axial vector currents, and is
approximately 1\%--2\%.  This scale ambiguity is compatible with and
smaller than the net $O(\alpha_s^2)$ uncertainty we estimate another 
way below. 

Since the vector and axial-vector currents are conserved (or partially
conserved) the procedure described so far completes the matching in
these cases. On the other hand the tensor currents must be matched to
the scheme and scale used to compute the Standard Model Wilson
coefficients, the $\overline{\mathrm{MS}}$ scheme at the scale $m_b$.
The matching coefficients in Table~\ref{tab:renormparam} are already
given for $\mu = m_b$.  A subtlety lies in the choice of coupling
constant and its scale.  One choice we could make would be to perform
the matching consistently using this scale; i.e., we would use
$\alpha_{\overline{\subrm{MS}}}(m_b) = 0.21$ in (\ref{eq:LOmatch}) and
(\ref{eq:NLOmatch}).  However past experience in evaluating $q^*$ 
using the BLM procedure suggests this scale is
higher than optimal and would lead to enhanced $\alpha_s^2$
corrections.  Instead we use the same values for $\alpha_V(q^*)$ as
for the vector and axial-vector current matching.  This means that we
truncate terms like $\alpha_s^2 \log(q^*/m_b)$.  Nevertheless,
choosing $q^*=2/a$ instead of $m_b$ as the matching scale should minimize
the overall $O(\alpha_s^2)$ contribution, including those terms.  To
estimate the systematic uncertainty due to choice of scheme and scale
for $\alpha_s$ in the 1-loop matching, we note that $\delta \alpha_s =
\alpha_V(q^*) - \alpha_{\overline{\subrm{MS}}}(m_b)$ is 0.09 and 0.03
for the coarse and the fine lattice spacings, respectively.
Multiplying $\delta \alpha_s$ by the tensor current renormalization
constants in Table~\ref{tab:renormparam} indicates that this ambiguity in
$\alpha_s$ has about a 1\% effect; again this is compatible with and
smaller than the net $\alpha_s^2$-truncation error estimated as
follows.
 
We assume the missing $O(\alpha_s^2)$ contributions in the matching
to have coefficients of approximately the same order as the generic
1-loop coefficients; the largest of these in Table~\ref{tab:renormparam}
is $\rho^{(k)}$.  Therefore we estimate the $\alpha_s^2$ corrections
to be suppressed by a factor of $\alpha_s$ compared to the $\alpha_s$
terms, possibly with a coefficient up to 2 times $\rho^{(k)}$.  This
yields an estimate of 4\% for the total uncertainty due to higher-loop
contributions or ambiguities in the current matching.  This is the
dominant systematic uncertainty in the form factor calculation
(see Sec.~\ref{sec:results} for further discussion).

\subsection{Data analysis details}
\label{ssec:analysis}

\begin{table*}[ht]
\caption{\label{tab:MesonMasses}Meson masses (statistical uncertainties only).
  Physical values, given for reference, neglect isospin splittings as
  we do in the Monte Carlo computations. Isospin-breaking effects in
  the form factors are negligible at the present level of precision.
  The $\eta_s$ is a fictional, pure $\bar{s}s$ pseudoscalar meson
  whose ``physical'' mass is defined using chiral perturbation theory
  and lattice data \cite{Sharpe:2000bc,Davies:2009tsa}.}
\begin{center}
\begin{tabular}{ccccccccc}
\hline \hline
Ensemble & $m_B$ (GeV) & $m_{B_s}$ (GeV) & $m_\pi$ (MeV) & $m_K$ (MeV) &
$m_{\eta_s}$ (MeV) & $m_\rho$ (MeV) & $m_{K^*}$ (MeV) & $m_\phi$ (MeV) \\ \hline
c007 & 5.5439(32) & 5.6233(7)  & 313.4(1) & 563.1(1) & 731.9(1) 
& 892(28) & 1045(6) & 1142(3) \\
c02 & 5.5903(44) & 5.6344(15) & 519.2(1) & 633.4(1) & 730.6(1) & 1050(7) &
1106(4) & 1162(3) \\ 
f0062 & 5.5785(22) & 5.6629(13) & 344.3(1) & 589.3(2) & 762.0(1) & 971(7) &
1035(4) & 1134(2) \\
physical & 5.279 & 5.366 & 140 & 495 & 686 & 775 & 892 & 1020
\\ \hline \hline
\end{tabular}
\end{center}
\end{table*}

One source of systematic uncertainty that affects the lattice
determination of any matrix element or energy is contributions to the 
correlation functions from excited states; the interpolating
operators create or annihilate all states with the corresponding quantum
numbers.  There is a trade-off between statistical error, which grows
as time separations are increased, and systematic error, if the time
separations are small enough that excited states contribute.  The
fitting of correlation functions which use naive or staggered fermions
is further complicated by the contributions of opposite-parity states
which give subdominant but non-negligible additive contributions of the
form $A^{\mathrm{osc}} (-1)^{\tau/a} \exp(-E^{\mathrm{osc}}\tau)$ to the
correlation functions (\ref{eq:BVcorr}).  We carried out two separate
analyses with different approaches in order to address these issues.

\subsubsection{Frequentist fits}
\label{sssec:freqfit}

In the frequentist approach, we restrict our fits to two exponentials
(one non-oscillating and one oscillating) for each propagating meson in
the correlation functions.  Thus, simultaneous fits to the three
correlators (\ref{eq:BVcorr}) involve 14 parameters: the 
energies $E_B^{\mathrm{sim}}$, $E_B^{\mathrm{osc}}$, $E_V$, $E_V^{\mathrm{osc}}$;
as well as two amplitudes for $C_{VV}$: $A_V$ and $A_V^{\mathrm{osc}}$; 
four amplitudes for the matrix fits to smeared and local correlators
$C_{BB}$; and four amplitudes $A_{ee}$, $A_{eo}$, $A_{oe}$, $A_{oo}$ for the
particular $B\to V$ three-point function $C_{\mathcal{J}}$.

We improve the precision of the fit results by including more precise
correlators which involve a zero-momentum pseudoscalar meson $P(\VEC{k}=0)$.
The $B\to P$ three-point function
\begin{equation}
C_{P,\gamma^0}(\VEC{0},\VEC{0},\tau,T) \;=\; \sum_{\VEC{y},\VEC{z}}
\langle \Phi_P(0) [\bar{\psi}_q \gamma^0 \Psi_b](y) \Phi_b^\dagger(z)
\rangle
\end{equation}
also depends on the energies $E_B^{\mathrm{sim}}$ and $E_B^{\mathrm{osc}}$.
Including this precise data in the simultaneous fit further constrains
those energies and allows a more stable determination of the other
12 fit parameters, even at the expense of introducing new parameters
to fit $C_{P,\gamma^0}$: $A_{Pee}$, $A_{Peo}$ and $E_P$ (there is no 
oscillating contribution from $P(\VEC{k}=0)$).  In fact a further
improvement is made by including the two-point function
\begin{equation}
C_{PP}(\VEC{0},\tau) \;=\; \sum_{\VEC{y}}\langle
\Phi_P(0)\Phi_P^\dagger(y) \rangle
\end{equation}
in order to further constrain $E_P$ with precise numerical data.

The $\chi^2$ and $Q$ statistics are used to judge goodness of fit and
correspondingly decide whether excited states contribute to the
numerical data being fit.
The goal is to find optimal values of cutoff 
separations $\{\tau_{\mathrm{min}}\}$ between meson sources and sinks 
while not discarding the most precise data.
With five correlators being fit for each combination of lattice spacing,
quark mass, and final state momentum, it is not practical to examine
every combination of $\{\tau_{\mathrm{min}}\}$ for each fit.  Therefore
we randomly sample the space of fit ranges.  For each source or sink,
we propose a range of reasonable values for $\tau_{\mathrm{min}}$, using
the whole set of correlation functions, with $T/a \in [11,26]$ on 
the coarse ensembles and $T/a \in [15,36]$ on the fine ensemble. We
randomly select among those ranges 500 sets of
$\tau_{\mathrm{min}}$ values which are then used in 500 fits for each
$C_{\mathcal{J}}$.  The results of those fits are ranked to find the most precise
fits which have a $Q$ value higher than 10\% of the maximal $Q$.
(It is not sufficient to choose the fit with the highest $Q$ since
this is usually the result of discarding all but the noisiest data.)

In order to incorporate the uncertainty in choosing among the top five or so
acceptable fit ranges, we vary these ranges as we perform a second set
of bootstrap fits.  These bootstrap fits are necessary to propagate
uncertainties taking into account correlations due to using the same
quark propagators to construct all the correlation functions.

\subsubsection{Bayesian fits}
\label{sssec:bayesfit}

Our Bayesian approach to fitting correlation functions follows
Refs.~\cite{Lepage:2001ym,Wingate:2002fh}.  The number of exponentials
included in the fit functions is increased so that we can fit data
closer to the meson sources and sinks.  Below we will label the number
of pure exponentials by $N$ and the number of oscillating exponentials
(those with a prefactor $(-1)^{\tau/a}$) by $\tilde{N}$.  Gaussian priors
are introduced in order to constrain those fit parameters which are
unconstrained by the numerical data.

For the two-point functions, we first perform a fit with
$N=1$ and $\tilde{N}=1$ (or 0) and a reasonably large
$\tau_{\mathrm{min}}/a$, as in Sec.~\ref{sssec:freqfit}.  We take the 
parameters from this fit as the mean values for the corresponding
Gaussian priors used in the multi-exponential fits (where we set 
$\tau_{\mathrm{min}}/a$ to be smaller).  The widths of the
priors are taken to be 5--10 times the uncertainties of the single-exponential
fit.  For the excited state exponentials, we use the logarithms of
energy differences as fit parameters in order to fix the ordering of
the states.  The priors for these parameters are typically set to have
mean $-1$ and width $1$.  

The three-point functions are fit simultaneously with the
corresponding two-point functions.  When $N$ and $\tilde{N}$ reach
4 or larger, the fit results for the ground state energies and
amplitudes stabilize.

In the Bayesian fits we do not typically fit to the data with all
values of $T/a$.  As more $T$ are included in the Bayesian fits, it
takes tremendous time to finish the bootstrap process (with 200
bootstrap samples, for example). On the other hand, the fit results
stabilize if three or more values of $T$ are included.  Therefore, we
typically use three to seven different $T$ values in our bootstrap analysis of
the simultaneous Bayesian fits. The range of $\tau$ is usually between
$\tau_{\mathrm{min}}/a=2$ and $\tau_{\mathrm{max}}/a=T/a-2$.  For the
light-light two-point functions, the fitted $\tau$ values are usually
between $\tau_{\mathrm{min}}/a = 2(4)$ and $\tau_{\mathrm{max}}/a = N_t-2(4)$
for the coarse (fine) lattice. For the heavy-light two-point function, we
use the data between $\tau_{\mathrm{min}}/a=4(6)$ and 
$\tau_{\mathrm{max}}/a=31(47)$ on the coarse (fine) lattice.

\subsubsection{Fit results}
\label{sssec:fitresults}

\begin{figure*}[t]
\begin{center}
\includegraphics[width=1.0\textwidth]{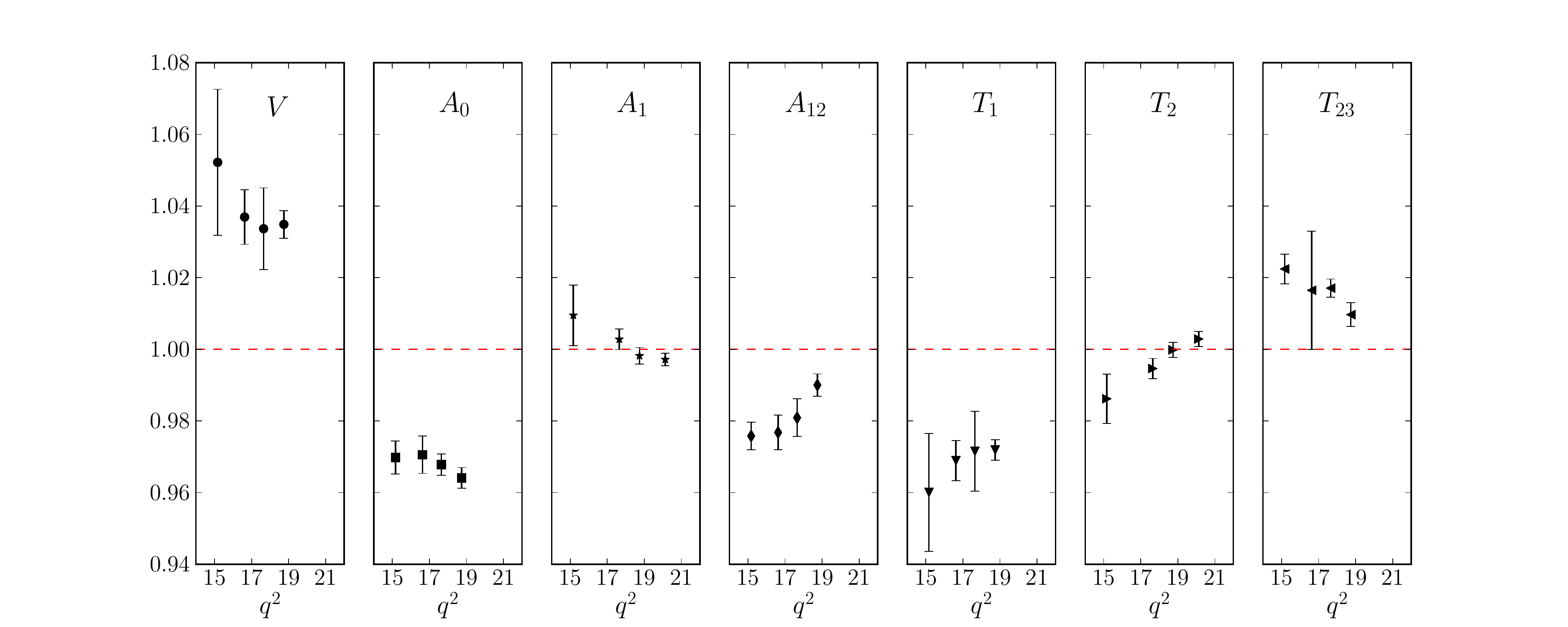}
\includegraphics[width=1.0\textwidth]{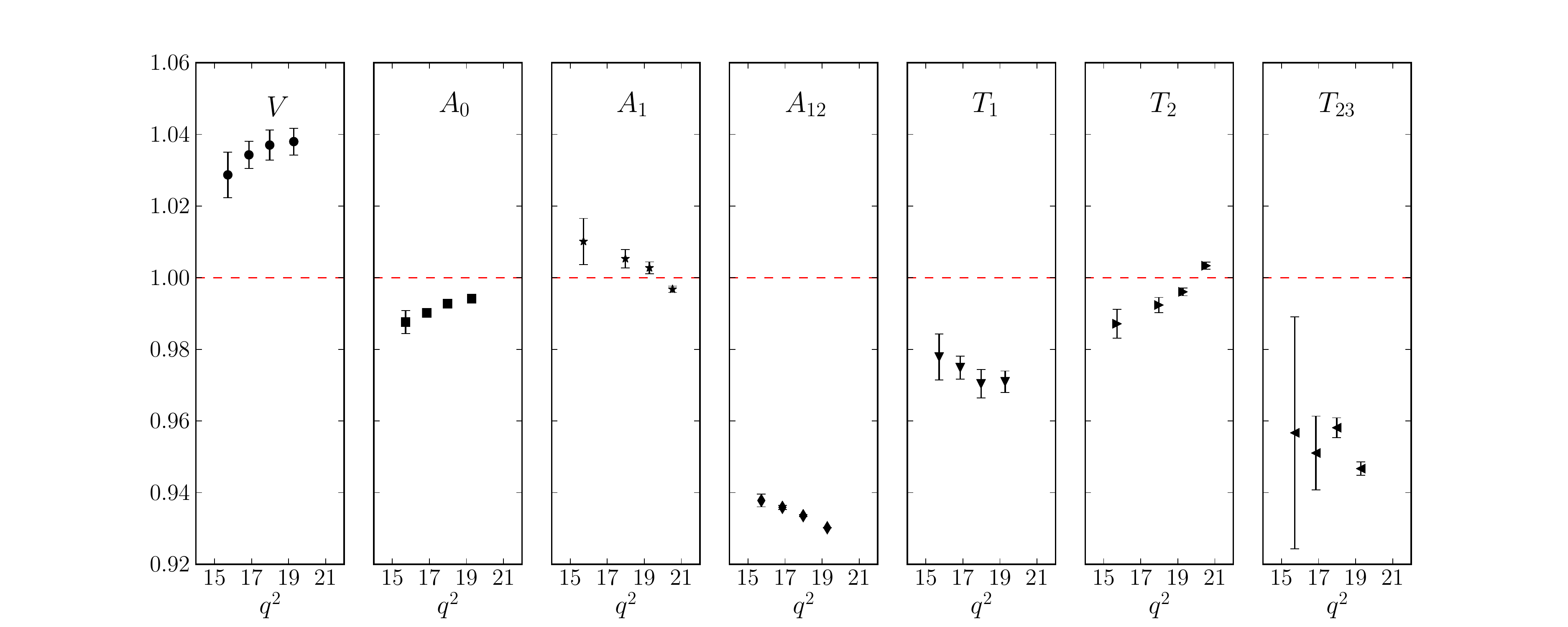}
\end{center}
\caption{\label{fig:NLOoverLO}Ratios of $B_s\to \phi$ matrix elements
  computed with matching done through next-to-leading order in
  $\Lambda_{\subrm{QCD}}/m_b$ (numerator) vs.\ leading order
  (denominator). The top figure is for the c007 lattice and the bottom
  for f0062.}
\end{figure*}

The comparison between the frequentist and Bayesian analyses described
above produced results which agree or differ within 1--2 times the
statistical and fitting uncertainties.  Where such differences appear,
it is usually the case that the energies determined by the frequentist
fits were about 1$\sigma$ lower than those from the Bayesian fits.
This suggests that the Bayesian fits, which were restricted to more
limited set of $T$ values, may have been more subject to excited state
contamination.  Given the flexibility of the frequentist fits to explore a 
wider variety of fit ranges in Euclidean time we take these as our main
results, with the Bayesian fits giving important cross-checks.

In Table~\ref{tab:MesonMasses} we give the meson masses 
resulting from fits to correlation functions as described
above.  Where comparisons can be made, these agree with calculations
done by other groups on the same lattices with the same parameters
\cite{Aubin:2004wf}.  The masses indicate that the valence strange and
bottom quark masses are not precisely tuned to their physical values.
We estimate the resulting systematic errors in Sec.~\ref{sec:results}.

In Fig.~\ref{fig:NLOoverLO} we compare the form factors computed with
currents matched at leading order in $\Lambda_{\subrm{QCD}}/m_b$ (LO)
to those which include next-to-leading order (NLO)
$\Lambda_{\subrm{QCD}}/m_b$-corrections in the currents.  (Both LO and NLO form
factors are matched at 1-loop in $\alpha_s$.)  The plots are shown for
$B_s \to \phi$ decays since the statistical errors are smallest, but
the ratios are of comparable size for $B \to K^*$ and $B_s\to K^*$
form factors, namely at most 7\% away from unity.  The statistical
errors in the ratio are very small because the ratio can be taken for
each bootstrap sample individually, taking correlations into account.
The significance of including NLO operators is reduced when
considering the absolute values of the form factors and the results of
fits to the form factor shapes as discussed in the next Section.  Most
results differ at or below the $1\sigma$ statistical-plus-fitting
uncertainty.  Nevertheless we take the NLO-matched fits as our final
results, so the largest terms truncated from the matching are
$O(\alpha_s^2, \alpha_s \Lambda_{\subrm{QCD}}/m_b,
(\Lambda_{\subrm{QCD}}/m_b)^2)$.

In the Appendix we provide tables of form factor results on each
lattice for several final-state momenta.  We also tabulate
corresponding values for useful kinematic variables.  In general these
raw lattice results still need to account for dependence on light
quark mass.  In the next section we describe our fits to the kinematic
shape and quark mass dependence of the lattice data.

\section{Form factor shape}
\label{sec:shape}

Since exclusive semileptonic branching fractions can precisely
determine CKM matrix elements, there is a sizeable body of work discussing
accurate parametrizations for form factor shapes
\cite{Bourrely:1980gp,Boyd:1994tt,Lellouch:1995yv,Becirevic:1997vw,Boyd:1997qw,Mannel:1998kp,Becirevic:1999kt,Flynn:2000gd,Ball:2004rg,Hill:2005ju,Becher:2005bg,Arnesen:2005ez,Flynn:2006vr,Flynn:2007qd,Bourrely:2008za,Flynn:2008zr,Bernard:2009ke,Bharucha:2010im}.  The method we use here is based on the simplified series expansion
\cite{Bourrely:2008za}, modified to account for lattice spacing and 
quark mass dependence \cite{Na:2010uf}.

Using $t = q^2$ and $t_\pm = (m_{B_{(s)}} \pm m_V)^2$, one constructs a dimensionless
variable which is small,
\begin{equation}
z(t,t_0) ~=~ \frac{\sqrt{t_+-t} - \sqrt{t_+-t_0}}{\sqrt{t_+-t} +
  \sqrt{t_+-t_0}} \,.
\end{equation}
The $t_0$ parameter simply shifts the origin and can be chosen to
minimize $|z|$ over the $q^2$ range of interest.  For simplicity we
use $t_0 = 12~\mathrm{GeV}^2$ throughout this paper.  One might try to
optimize the choice of $t_0$ to make the series expansion in $z$
converge most quickly \cite{Bharucha:2010im}; however, 
we see no discernible in our final results if $t_0$ is varied by
several GeV${}^2$.  After removing any poles due to bound-state
resonances, the form factors are represented by a power series in $z$.
One can introduce additional coefficients to account for lattice
spacing and mass dependence \cite{Na:2010uf}.  Therefore, we fit the
form factors $F = V, A_0, A_1, A_{12}, T_1, T_2, T_{23}$ to the
following form:
\begin{equation}
F(t) ~=~ \frac{1}{P(t;\Delta m)}[1 + b_1 (a E_F)^2 + \ldots]\sum_n a_n d_n z^n
\label{eq:fit_gen_form}
\end{equation}
where the pole factor is given as
\begin{equation}
P(t;\Delta m) ~=~ 1 - \frac{t}{(m_{B_{(s)}} + \Delta m)^2} \,.
\end{equation}
Changing the numerical value of $t_0$ by a few GeV${}^2$ simply
results in a compensating shift in the $a_n$, without significantly
affecting the values of the fit function $F(t)$.

\begin{table}
\caption{\label{tab:dmres} Mass differences (in MeV), between the
initial state and pertinent resonance, used in the function $P(t,\Delta m)$.}
\begin{center}
\begin{tabular}{cccc} \hline \hline
Form factor & $B \to K^*$ & $B_s \to \phi$ & $B_s \to K^*$ \\ \hline
$A_0$ &  87 & 0 & $-87$ \\
$V$, $T_1$ &  135 & 45 & $-42$ \\
$A_1$, $A_{12}$, $T_2$, $T_{23}$ & 550 & 440 & 350 \\
\hline \hline
\end{tabular}
\end{center}
\end{table}

The mass of the resonance used as input to the fits is taken to be a
fixed splitting (in physical units) above the initial state meson
$m_{\mathrm{res}} = m_{B_{(s)}} + \Delta m$.  The value of $\Delta m$ depends
on the lowest lying resonance contributing to a particular form
factor.  The values we use are given in Table~\ref{tab:dmres}.  Fits
have been redone, varying these $\Delta m$ values by 20\% and this has
no effect on the final results for the form factor curves (although
the fit parameters vary to compensate for the change in $\Delta m$).

The dependence of the form factor on the quark masses is taken into
account by the $d_n$ terms
\begin{align}
d_n = [1 &+ c_{n1} \Delta x + c_{n2} (\Delta x)^2 + \ldots \nonumber \\ 
 & +  c_{n1s} \Delta x_s + c_{n2s} (\Delta x_s)^2 + \ldots ]
\end{align}
with $\Delta x = (m_\pi^2 - m_{\pi,\mathrm{phys}}^2)/(4\pi f_\pi)^2$
and $\Delta x_s = (m_{\eta_s}^2 - m_{\eta_s,\mathrm{phys}}^2)/(4\pi
f_\pi)^2$ acting as proxies for the differences away from physical
$u/d$ and $s$ quark masses, respectively.  We use $f_\pi =
132~\mathrm{MeV}$ and the pseudoscalar meson masses displayed in
Table~\ref{tab:MesonMasses}. 

Our data for the separate $B \to K^*$, $B_s \to K^*$, and $B_s \to
\phi$ form factors are computed with constant, somewhat mistuned
values of the strange quark mass.  We can estimate the dependence on
the valence strange quark mass by performing a simultaneous fit which
treats the $B \to K^*$ and $B_s \to K^*$ form factors as calculations
of the $B_s \to \phi$ form factors using a very mistuned spectator or
offspring quark mass.  Departures from the physical strange mass are
parametrized in terms of the corresponding pseudoscalar meson mass:
\begin{eqnarray}
\Delta y &=& \frac{1}{(4\pi f_\pi)^2} \left(m_{\mathrm{offspr}}^2 - 
m_{\eta_s,\mathrm{phys}}^2\right) \nonumber \\
\Delta w &=& \frac{1}{(4\pi f_\pi)^2} \left(m_{\mathrm{spect}}^2 - 
m_{\eta_s,\mathrm{phys}}^2\right) \,.
\end{eqnarray}
For example, $\Delta y = \Delta w \approx 2\% (4\%)$ for the $B_s \to
\phi$ form factors on the coarse (fine) lattice.  In the case of $B
\to K^*$ form factors, $\Delta y \approx 2\% (4\%)$ for the coarse
(fine) lattice and $\Delta w \approx -8\%$ to $-13\%$ depending on the
``light'' quark mass.  These values are swapped when considering
$B_s \to K^*$ decays.

We obtain good fits to all the data for a particular form factor $F$
using the following ansatz:
\begin{align}
F(t;\Delta y, \Delta w) \;=\; & \frac{1}{P(t)}\left[a_0\left( 1 + f_{01}\Delta y
+ g_{01}\Delta w\right) + a_1 z \right] \,.
\label{eq:3ff_fit}
\end{align}
Results of the fits for the form factors $V$, $A_0$, $A_1$, $A_{12}$,
and $T_{23}$ are given in Table~\ref{tab:3ff}. (The tables of fit
results also give matrix elements $C(p,q)$ of the correlation
matrix. These are related to covariance matrix elements $\sigma(p,q)$
by $C(p,q) = \sigma(p,q)/(\sigma_p \sigma_q)$, with $\sigma(p,p) =
\sigma_p^2$.) Fits were also performed allowing $a_1 z$ to
be multiplied by a factor $(1 + f_{11}\Delta y + g_{11}\Delta w)$;
however, the data do not constrain the parameters $f_{11}$ and $g_{11}$.

Our final results for the form factors are obtained by separately considering
the form factors with specific initial and final state combinations.
In each case we use the following function to fit form factor $F(t)$
\begin{equation}
  F(t) ~=~ \frac{1}{P(t)}[a_0 (1 + c_{01} \Delta x 
  + c_{01s} \Delta x_s) + a_1 z] \,.
\label{eq:sse_3par}
\end{equation}
The parameter describing the strange-quark mass dependence $c_{01s}$
is included in the fit with a Gaussian prior using the
results in Table~\ref{tab:3ff}: $c_{01s} = f_{01}$ for the
$B\to K^*$ form factors, $c_{01s} = g_{01}$ for $B_s\to K^*$, and
$c_{01s} = f_{01} + g_{01}$ for $B_s\to \phi$.  In the last case, the
width for the $c_{01s}$ prior is taken by combining the $f_{01}$ and 
$g_{01}$ uncertainties in quadrature.  The parameters $a_0$, $a_1$,
and $c_{01}$ are not constrained by priors.  We find the lattice
spacing dependence to be negligible when we include the parameter
$b_1$ in fits of the form (\ref{eq:fit_gen_form}), therefore we do not
include this parameter in our final fits.  The results, including
correlation matrices, are given in Table~\ref{tab:sse_va_3par_sl} for
5 of the $B \to K^*$ form factors and Table~\ref{tab:sse_va_3par_ss}
for 5 of the $B_s\to \phi$ form factors.  (See the Appendix for $B_s
\to K^*$ form factors.)

We have an extra piece of information about the $T_1(q^2)$ and
$T_2(q^2)$ form factors, namely the kinematic constraint that they
equal each other at $q^2 = 0$.  We implement this by performing a
combined eight-parameter fit, with each form factor parametrized as in
(\ref{eq:sse_3par}), and adding to the $\chi^2$ function a term
$[a_0^{T_1} - a_0^{T_2} + (a_1^{T_1} - a_1^{T_2})z_0]^2/10^{-8}$,
where $z_0 = z(0, 12~\mathrm{GeV}^2)$.  As with the other form
factors, we first determine the parameters governing the strange-quark
mass dependence by a joint fit to $B_s \to \phi$, $B\to K^*$, and
$B_s\to K^*$ form factors (Table~\ref{tab:pt1t2_3ff}).  Those
parameters are then included in the eight-parameter fit, the results of
which appear in Tables~\ref{tab:sse_t_6par_sl} and
\ref{tab:sse_t_6par_ss} (and the Appendix).

We obtain values for $\chi^2$ per degree-of-freedom close to 1 for all
the fits to form factor shape.  Nevertheless, we have experimented
with including terms corresponding to the parameters $c_{11}$ and
$c_{02}$, using Gaussian priors to prevent the fits from diverging.
As we found with including a $b_1$ parameter in the fit, the data
clearly do not constrain these parameters.  The fit returns a value and
error for these parameters corresponding to the prior mean and width, for
narrow and wide Gaussians.  The $\chi^2$ of the fit is unaffected by 
including or excluding terms in this way.

\begin{table}
\caption{\label{tab:3ff}Fit results (with correlation matrices)
  determining the dependence of form factors on the strange quark mass.}
\begin{center}
\begin{tabular}{c|c|rrr} \hline\hline
\multicolumn{5}{c}{$P(t)V(t)$} \\ \hline
$p$ & Value  &  \multicolumn{1}{c}{$C(p,a_0)$}  &  
\multicolumn{1}{c}{$C(p,a_1)$}  &  \multicolumn{1}{c}{$C(p,f_{01})$}  \\ \hline
$a_0$ & $0.5386(196)$ \\
$a_1$ & $-1.85(35)$ & $0.95$ \\
$f_{01}$ & $1.069(112)$ & $-0.66$ & $-0.60$ \\
$g_{01}$ & $-0.047(134)$ & $-0.34$ & $-0.21$ & $0.17$ \\ \hline
\multicolumn{5}{c}{$P(t)A_0(t)$} \\ \hline
$p$ & Value  &  \multicolumn{1}{c}{$C(p,a_0)$}  &  
\multicolumn{1}{c}{$C(p,a_1)$}  &  \multicolumn{1}{c}{$C(p,f_{01})$}  \\ \hline
$a_0$ & $0.5538(181)$ \\
$a_1$ & $-1.48(32)$ & $0.96$ \\
$f_{01}$ & $0.420(80)$ & $-0.44$ & $-0.41$ \\
$g_{01}$ & $-0.163(144)$ & $-0.44$ & $-0.30$ & $0.42$ \\ \hline
\multicolumn{5}{c}{$P(t)A_1(t)$} \\ \hline
$p$ & Value  &  \multicolumn{1}{c}{$C(p,a_0)$}  &  \multicolumn{1}{c}{$C(p,a_1)$}  &  \multicolumn{1}{c}{$C(p,f_{01})$}  \\ \hline
$a_0$ & $0.2984(70)$ \\
$a_1$ & $0.158(99)$ & $0.97$ \\
$f_{01}$ & $0.841(52)$ & $-0.67$ & $-0.65$ \\
$g_{01}$ & $-0.053(78)$ & $-0.20$ & $-0.09$ & $0.17$ \\ \hline
\multicolumn{5}{c}{$P(t)A_{12}(t)$} \\ \hline
$p$ & Value  &  \multicolumn{1}{c}{$C(p,a_0)$}  &  
\multicolumn{1}{c}{$C(p,a_1)$}  &  \multicolumn{1}{c}{$C(p,f_{01})$}  \\ \hline
$a_0$ & $0.2057(71)$ \\
$a_1$ & $0.406(124)$ & $0.97$ \\
$f_{01}$ & $0.151(91)$ & $-0.35$ & $-0.36$ \\
$g_{01}$ & $-0.599(137)$ & $-0.19$ & $-0.10$ & $0.29$ \\ \hline
\multicolumn{5}{c}{$P(t)T_{23}(t)$} \\ \hline
$p$ & Value  &  \multicolumn{1}{c}{$C(p,a_0)$}  &  
\multicolumn{1}{c}{$C(p,a_1)$}  &  \multicolumn{1}{c}{$C(p,f_{01})$}  \\ \hline
$a_0$ & $0.5167(113)$ \\
$a_1$ & $0.389(187)$ & $0.97$ \\
$f_{01}$ & $0.476(44)$ & $-0.59$ & $-0.62$ \\
$g_{01}$ & $-0.321(102)$ & $-0.25$ & $-0.12$ & $0.03$ \\ \hline\hline
\end{tabular}
\end{center}
\end{table}

\section{Discussion of results and systematic uncertainties}
\label{sec:results}

In this section we present our main results and discuss the systematic
uncertainties in our calculations of the form factors.  We compare our
results to other determinations.

The results of the fits for the meson masses in
Table~\ref{tab:MesonMasses} indicate that the heavy quark mass has
been tuned so that the $B$ and $B_s$ masses are 5\% too heavy.  In the
$m_B\to\infty$ limit the form factors scale like \cite{Isgur:1990kf} 
(also \cite{Grinstein:2004vb})
\begin{eqnarray}
& V, A_0, T_1, T_{23} \propto m_B^{1/2} & \nonumber \\
&A_1, A_{12}, T_2 \propto m_B^{-1/2} \,. &
\label{eq:HQETscaling}
\end{eqnarray}
Therefore, we compensate for this error due to the $m_b$ mistuning
by scaling the central values of the form factors by $0.976$ ($V$,
$A_0$, $T_1$, $T_{23}$) and $1.025$ ($A_1$, $A_{12}$, $T_2$).  The remaining
error is suppressed compared to (\ref{eq:HQETscaling}) by a factor of
$\Lambda_{\subrm{QCD}}/m_b$; i.e.\ the remaining $m_b$ mistuning error is
well below 1\% and is treated as negligible compared to other uncertainties.

\begin{figure*}
\begin{center}
\includegraphics[width=0.495\textwidth]{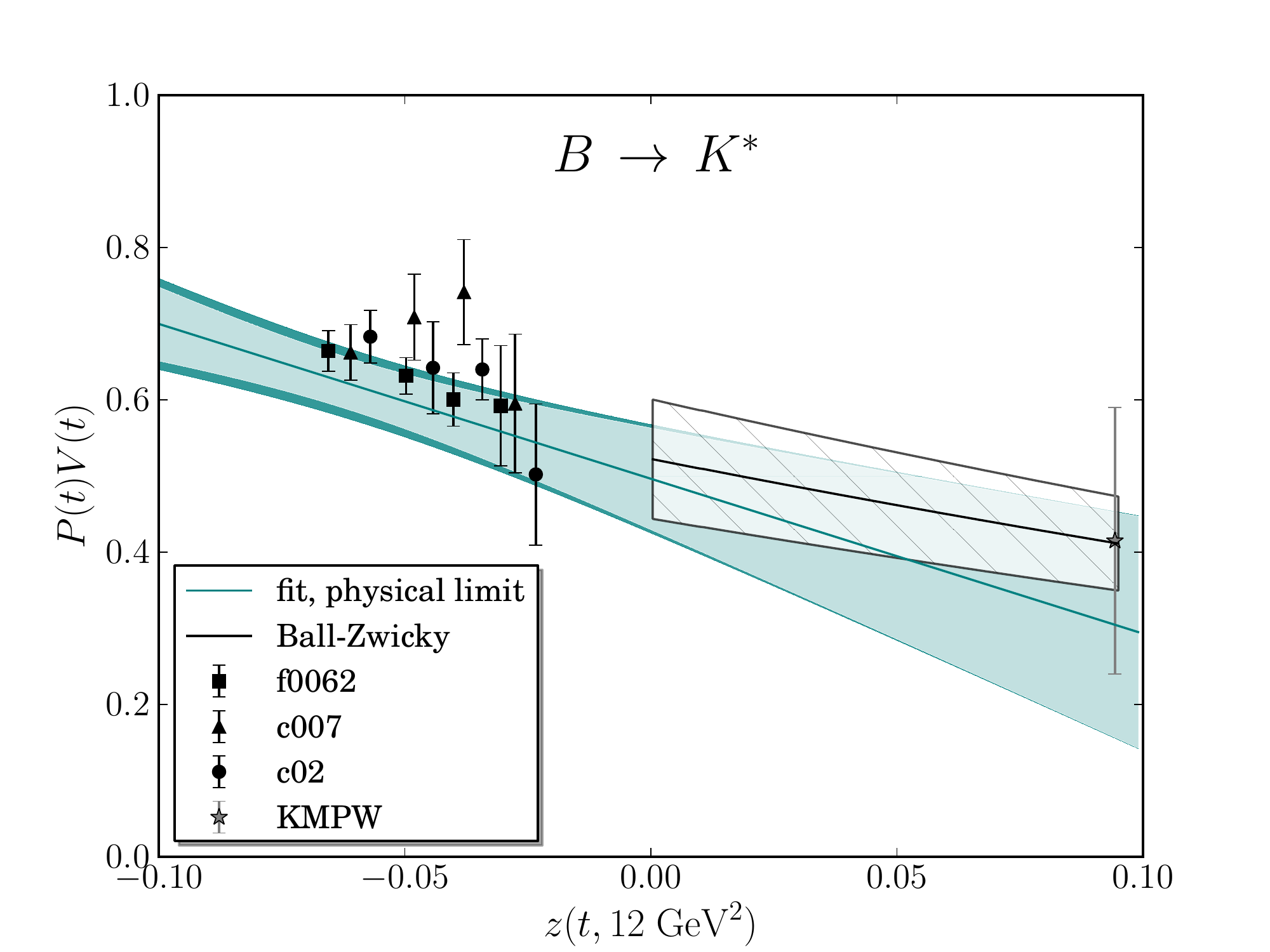}
\includegraphics[width=0.495\textwidth]{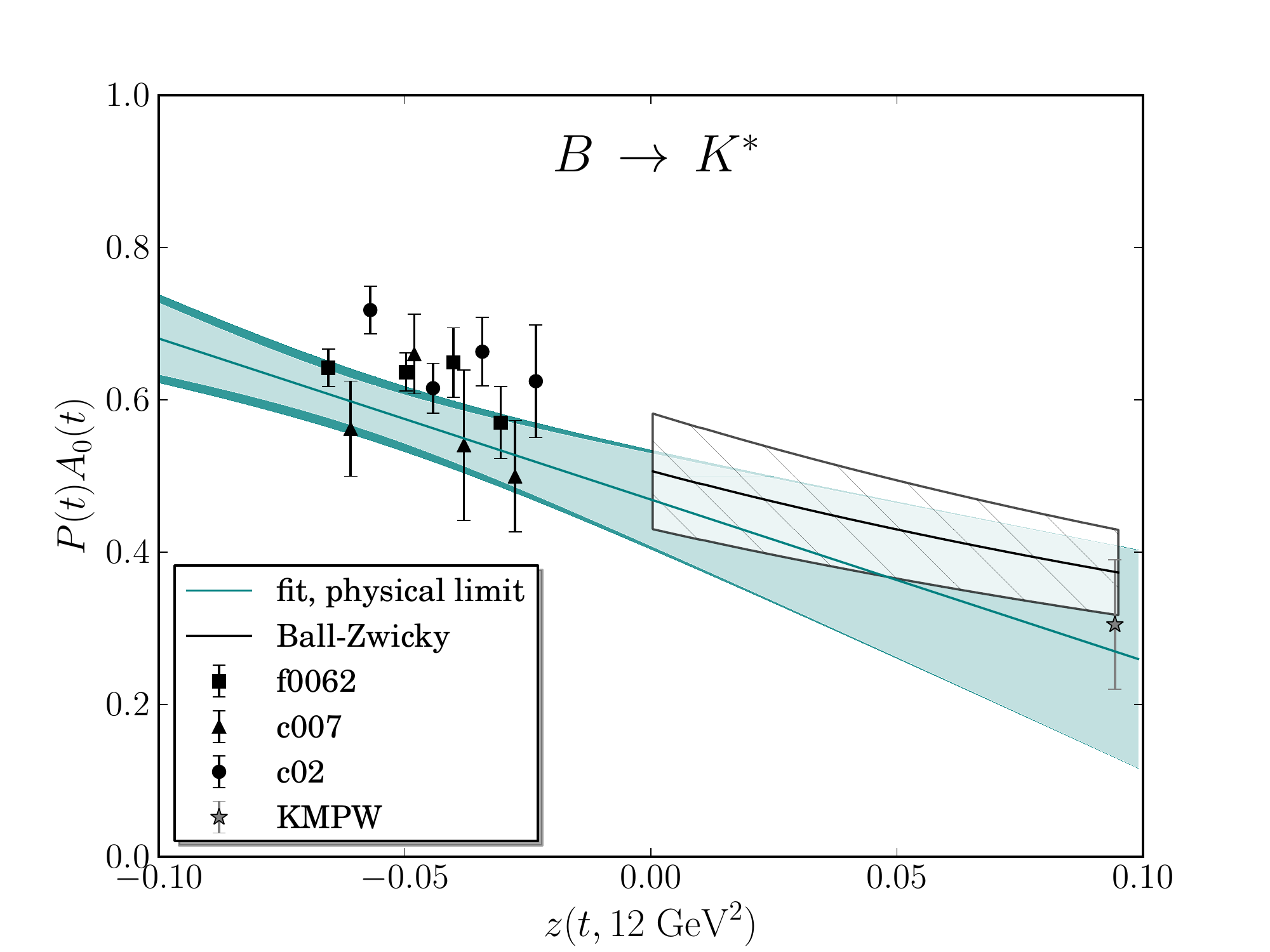}
\includegraphics[width=0.495\textwidth]{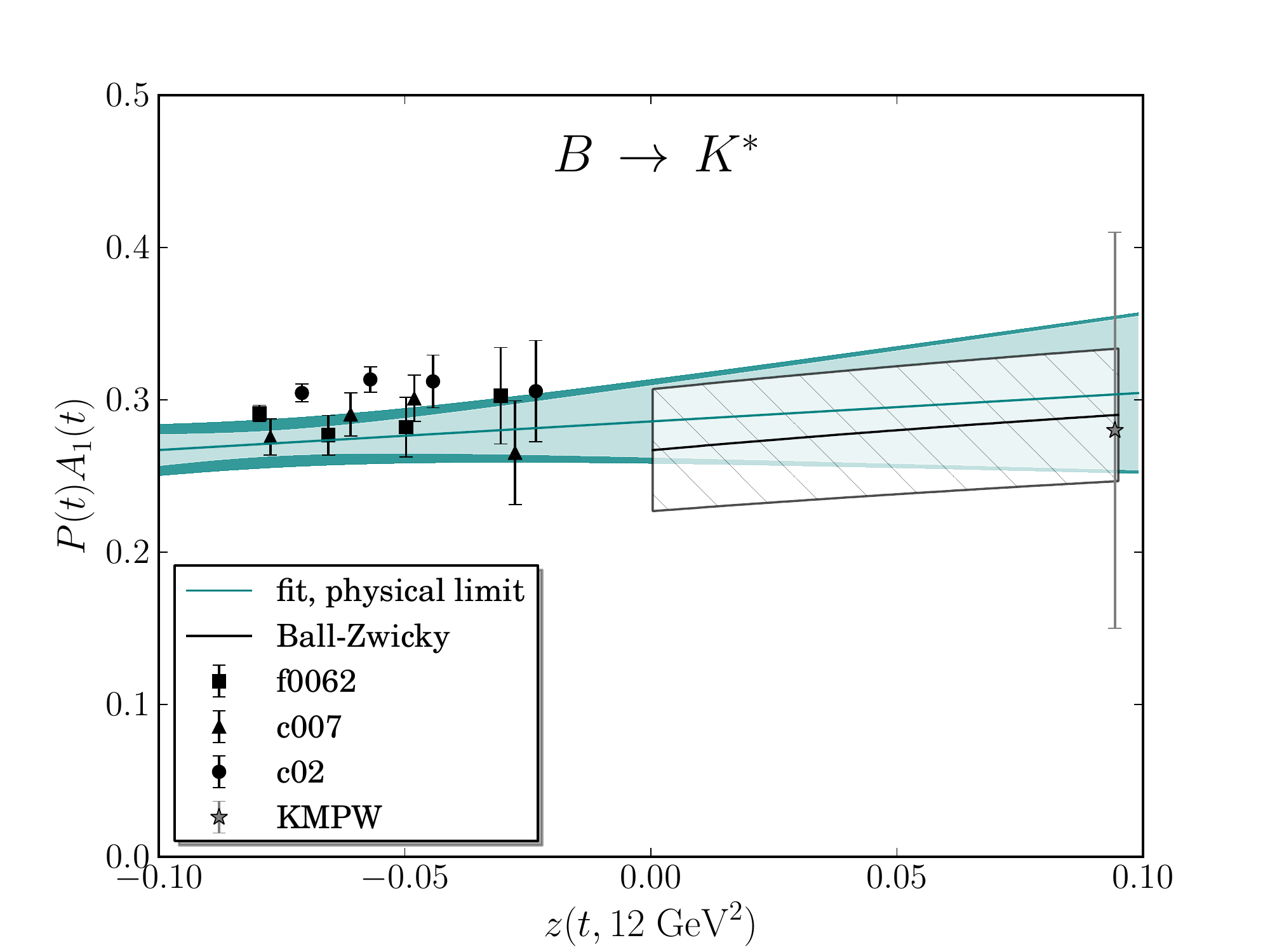}
\includegraphics[width=0.495\textwidth]{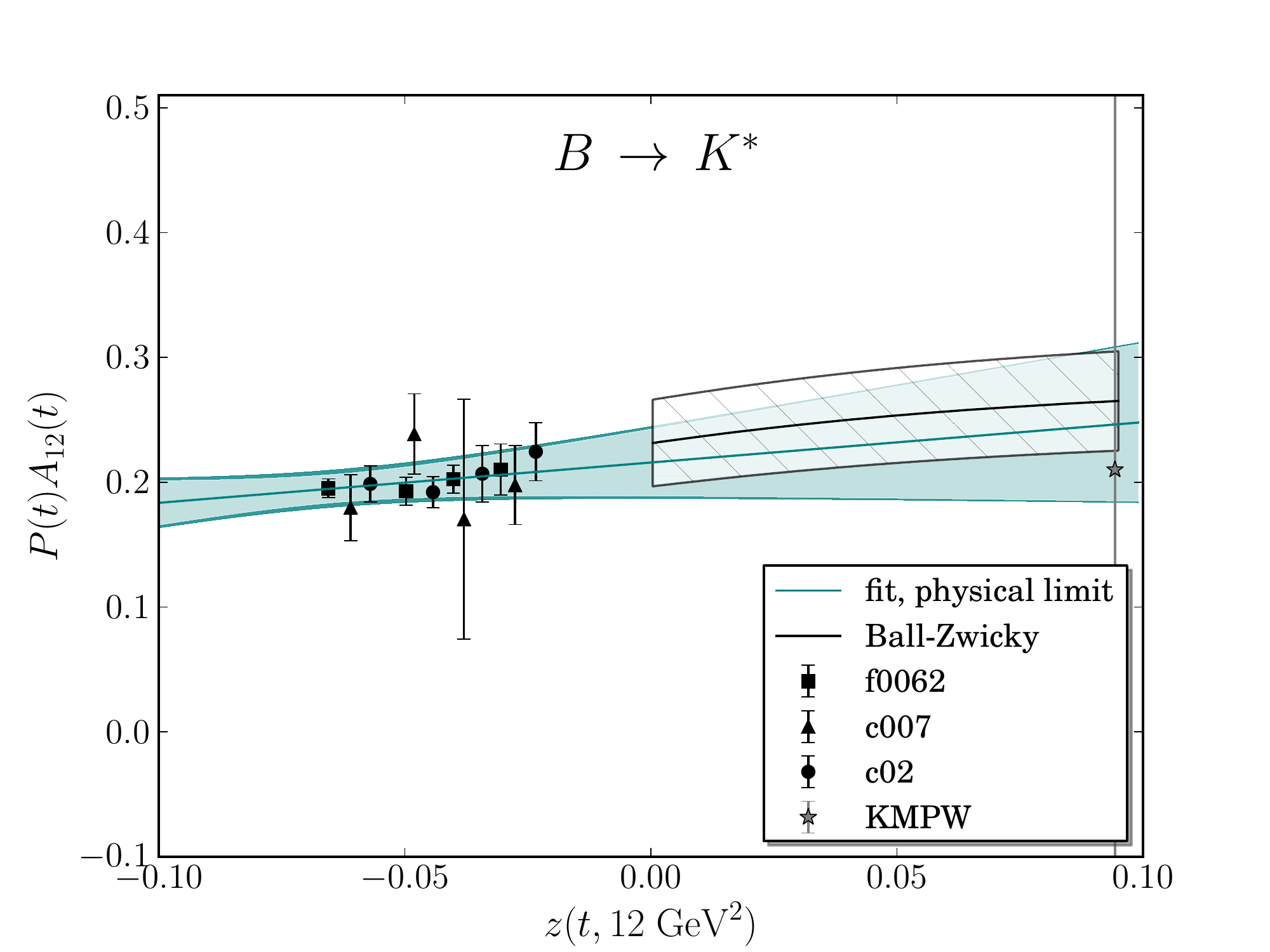}
\includegraphics[width=0.495\textwidth]{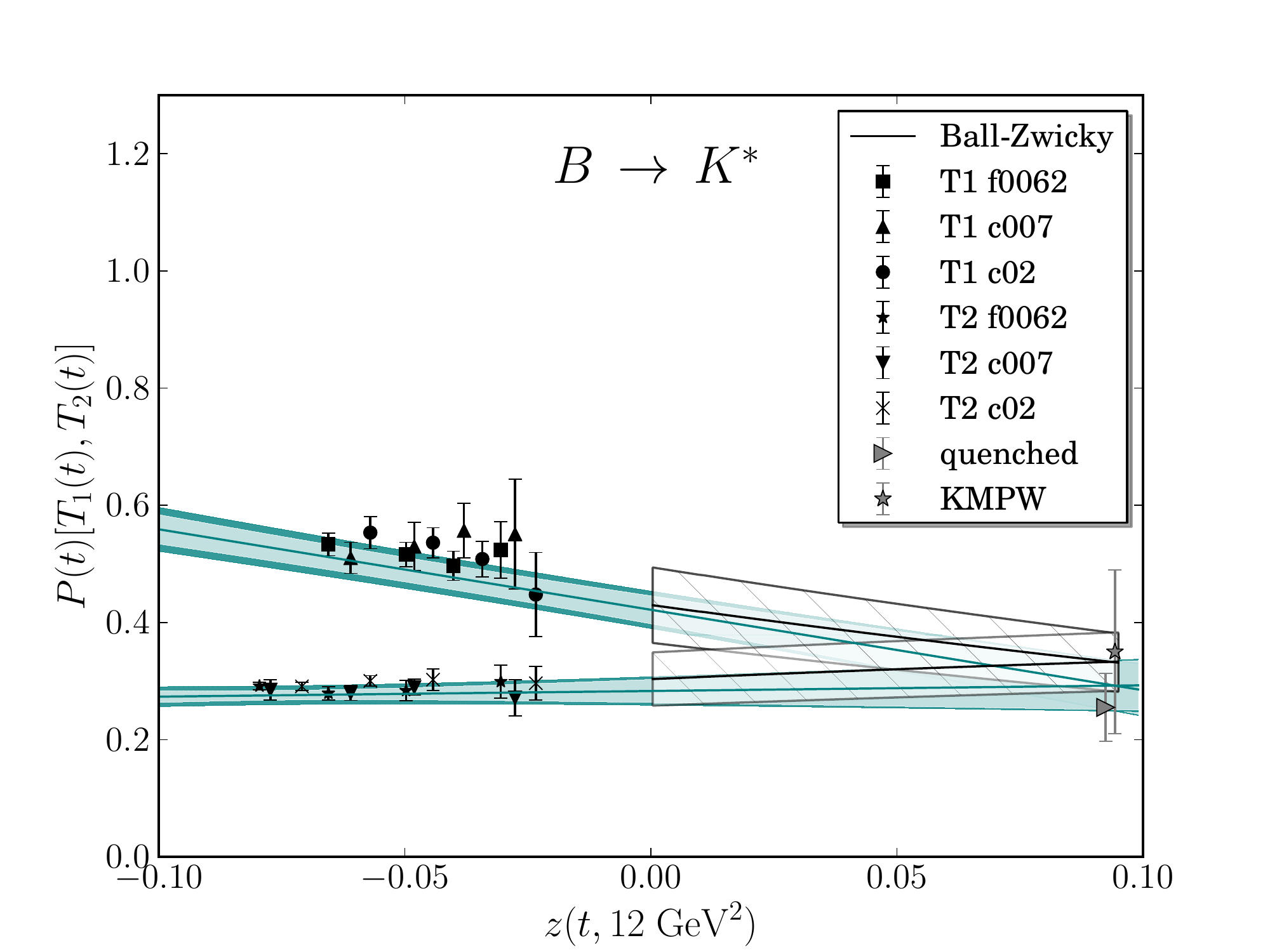}
\includegraphics[width=0.495\textwidth]{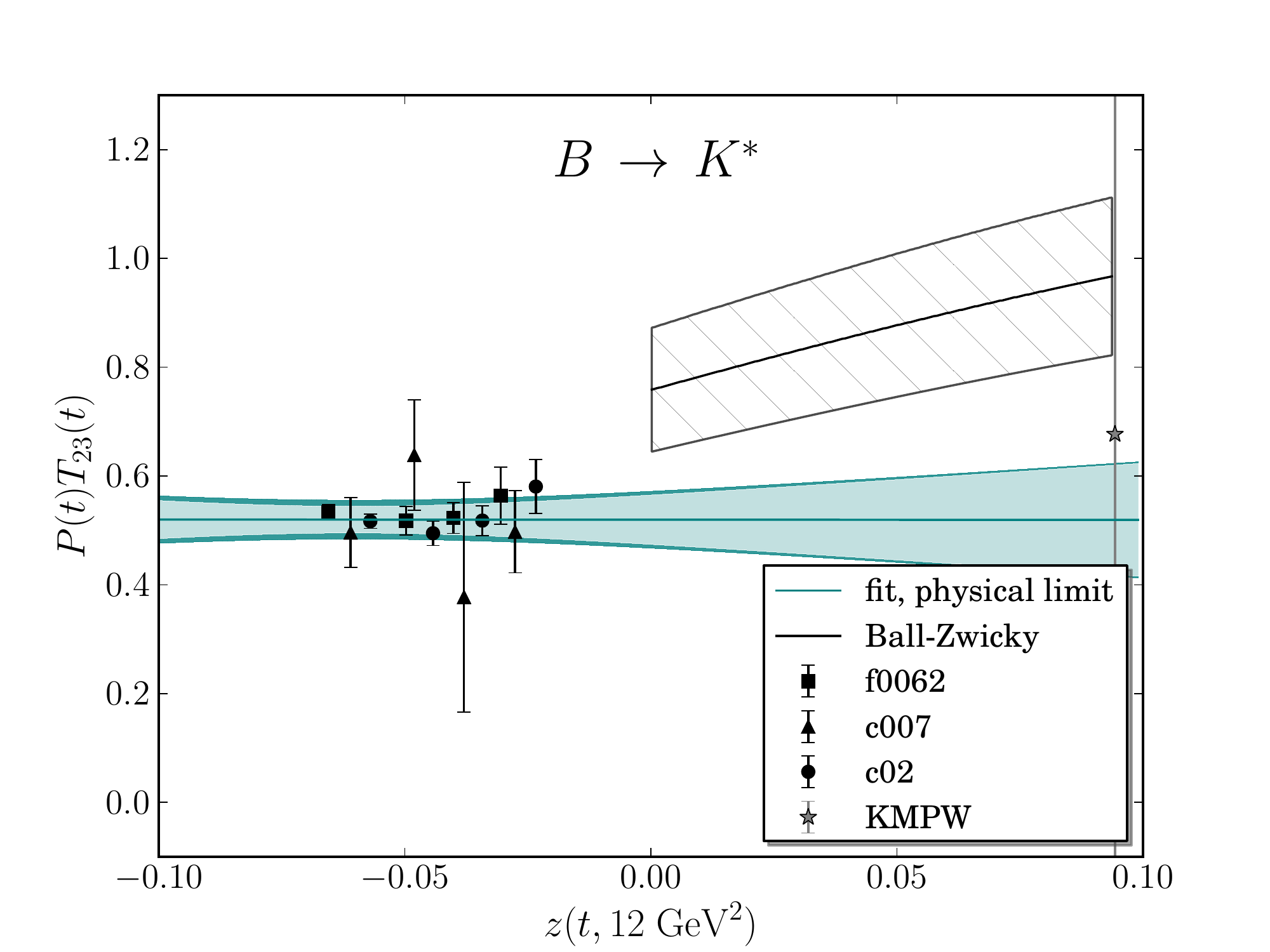}
\end{center}
\caption{\label{fig:pff_sl}$B \to K^*$ form factors.  Data points for
  the three ensembles of lattice gauge fields appear in black. The fit of
  the lattice data to the function of Eq.~(\ref{eq:sse_3par}),
  extrapolated to the physical quark mass limit, is shown as a solid
  curve, with statistical (pale) and total (dark) error bands.  For
  comparison, the LCSR results of \cite{Ball:2004rg} are shown with a
  15\% uncertainty (hatched band) \cite{Bobeth:2010wg}.  We also
  display $q^2=0$ LCSR results from \cite{Khodjamirian:2010vf} as gray
  stars (central values shifted slightly so that error bars are
  symmetric); the errors have been propagated as uncorrelated for
  $A_{12}$ and $T_{23}$ possibly resulting in overestimates of the
  corresponding uncertainties.  In the lower left plot, the quenched
  lattice QCD result for $T_1(0)=T_2(0)$ \cite{Becirevic:2006nm} is
  displayed as a gray triangle.}
\end{figure*}

\begin{figure*}
\begin{center}
\includegraphics[width=0.495\textwidth]{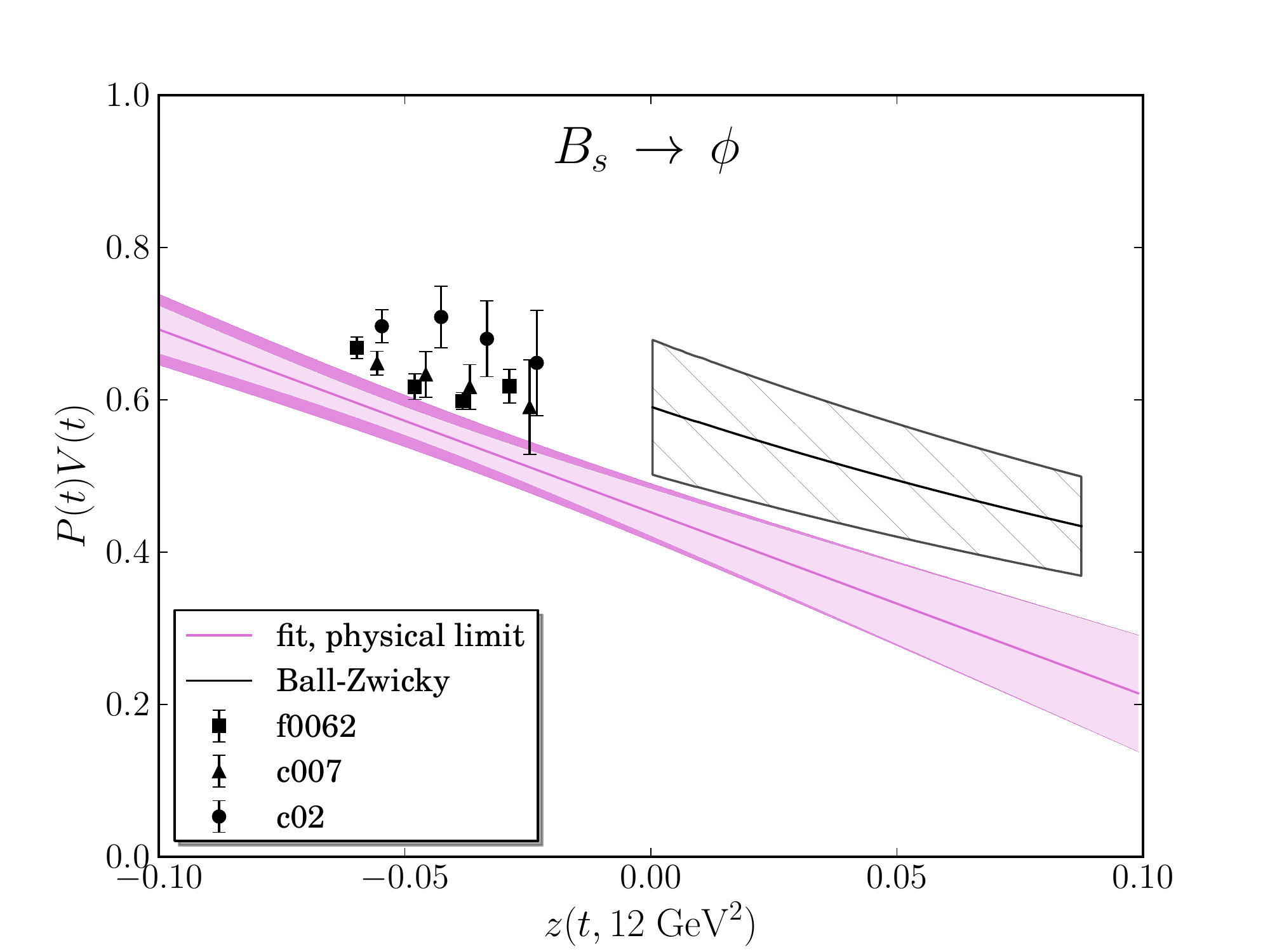}
\includegraphics[width=0.495\textwidth]{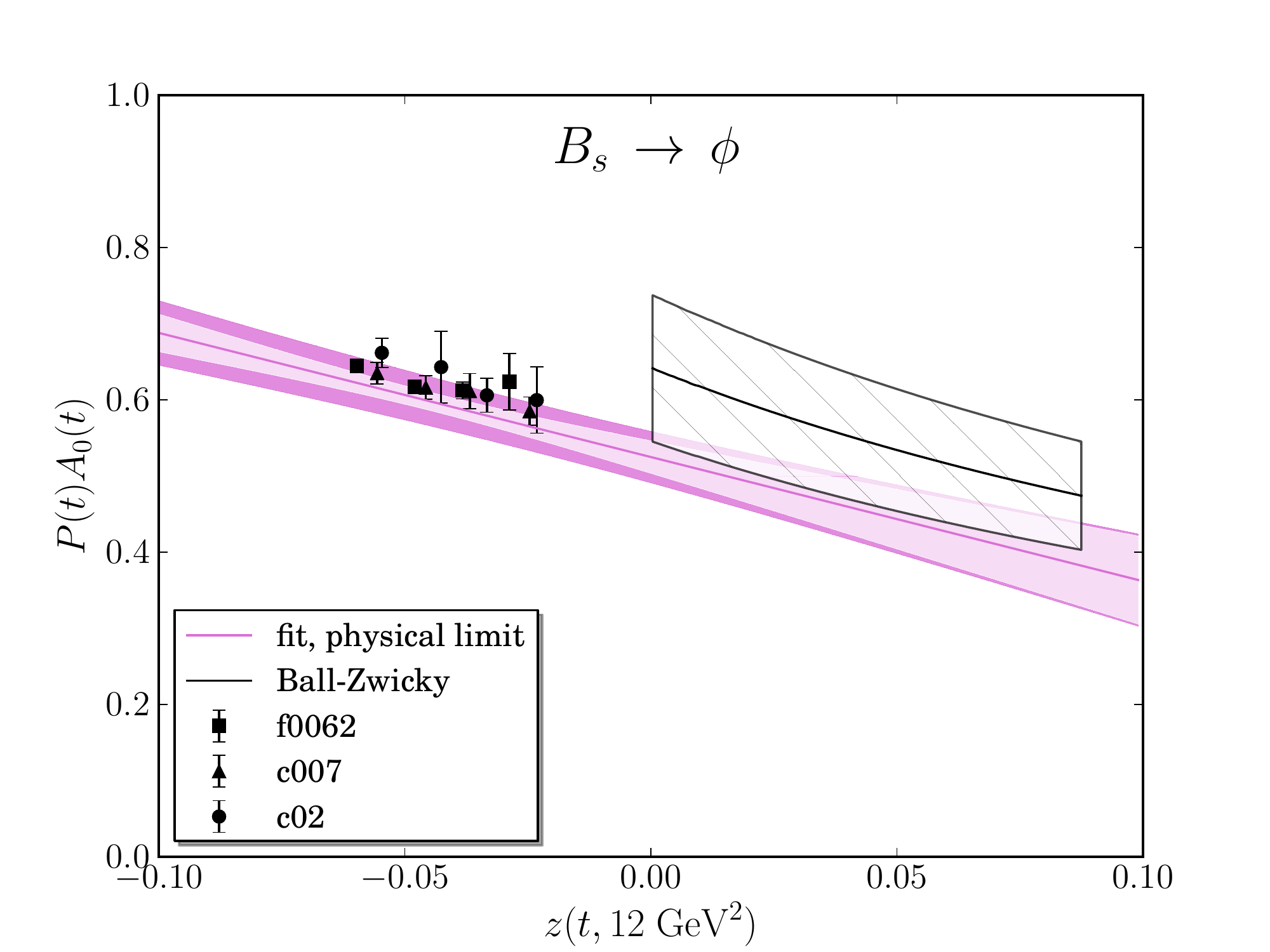}
\includegraphics[width=0.495\textwidth]{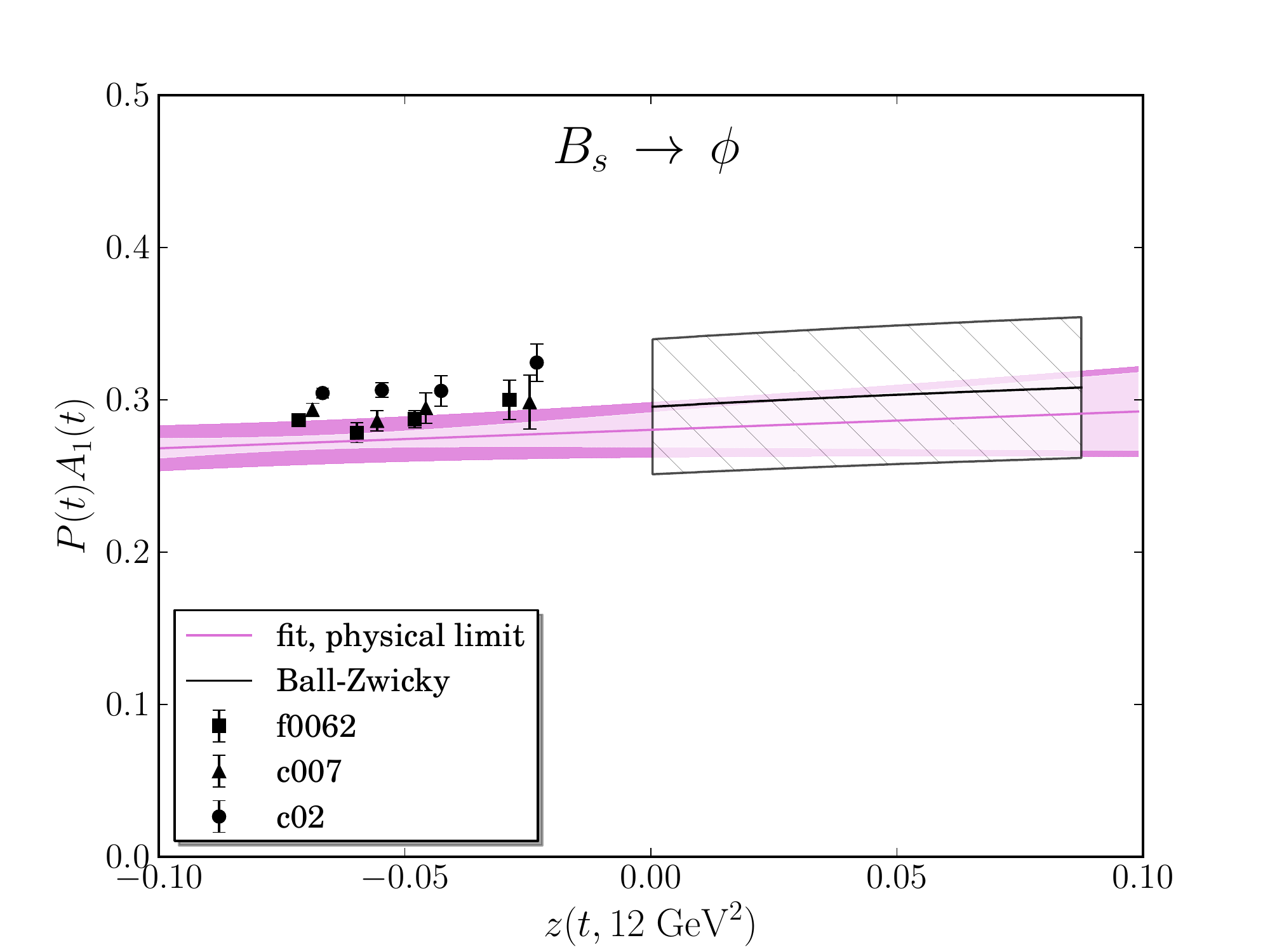}
\includegraphics[width=0.495\textwidth]{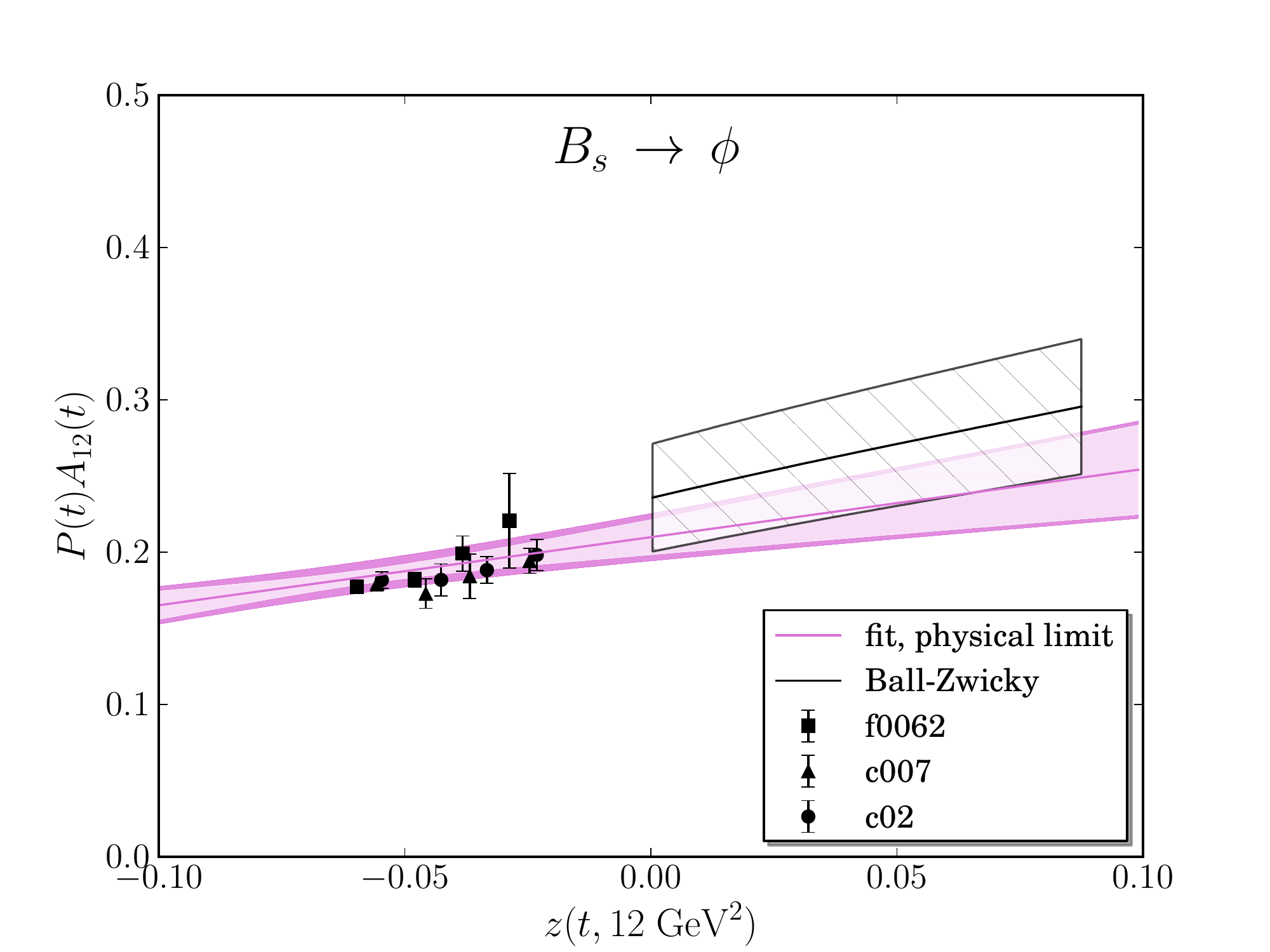}
\includegraphics[width=0.495\textwidth]{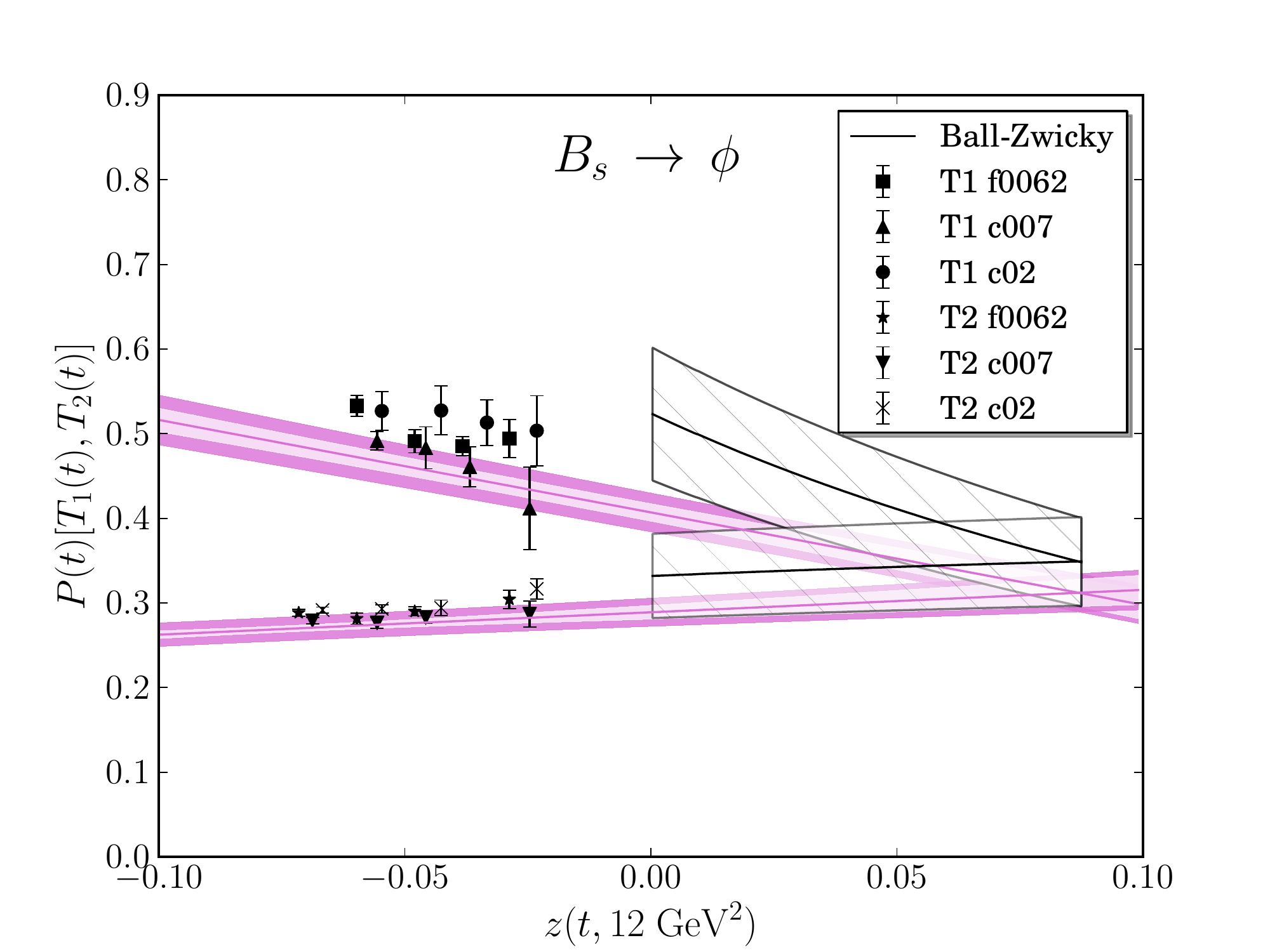}
\includegraphics[width=0.495\textwidth]{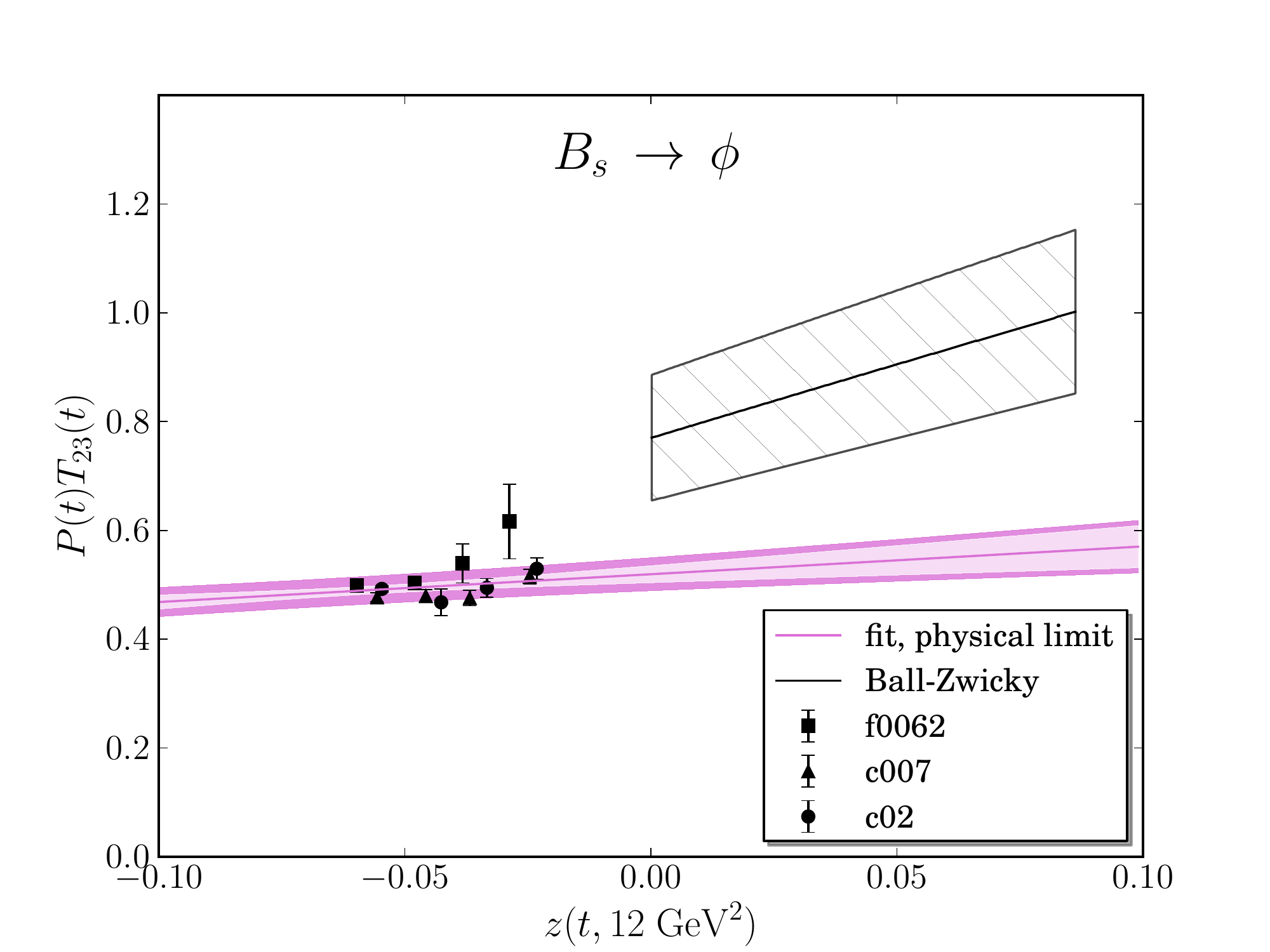}
\end{center}
\caption{\label{fig:pff_ss}$B_s \to \phi$ form factors, as in
Fig.~\ref{fig:pff_sl}.}
\end{figure*}

The $B\to K^*$, $B_s \to \phi$, and $B_s \to K^*$ form factors have
been fit to the form given in Eq.~(\ref{eq:sse_3par}).  All fits have
been done after compensating for the mistuning of the heavy quark
mass. The results for $B\to K^*$ and $B_s \to \phi$ are tabulated in
Tables \ref{tab:sse_va_3par_sl}--\ref{tab:sse_t_6par_ss}, along with
their correlation matrices.  Figures \ref{fig:pff_sl} and
\ref{fig:pff_ss} show the form factor fits, along with the lattice
data.  These tables and curves constitute our final results.  Data
points corresponding to the physical limit can be obtained from the
fits by setting $\Delta x = 0$ and $\Delta x_s = 0$ in
Eq.~(\ref{eq:sse_3par}). The corresponding tables and plots for $B_s
\to K^*$ form factors appear in the Appendix.  In order to aid
comparison with other work, a table in the Appendix gives numerical
results for the form factors in the physical limit at a few fiducial
values of $q^2$.

In Figs.~\ref{fig:pff_sl} and \ref{fig:pff_ss} we also plot results
from light-cone sum rules (LCSR) \cite{Ball:2004rg} with a uniform
15\% error band \cite{Bobeth:2010wg}.  There have also been LCSR
calculations of the $B\to K^*$ form factors using $B$ meson
distribution amplitudes \cite{Khodjamirian:2010vf}; these are in
agreement but quote larger uncertainties so we only display their
$q^2=0$ points.  The agreement between the LQCD and LCSR results is
generally good.  The $T_{23}$ form factors extracted from Ball and
Zwicky are notable exceptions.  It is likely that 15\% is an
underestimate of the LCSR uncertainty in this linear combination of
the $T_2$ and $\tilde{T}_3$ form factors calculated by Ball and
Zwicky.  We, however, are not in a place to better estimate their
uncertainties, so we interpret the 15\% error band in our paper in a
weaker sense than a $1\sigma$ range.  Propagating the uncertainties in
$T_2$ and $T_3$ determined by Khodjamirian \textit{et al.}\ as
uncorrelated leads to large uncertainties in $T_{23}$, even if the
central values are in good agreement with our results.  The $V$ form
factors for $B_s\to \phi$ and $B_s \to K^*$ also appear to disagree
with extrapolations of our lattice data to large recoil.  One
possibility is that a $z^2$ term is necessary to fit both results.

Our extrapolation of $B \to K^*$ form factors $T_1$
and $T_2$ agrees with the latest quenched lattice QCD results for the
$q^2=0$ value of $T_1(0)=T_2(0)$ \cite{Becirevic:2006nm} (see
Fig.~\ref{fig:pff_sl}).  Form factors have also been predicted using a
relativistic quark model \cite{Ebert:2010dv,Faustov:2013pca}.

\begin{figure*}
\begin{center}
\includegraphics[width=0.495\textwidth]{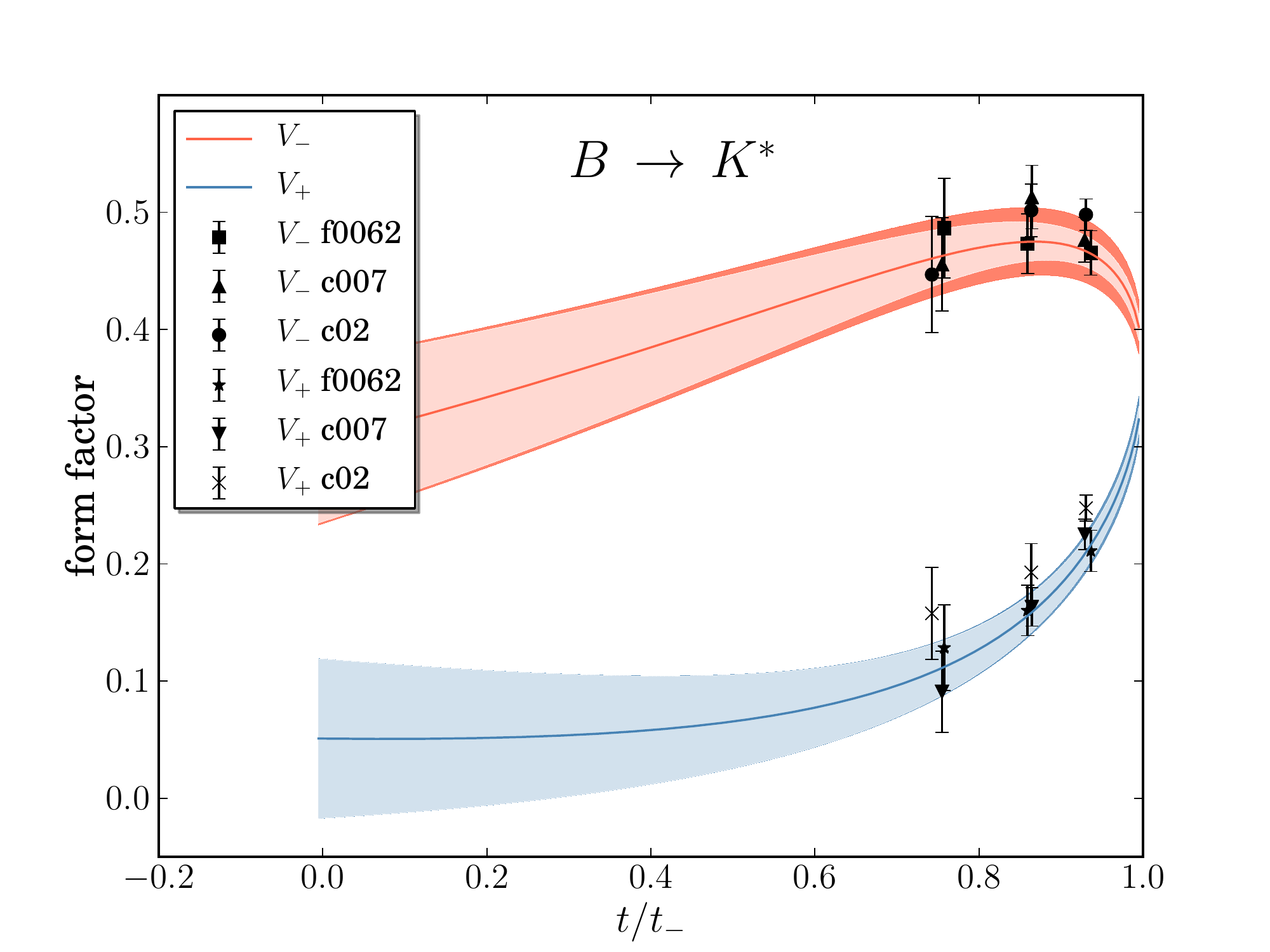}
\includegraphics[width=0.495\textwidth]{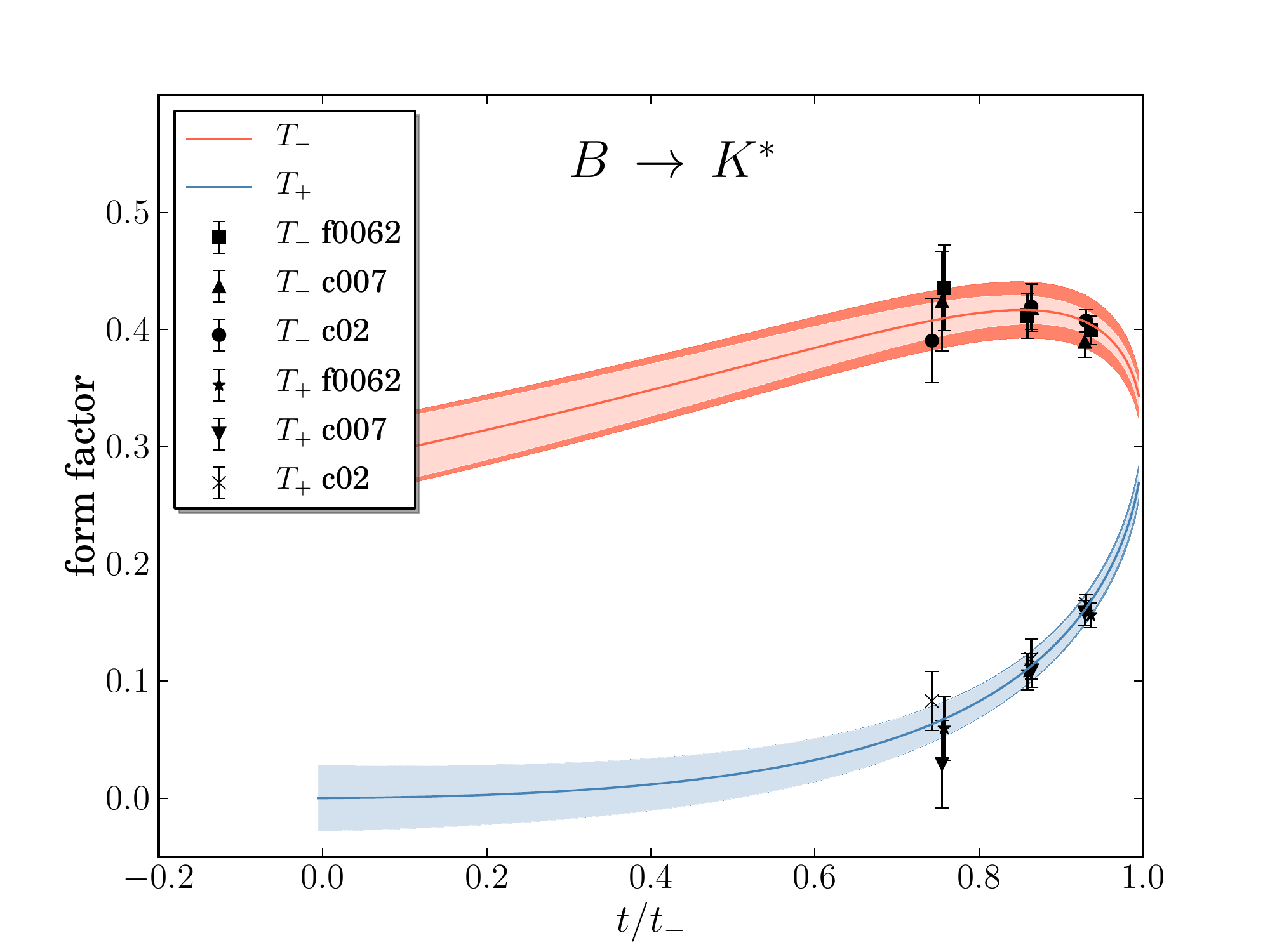}
\includegraphics[width=0.495\textwidth]{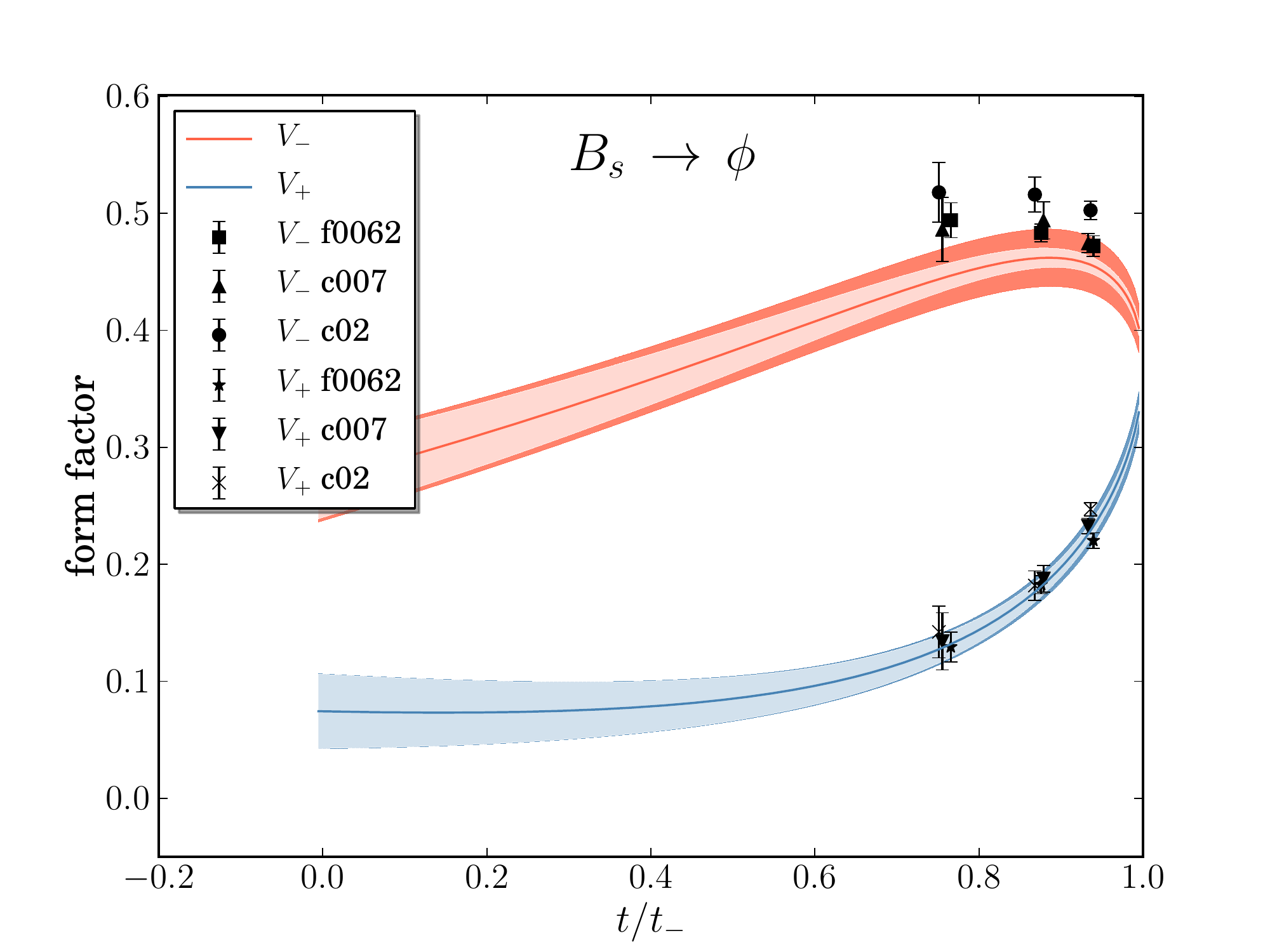}
\includegraphics[width=0.495\textwidth]{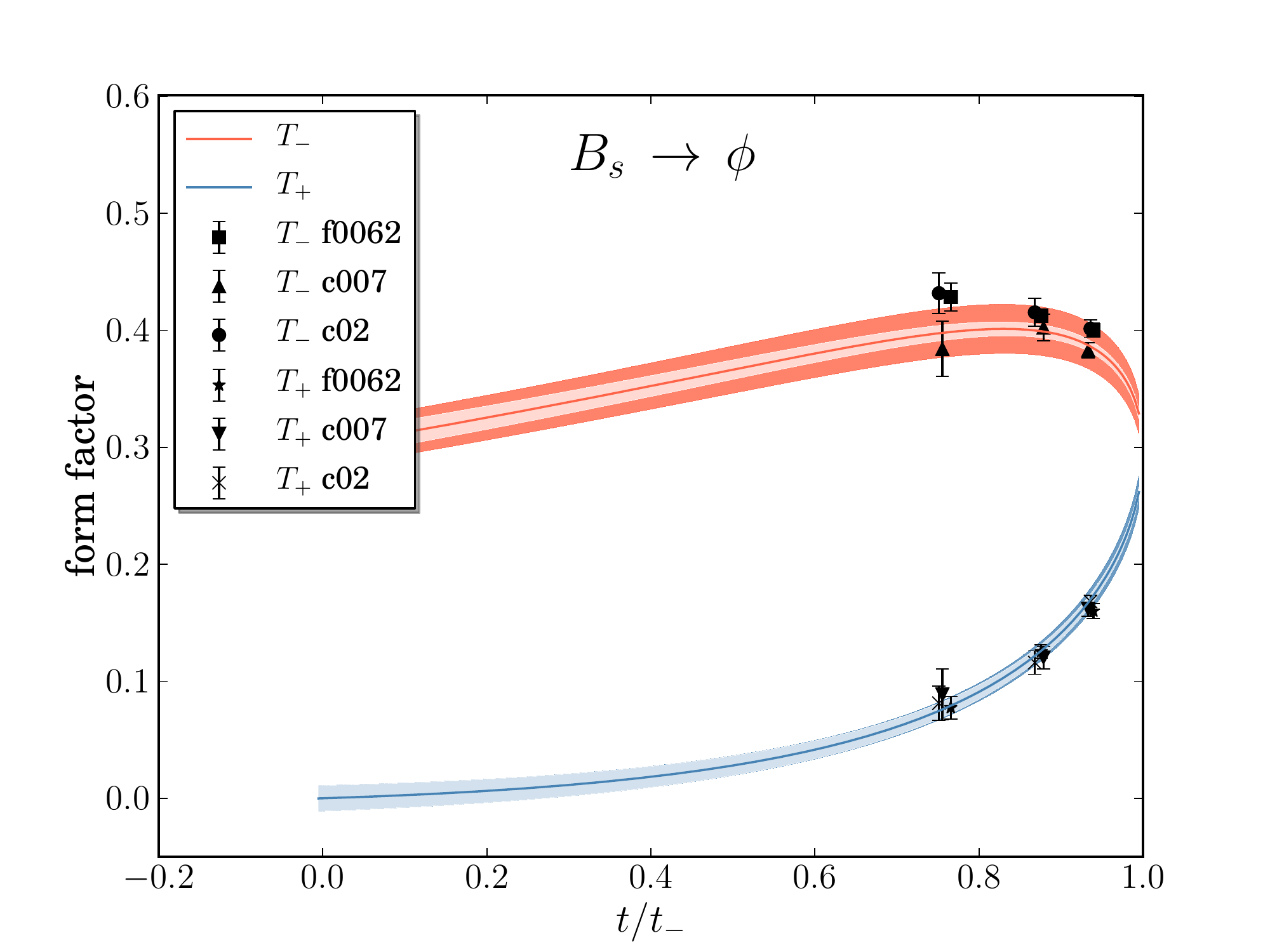}
\end{center}
\caption{\label{fig:pm_sl_ss}Curves describing helicity basis form factors 
 $V_\pm(q^2)$ and $T_\pm(q^2)$ form factors for $B \to K^*$ (top) and 
$B_s \to \phi$ (bottom).  Data points represent the linear combinations
(\ref{eq:helicity_vpm_tpm}) computed on each ensemble.
The curves are obtained by using the
fit results in Tables~\ref{tab:sse_va_3par_sl}--\ref{tab:sse_t_6par_ss}
in the physical quark mass limit, with statistical (pale) and total
(dark) error bands shown.}
\end{figure*}

Fig.~\ref{fig:pm_sl_ss} shows the results for the form factors $V_\pm$
and $T_\pm$.  Note in these cases, the fits are not done directly to
the data; instead the curves are the results of linearly combining the
fits in Tables \ref{tab:sse_va_3par_sl}--\ref{tab:sse_t_6par_ss}.  

\begin{table}[t]
\caption{\label{tab:sse_va_3par_sl}Results and correlation matrices of
 fits to $B\to K^*$ form factors. }
\begin{center}
\begin{tabular}{c|c|rrr} \hline \hline
\multicolumn{5}{c}{$P(t; 135\mathrm{MeV}) V(t)$} \\ \hline
$p$ & Value  & \multicolumn{1}{c}{$C(p,a_0)$}  & \multicolumn{1}{c}{$C(p,a_1)$}  & \multicolumn{1}{c}{$C(p,c_{01})$} \\ \hline
$a_0$ & $0.496(67)$ \\
$a_1$ & $-2.03(92)$ & $0.86$ \\
$c_{01}$ & $1.38(1.49)$ & $-0.79$ & $-0.41$ \\
$c_{01s}$ & $1.066(112)$ & $-0.04$ & $-0.00$ & $0.02$
\\ \hline
\multicolumn{5}{c}{$P(t; 87\mathrm{MeV}) A_0(t)$} \\ \hline
$p$ & Value  & \multicolumn{1}{c}{$C(p,a_0)$}  & \multicolumn{1}{c}{$C(p,a_1)$}  & \multicolumn{1}{c}{$C(p,c_{01})$} \\ \hline
$a_0$ & $0.469(61)$ \\
$a_1$ & $-2.11(88)$ & $0.86$ \\
$c_{01}$ & $2.46(1.27)$ & $-0.76$ & $-0.39$ \\
$c_{01s}$ & $0.421(80)$ & $-0.02$ & $0.00$ & $0.02$
\\ \hline
\multicolumn{5}{c}{$P(t; 550\mathrm{MeV}) A_1(t)$} \\ \hline
$p$ & Value  & \multicolumn{1}{c}{$C(p,a_0)$}  & \multicolumn{1}{c}{$C(p,a_1)$}  & \multicolumn{1}{c}{$C(p,c_{01})$} \\ \hline
$a_0$ & $0.286(24)$ \\
$a_1$ & $0.19(28)$ & $0.94$ \\
$c_{01}$ & $1.07(53)$ & $-0.68$ & $-0.42$ \\
$c_{01s}$ & $0.841(52)$ & $-0.02$ & $0.01$ & $0.02$
\\ \hline
\multicolumn{5}{c}{$P(t; 550\mathrm{MeV}) A_{12}(t)$} \\ \hline
$p$ & Value  & \multicolumn{1}{c}{$C(p,a_0)$}  & \multicolumn{1}{c}{$C(p,a_1)$}  & \multicolumn{1}{c}{$C(p,c_{01})$} \\ \hline
$a_0$ & $0.216(27)$ \\
$a_1$ & $0.32(38)$ & $0.91$ \\
$c_{01}$ & $-0.12(97)$ & $-0.65$ & $-0.33$ \\
$c_{01s}$ & $0.151(91)$ & $-0.03$ & $0.00$ & $0.02$
\\ \hline
\multicolumn{5}{c}{$P(t; 550\mathrm{MeV}) T_{23}(t)$} \\ \hline
$p$ & Value  & \multicolumn{1}{c}{$C(p,a_0)$}  & \multicolumn{1}{c}{$C(p,a_1)$}  & \multicolumn{1}{c}{$C(p,c_{01})$} \\ \hline
$a_0$ & $0.520(45)$ \\
$a_1$ & $-0.00(63)$ & $0.84$ \\
$c_{01}$ & $-0.07(65)$ & $-0.59$ & $-0.11$ \\
$c_{01s}$ & $0.474(44)$ & $-0.02$ & $0.00$ & $0.02$
\\ \hline\hline
\end{tabular}
\end{center}
\end{table}

\begin{table}[t]
\caption{\label{tab:sse_va_3par_ss}Results and correlation matrices of
 fits to $B_s\to \phi$ form factors. }
\begin{center}
\begin{tabular}{c|c|rrr} \hline \hline
\multicolumn{5}{c}{$P(t; 45\mathrm{MeV}) V(t)$} \\ \hline
$p$ & Value  & \multicolumn{1}{c}{$C(p,a_0)$}  & \multicolumn{1}{c}{$C(p,a_1)$}  & \multicolumn{1}{c}{$C(p,c_{01})$} \\ \hline
$a_0$ & $0.452(30)$ \\
$a_1$ & $-2.40(50)$ & $0.80$ \\
$c_{01}$ & $2.80(1.04)$ & $-0.69$ & $-0.18$ \\
$c_{01s}$ & $0.998(175)$ & $-0.14$ & $-0.03$ & $0.07$
\\ \hline
\multicolumn{5}{c}{$P(t; 0\mathrm{MeV}) A_0(t)$} \\ \hline
$p$ & Value  & \multicolumn{1}{c}{$C(p,a_0)$}  & \multicolumn{1}{c}{$C(p,a_1)$}  & \multicolumn{1}{c}{$C(p,c_{01})$} \\ \hline
$a_0$ & $0.525(21)$ \\
$a_1$ & $-1.63(39)$ & $0.83$ \\
$c_{01}$ & $0.81(52)$ & $-0.38$ & $0.13$ \\
$c_{01s}$ & $0.248(164)$ & $-0.13$ & $0.04$ & $0.06$
\\ \hline
\multicolumn{5}{c}{$P(t; 440\mathrm{MeV}) A_1(t)$} \\ \hline
$p$ & Value  & \multicolumn{1}{c}{$C(p,a_0)$}  & \multicolumn{1}{c}{$C(p,a_1)$}  & \multicolumn{1}{c}{$C(p,c_{01})$} \\ \hline
$a_0$ & $0.2803(113)$ \\
$a_1$ & $0.121(150)$ & $0.92$ \\
$c_{01}$ & $1.01(29)$ & $-0.64$ & $-0.33$ \\
$c_{01s}$ & $0.768(94)$ & $-0.10$ & $-0.01$ & $0.07$
\\ \hline
\multicolumn{5}{c}{$P(t; 440\mathrm{MeV}) A_{12}(t)$} \\ \hline
$p$ & Value  & \multicolumn{1}{c}{$C(p,a_0)$}  & \multicolumn{1}{c}{$C(p,a_1)$}  & \multicolumn{1}{c}{$C(p,c_{01})$} \\ \hline
$a_0$ & $0.2098(115)$ \\
$a_1$ & $0.447(186)$ & $0.92$ \\
$c_{01}$ & $-0.18(41)$ & $-0.57$ & $-0.25$ \\
$c_{01s}$ & $-0.435(163)$ & $-0.10$ & $0.03$ & $0.06$
\\ \hline
\multicolumn{5}{c}{$P(t; 440\mathrm{MeV}) T_{23}(t)$} \\ \hline
$p$ & Value  & \multicolumn{1}{c}{$C(p,a_0)$}  & \multicolumn{1}{c}{$C(p,a_1)$}  & \multicolumn{1}{c}{$C(p,c_{01})$} \\ \hline
$a_0$ & $0.5194(157)$ \\
$a_1$ & $0.51(25)$ & $0.88$ \\
$c_{01}$ & $0.02(24)$ & $-0.48$ & $-0.08$ \\
$c_{01s}$ & $0.186(110)$ & $-0.13$ & $0.03$ & $0.10$
\\ \hline\hline
\end{tabular}
\end{center}
\end{table}

\begin{table*}
\caption{\label{tab:pt1t2_3ff}Fit results (with correlation matrix)
  determining the dependence of $T_1$ and $T_2$ form factors on the
  strange quark mass.}
\begin{center}
\begin{tabular}{c|c|rrrrrrr} \hline\hline
$p$ & Value  & \multicolumn{1}{c}{$C(p,a_0^{T_1})$}  & 
\multicolumn{1}{c}{$C(p,a_1^{T_1})$} & \multicolumn{1}{c}{$C(p,f_{01}^{T_1})$}  
& \multicolumn{1}{c}{$C(p,g_{01}^{T_1})$} &\multicolumn{1}{c}{$C(p,a_0^{T_2})$}  
& \multicolumn{1}{c}{$C(p,a_1^{T_2})$}  & \multicolumn{1}{c}{$C(p,f_{01}^{T_2})$} 
\\ \hline
$a_0^{T_1}$ & $0.4434(50)$ \\
$a_1^{T_1}$ & $-1.140(61)$ & $0.74$ \\
$f_{01}^{T_1}$ & $1.224(61)$ & $-0.28$ & $-0.01$ \\
$g_{01}^{T_1}$ & $-0.249(85)$ & $-0.18$ & $0.21$ & $0.07$ \\
$a_0^{T_2}$ & $0.3039(43)$ & $0.94$ & $0.91$ & $-0.18$ & $0.00$ \\
$a_1^{T_2}$ & $0.390(62)$ & $0.91$ & $0.95$ & $-0.12$ & $0.05$ & $0.98$ \\
$f_{01}^{T_2}$ & $0.750(38)$ & $-0.49$ & $-0.38$ & $0.21$ & $-0.18$ & $-0.49$ & $-0.43$ \\
$g_{01}^{T_2}$ & $-0.093(47)$ & $0.04$ & $0.12$ & $0.13$ & $0.35$ & $0.04$ & $0.13$ & $-0.07$ \\ \hline\hline
\end{tabular}
\end{center}
\end{table*}

\begin{table*}
\caption{\label{tab:sse_t_6par_sl}Results and correlation matrix of the
fit to $B\to K^*$ form factors $P(t; 135\mathrm{MeV}) T_1(t)$ and 
$P(t;550\mathrm{MeV}) T_2(t)$.  The fit
implements the constraint that $T_1(0) = T_2(0)$.}
\begin{center}
\begin{tabular}{c|c|rrrrrrr} \hline \hline
$p$ & Value  & \multicolumn{1}{c}{$C(p,a_0^{T_1})$} 
& \multicolumn{1}{c}{$C(p,a_1^{T_1})$}  & \multicolumn{1}{c}{$C(p,c_{01}^{T_1})$} 
& \multicolumn{1}{c}{$C(p,a_0^{T_2})$}  & \multicolumn{1}{c}{$C(p,a_1^{T_2})$}  
& \multicolumn{1}{c}{$C(p,c_{01}^{T_2})$} & \multicolumn{1}{c}{$C(p,c_{01s}^{T_1})$} \\ \hline
$a_0^{T_1}$ & $0.422(24)$ \\
$a_1^{T_1}$ & $-1.37(25)$ & $0.48$ \\
$c_{01}^{T_1}$ & $0.71(85)$ & $-0.75$ & $0.05$ \\
$a_0^{T_2}$ & $0.2830(197)$ & $0.86$ & $0.81$ & $-0.43$ \\
$a_1^{T_2}$ & $0.10(24)$ & $0.82$ & $0.86$ & $-0.37$ & $0.91$ \\
$c_{01}^{T_2}$ & $0.45(46)$ & $-0.51$ & $-0.32$ & $0.39$ & $-0.64$ & $-0.32$ \\
$c_{01s}^{T_1}$ & $1.223(61)$ & $-0.03$ & $0.02$ & $0.01$ & $-0.01$ & $-0.01$ & $0.00$ \\
$c_{01s}^{T_2}$ & $0.750(38)$ & $-0.00$ & $0.00$ & $0.00$ & $-0.01$ & $0.01$ & $0.01$ & $0.00$
\\
\hline\hline
\end{tabular}
\end{center}
\end{table*}

\begin{table*}
\caption{\label{tab:sse_t_6par_ss}Results and correlation matrix of the
fit to $B_s\to \phi$ form factors $P(t; 45\mathrm{MeV}) T_1(t)$ and 
$P(t;440\mathrm{MeV}) T_2(t)$.  The fit implements the constraint that 
$T_1(0) = T_2(0)$.}
\begin{center}
\begin{tabular}{c|c|rrrrrrr} \hline \hline
$p$ & Value  & \multicolumn{1}{c}{$C(p,a_0^{T_1})$}  & 
\multicolumn{1}{c}{$C(p,a_1^{T_1})$}  & \multicolumn{1}{c}{$C(p,c_{01}^{T_1})$}  
& \multicolumn{1}{c}{$C(p,a_0^{T_2})$} & \multicolumn{1}{c}{$C(p,a_1^{T_2})$} 
& \multicolumn{1}{c}{$C(p,c_{01}^{T_2})$} & \multicolumn{1}{c}{$C(p,c_{01s}^{T_1})$} \\ \hline
$a_0^{T_1}$ & $0.4070(104)$ \\
$a_1^{T_1}$ & $-1.093(119)$ & $0.22$ \\
$c_{01}^{T_1}$ & $1.48(59)$ & $-0.67$ & $0.36$ \\
$a_0^{T_2}$ & $0.2890(81)$ & $0.78$ & $0.73$ & $-0.22$ \\
$a_1^{T_2}$ & $0.265(97)$ & $0.74$ & $0.78$ & $-0.18$ & $0.89$ \\
$c_{01}^{T_2}$ & $0.66(24)$ & $-0.48$ & $-0.33$ & $0.25$ & $-0.69$ & $-0.33$ \\
$c_{01s}^{T_1}$ & $0.974(105)$ & $-0.12$ & $0.06$ & $0.02$ & $-0.03$ & $-0.04$ & $0.00$ \\
$c_{01s}^{T_2}$ & $0.658(48)$ & $-0.01$ & $-0.02$ & $-0.00$ & $-0.05$ & $0.01$ & $0.03$ & $0.00$
\\ \hline\hline
\end{tabular}
\end{center}
\end{table*}

\begin{table}
\caption{\label{tab:systematics}Estimates of systematic uncertainties.
Effects due to light quark mass dependence and lattice spacing dependence
are included in the statistical fitting uncertainties.}
\begin{center}
\begin{tabular}{lr} \hline\hline
Source & Size \\ \hline
Truncation of $O(\alpha_s^2)$ terms & 4\% \\
Truncation of $O(\alpha_s\Lambda_{\subrm{QCD}}/m_b)$ terms & 2\% \\
Truncation of $O(\Lambda_{\subrm{QCD}}^2/m_b^2)$ terms & 1\% \\
Mistuning of $m_b$ & $<1 \%$ \\
\hline 
Net systematic uncertainty & 5\% \\
\hline\hline
\end{tabular}
\end{center}
\end{table}

The uncertainties in the fit parameters reflect both statistical
fluctuations and effects due to quark mass extrapolation.  The
inclusion in the fits of a term to account for finite lattice spacing
effects had no significant effect on the results.  It is evident from
the figures that the data from ensemble f0062 and c007 show little
systematic difference.  Therefore we assume that errors due to
discretization are not significant compared to the statistical and
fitting uncertainties.

In Table~\ref{tab:systematics} we summarize our estimates for other
sources of systematic uncertainties.  The largest of these, at 4\%, 
is due to the truncation of $O(\alpha_s^2)$ terms in the perturbative
matching from lattice NRQCD to the continuum, as  discussed in
detail in Sec.~\ref{ssec:matching}.

In the previous section we have already discussed our determination
of the strange quark mass dependence of the form factors.  Whether or
not we interpolate the form factors to the physical strange quark mass
we obtain fit results consistent within errors.  Since we use the
interpolated values, the remaining uncertainty due to mistuned strange
quark mass is negligible compared to other sources of uncertainty.

Partial quenching effects due to different sea and valence strange quark 
masses should also be negligible. This is clear in the pseudoscalar sector
where one can use partially quenched chiral perturbation theory
to predict the size of such effects; sea quark mass effects arise
at one-loop order while valence quark masses affect tree-level
diagrams.  Given that even the leading order linear mass dependence 
in the $B\to V$ form factors is barely significant statistically the
form factors are assumed to be insensitive to the ``loop-suppressed''
effects of partial quenching for the strange quark mass.  

Our calculation of $B_s \to \phi$ form factors neglects disconnected
contributions to the $\phi$ propagator.  The OZI rule suggests
such effects, due to pair annihilation and creation of the valence
strange quark--antiquark pair, are small for vector and axial-vector
mesons.

\begin{table}
\caption{\label{tab:ffratios_sl}$B\to K^*$ form factor ratios.  Statistical
uncertainties were determined by bootstrap analysis.}
\begin{center}
\begin{tabular}{rcrrrr} \hline \hline
Ensemble & $|\VEC{n}|^2$ & \multicolumn{1}{c}{$V/A_1$} & 
\multicolumn{1}{c}{$A_{12}/A_1$} & \multicolumn{1}{c}{$T_1/T_2$} 
& \multicolumn{1}{c}{$T_{23}/T_2$} \\ \hline 
f0062 & 1 & 2.83(17) & 0.70(4) & 2.25(10) & 1.89(7) \\
~ & 2 & 2.62(19) & 0.69(6) & 2.13(14) & 1.84(12) \\
~ & 4 & 2.2(3) & 0.69(10) & 2.0(2) & 1.9(3) \\
c007 & 1 & 2.70(13) & 0.62(9) & 2.16(12) & 1.8(2) \\
~ & 2 & 2.74(20) & 0.79(11) & 2.13(17) & 2.2(3) \\
~ & 4 & 2.5(4) & 0.75(16) & 2.2(4) & 1.9(3) \\
c02 & 1 & 2.57(13) & 0.64(4) & 2.16(11) & 1.73(5) \\
~ & 2 & 2.4(3) & 0.62(5) & 2.03(14) & 1.67(13) \\
~ & 4 & 1.8(4) & 0.73(11) & 1.7(3) & 2.0(2) \\
\hline\hline
\end{tabular}
\end{center}
\end{table}

\begin{table}
\caption{\label{tab:ffratios_ss}$B_s\to \phi$ form factor ratios.  
Statistical uncertainties were determined by bootstrap analysis.}
\begin{center}
\begin{tabular}{rcrrrr} \hline \hline
Ensemble & $|\VEC{n}|^2$ & \multicolumn{1}{c}{$V/A_1$} & 
\multicolumn{1}{c}{$A_{12}/A_1$} & \multicolumn{1}{c}{$T_1/T_2$} 
& \multicolumn{1}{c}{$T_{23}/T_2$} \\ \hline 
f0062 & 1 & 2.83(7) & 0.635(18) & 2.23(7) & 1.77(4) \\
~ & 2 & 2.49(9) & 0.636(19) & 1.96(6) & 1.74(3) \\
~ & 4 & 2.28(12) & 0.73(10) & 1.80(8) & 2.0(3) \\
c007 & 1 & 2.67(7) & 0.627(19) & 2.10(5) & 1.72(4) \\
~ & 2 & 2.48(12) & 0.59(3) & 1.97(11) & 1.69(4) \\
~ & 4 & 2.2(3) & 0.65(5) & 1.61(19) & 1.80(11) \\
c02 & 1 & 2.67(8) & 0.597(18) & 2.11(8) & 1.69(2) \\
~ & 2 & 2.65(16) & 0.60(3) & 2.06(11) & 1.61(9) \\
~ & 4 & 2.2(2) & 0.62(3) & 1.76(15) & 1.68(8) \\
\hline\hline
\end{tabular}
\end{center}
\end{table}

\begin{table}
\caption{\label{tab:ratios_3ff}Fit results and correlation matrices
determining dependence of form factor ratios on strange quark mass.}
\begin{center}
\begin{tabular}{c|c|rrrrr} \hline\hline
\multicolumn{5}{c}{$V/A_1 \times P(t, \Delta m_V)/P(t, \Delta m_{A_1})$} 
\\ \hline
$p$ & Value  &  \multicolumn{1}{c}{$C(p,a_0)$}  &  
\multicolumn{1}{c}{$C(p,a_1)$}  &  \multicolumn{1}{c}{$C(p,f_{01})$}  \\ \hline
$a_0$ & $1.679(97)$ \\
$a_1$ & $-10.38(1.73)$ & $0.96$ \\
$f_{01}$ & $0.321(172)$ & $-0.16$ & $-0.16$ \\
$g_{01}$ & $0.31(23)$ & $-0.11$ & $-0.00$ & $0.13$ \\ \hline
\multicolumn{5}{c}{$A_{12}/A_1$}\\ \hline
$p$ & Value  &  \multicolumn{1}{c}{$C(p,a_0)$}  &  
\multicolumn{1}{c}{$C(p,a_1)$}  &  \multicolumn{1}{c}{$C(p,f_{01})$}  \\ \hline
$a_0$ & $0.687(39)$ \\
$a_1$ & $0.92(68)$ & $0.98$ \\
$f_{01}$ & $-0.590(133)$ & $-0.09$ & $-0.10$ \\
$g_{01}$ & $-0.41(22)$ & $-0.11$ & $-0.03$ & $0.24$ \\ \hline
\multicolumn{5}{c}{$T_1/T_2 \times P(t, \Delta m_{T_1})/P(t, \Delta m_{T_2})$}
\\ \hline
$p$ & Value  &  \multicolumn{1}{c}{$C(p,a_0)$}  &  
\multicolumn{1}{c}{$C(p,a_1)$}  &  \multicolumn{1}{c}{$C(p,f_{01})$}  \\ \hline
$a_0$ & $1.4847(128)$ \\
$a_1$ & $-5.315(141)$ & $-1.00$ \\
$f_{01}$ & $0.230(145)$ & $0.18$ & $-0.18$ \\
$g_{01}$ & $-0.232(198)$ & $-0.09$ & $0.09$ & $0.07$ \\ \hline
\multicolumn{5}{c}{$T_{23}/T_2$} \\ \hline
$p$ & Value  &  \multicolumn{1}{c}{$C(p,a_0)$}  &  
\multicolumn{1}{c}{$C(p,a_1)$}  &  \multicolumn{1}{c}{$C(p,f_{01})$}  \\ \hline
$a_0$ & $1.650(75)$ \\
$a_1$ & $-1.75(1.36)$ & $0.98$ \\
$f_{01}$ & $-0.264(79)$ & $-0.43$ & $-0.40$ \\
$g_{01}$ & $-0.244(163)$ & $-0.16$ & $-0.05$ & $0.29$ \\ \hline \hline
\end{tabular}
\end{center}
\end{table}

\begin{table}
\caption{\label{tab:sse_v_over_a_3par_sl}Results and correlation matrices of
  fits to $B\to K^*$ form factor ratios. The fit of $T_1/T_2$ 
has been constrained to enforce $T_1(0)/T_2(0) = 1$.}
\begin{center}
\begin{tabular}{c|c|rrr} \hline\hline
\multicolumn{5}{c}{$V/A_1 \times P(t, 135\mathrm{MeV})/P(t, 550\mathrm{MeV})$}\\ \hline
$p$ & Value  & \multicolumn{1}{c}{$C(p,a_0)$}  & \multicolumn{1}{c}{$C(p,a_1)$}  & \multicolumn{1}{c}{$C(p,c_{01})$} \\ \hline
$a_0$ & $1.89(28)$ \\
$a_1$ & $-8.7(4.4)$ & $0.83$ \\
$c_{01}$ & $-1.33(1.23)$ & $-0.35$ & $0.18$ \\
$c_{01s}$ & $0.321(172)$ & $-0.04$ & $-0.00$ & $0.01$
\\ \hline
\multicolumn{5}{c}{$A_{12}/A_1$}\\ \hline
$p$ & Value  & \multicolumn{1}{c}{$C(p,a_0)$}  & \multicolumn{1}{c}{$C(p,a_1)$}  & \multicolumn{1}{c}{$C(p,c_{01})$} \\ \hline
$a_0$ & $0.848(154)$ \\
$a_1$ & $1.5(2.2)$ & $0.93$ \\
$c_{01}$ & $-1.59(87)$ & $-0.46$ & $-0.14$ \\
$c_{01s}$ & $-0.589(133)$ & $-0.02$ & $0.01$ & $0.02$
\\ \hline
\multicolumn{5}{c}{$T_1/T_2 \times P(t, 135\mathrm{MeV})/P(t, 550\mathrm{MeV})$}\\ \hline
$p$ & Value  & \multicolumn{1}{c}{$C(p,a_0)$}  & \multicolumn{1}{c}{$C(p,a_1)$}  & \multicolumn{1}{c}{$C(p,c_{01})$} \\ \hline
$a_0$ & $1.530(52)$ \\
$a_1$ & $-5.62(55)$ & $-1.00$ \\
$c_{01}$ & $-0.18(87)$ & $-0.82$ & $0.82$ \\
$c_{01s}$ & $0.231(145)$ & $-0.09$ & $0.09$ & $0.01$
\\ \hline
\multicolumn{5}{c}{$T_{23}/T_2$}\\ \hline
$p$ & Value  & \multicolumn{1}{c}{$C(p,a_0)$}  & \multicolumn{1}{c}{$C(p,a_1)$}  & \multicolumn{1}{c}{$C(p,c_{01})$} \\ \hline
$a_0$ & $2.15(26)$ \\
$a_1$ & $1.9(3.8)$ & $0.93$ \\
$c_{01}$ & $-1.52(49)$ & $-0.36$ & $-0.03$ \\
$c_{01s}$ & $-0.263(198)$ & $-0.06$ & $0.02$ & $0.07$ 
\\ \hline\hline
\end{tabular}
\end{center}
\end{table}

\begin{table}
\caption{\label{tab:sse_v_over_a_3par_ss}Results and correlation
  matrices fits to $B_s\to \phi$ form factor
  ratios. The fit of $T_1/T_2$ has been constrained to enforce
  $T_1(0)/T_2(0) = 1$.}
\begin{center}
\begin{tabular}{c|c|rrr} \hline\hline
\multicolumn{5}{c}{$V/A_1 \times P(t, 45\mathrm{MeV})/P(t, 440\mathrm{MeV})$}\\ \hline
$p$ & Value  & \multicolumn{1}{c}{$C(p,a_0)$}  & \multicolumn{1}{c}{$C(p,a_1)$}  & \multicolumn{1}{c}{$C(p,c_{01})$} \\ \hline
$a_0$ & $1.646(165)$ \\
$a_1$ & $-10.7(2.6)$ & $0.90$ \\
$c_{01}$ & $0.34(82)$ & $-0.53$ & $-0.16$ \\
$c_{01s}$ & $0.64(29)$ & $-0.10$ & $0.01$ & $0.05$
\\ \hline
\multicolumn{5}{c}{$A_{12}/A_1$}\\ \hline
$p$ & Value  & \multicolumn{1}{c}{$C(p,a_0)$}  & \multicolumn{1}{c}{$C(p,a_1)$}  & \multicolumn{1}{c}{$C(p,c_{01})$} \\ \hline
$a_0$ & $0.683(54)$ \\
$a_1$ & $0.26(87)$ & $0.92$ \\
$c_{01}$ & $-0.86(49)$ & $-0.38$ & $-0.04$ \\
$c_{01s}$ & $-0.91(25)$ & $-0.07$ & $0.05$ & $0.05$
\\ \hline
\multicolumn{5}{c}{$T_1/T_2 \times P(t, 45\mathrm{MeV})/P(t, 440\mathrm{MeV})$}\\ \hline
$p$ & Value  & \multicolumn{1}{c}{$C(p,a_0)$}  & \multicolumn{1}{c}{$C(p,a_1)$}  & \multicolumn{1}{c}{$C(p,c_{01})$} \\ \hline
$a_0$ & $1.472(34)$ \\
$a_1$ & $-5.43(39)$ & $-1.00$ \\
$c_{01}$ & $0.15(66)$ & $-0.85$ & $0.85$ \\
$c_{01s}$ & $-0.01(24)$ & $-0.24$ & $0.24$ & $0.05$
\\ \hline
\multicolumn{5}{c}{$T_{23}/T_2$}\\ \hline
$p$ & Value  & \multicolumn{1}{c}{$C(p,a_0)$}  & \multicolumn{1}{c}{$C(p,a_1)$}  & \multicolumn{1}{c}{$C(p,c_{01})$} \\ \hline
$a_0$ & $1.730(95)$ \\
$a_1$ & $-1.26(1.68)$ & $0.92$ \\
$c_{01}$ & $-0.56(33)$ & $-0.25$ & $0.11$ \\
$c_{01s}$ & $-0.448(178)$ & $-0.08$ & $0.05$ & $0.09$
\\ \hline\hline
\end{tabular}
\end{center}
\end{table}

\begin{table}
\caption{\label{tab:ffratios2}Form factors ratios subject to Isgur-Wise
relations.}
\begin{center}
\begin{tabular}{rcrrrr} \hline\hline
& & \multicolumn{2}{c}{$B\to K^*$} & \multicolumn{2}{c}{$B_s\to \phi$} \\
Ensemble & $|\VEC{n}|^2$ & \multicolumn{1}{c}{$V/T_1$}
& \multicolumn{1}{c}{$A_1/T_2$} & \multicolumn{1}{c}{$V/T_1$} 
& \multicolumn{1}{c}{$A_1/T_2$} \\ \hline 
f0062 & 0 & \multicolumn{1}{c}{$\cdots$} & 0.999(12) & \multicolumn{1}{c}{$\cdots$} & 0.994(10) \\
~ & 1 & 1.25(5) & 0.99(3) & 1.26(2) & 0.988(17) \\
~ & 2 & 1.22(5) & 0.99(6) & 1.26(4) & 0.991(17) \\
~ & 3 & 1.21(4) & \multicolumn{1}{c}{$\cdots$} & 1.23(2) & \multicolumn{1}{c}{$\cdots$} \\
~ & 4 & 1.12(11) & 1.01(7) & 1.25(7) & 0.99(3) \\
c007 & 0 & \multicolumn{1}{c}{$\cdots$} & 0.97(5) & \multicolumn{1}{c}{$\cdots$} & 1.051(10) \\
~ & 1 & 1.30(4) & 1.04(4) & 1.318(19) & 1.036(15) \\
~ & 2 & 1.34(10) & 1.04(3) & 1.31(5) & 1.040(18) \\
~ & 3 & 1.33(9) & \multicolumn{1}{c}{$\cdots$} & 1.34(5) & \multicolumn{1}{c}{$\cdots$} \\
~ & 4 & 1.12(20) & 0.98(5) & 1.42(16) & 1.04(4) \\
c02 & 0 & \multicolumn{1}{c}{$\cdots$} & 1.046(15) & \multicolumn{1}{c}{$\cdots$} & 1.044(9) \\
~ & 1 & 1.24(8) & 1.04(3) & 1.33(6) & 1.046(11) \\
~ & 2 & 1.22(9) & 1.03(5) & 1.34(9) & 1.04(3) \\
~ & 3 & 1.25(10) & \multicolumn{1}{c}{$\cdots$} & 1.32(10) & \multicolumn{1}{c}{$\cdots$} \\
~ & 4 & 1.10(17) & 1.03(9) & 1.29(12) & 1.02(3) \\
\hline\hline
\end{tabular}
\end{center}
\end{table}

One question is still left to be addressed: are there any significant quark
mass-dependent effects in the form factors as the quark masses are tuned to
their physical values so that the vector mesons become unstable in the lattice
calculations?  In the case which can be studied in heavy-meson chiral
perturbation theory, the $B\to D^*$ form factor at zero recoil $h_{A_1}(1)$, 
one finds a cusp at the quark mass corresponding to the $D \pi$ threshold
which is about a 2\% effect \cite{Randall:1993qg,Hashimoto:2001nb}; the cusp
is even smaller taking into account staggered quark effects
\cite{Laiho:2005ue}.  

Of course this observation does not constitute a
reliable estimate of the systematic uncertainty due to $K\pi$ or
$KK$ thresholds.  However, we do note that the form factors, extrapolated
to low $q^2$, generally agree with determinations from light-cone sum rules
which have systematic errors of a different nature.  Given that the
$\phi$ is relatively narrow compared to the $K^*$, one might expect the
threshold effects to be smaller for $B_s \to \phi$ form factors than
for $B_{(s)} \to K^*$.  In order to make progress, more theoretical
work is necessary to understand how to use LQCD to compute matrix elements
involving unstable resonances.

In some phenomenological studies ratios of form factors have been
used or extracted from data \cite{Hambrock:2012dg,Hambrock:2013zya}.
Here we provide tables of the lattice data and simplified series
expansion fits directly to several ratios, so that correlations may
properly be taken into account.  For the ratios, the fit function
is simply generalized from (\ref{eq:sse_3par}) in order to take into
account the poles in both numerator and denominator:
\begin{equation}
\frac{F_1(t)}{F_2(t)} ~=~ \frac{P_2(t)}{P_1(t)}
[a_0 (1 + c_{01} \Delta x + c_{01s} \Delta x_s) + a_1 z] \,.
\label{eq:sse_ratio_3par}
\end{equation}
Monte Carlo data for form factor ratios are given in
Tables~\ref{tab:ffratios_sl} and \ref{tab:ffratios_ss} for $B\to K^*$
and $B_s\to\phi$, respectively.  The fit results describing the
dependence on the strange quark mass appear in
Table~\ref{tab:ratios_3ff}.  The fits to the shapes of the form factor
ratios are given in Tables~\ref{tab:sse_v_over_a_3par_sl} and
\ref{tab:sse_v_over_a_3par_ss} for $B\to K^*$ and $B_s\to\phi$,
respectively.

\section{Remarks and conclusions}
\label{sec:concl}

As Figs.~\ref{fig:pff_sl} and \ref{fig:pff_ss} show, the form factors
calculated from lattice QCD at low recoil appear broadly consistent
with light cone sum rule determinations at large recoil.  The fits to
the lattice data included only constant and linear terms in $z$, after
removing the pole factor.  The presence of a $z^2$ term, not necessary
to fit the lattice data, would affect the extrapolation of lattice
results to low $q^2$.  One possibility would be to fit both lattice
and sum rule results to obtain a parametrization of the form factors
over the whole physical range of $q^2$ (e.g.\ as in
Ref.~\cite{Bharucha:2010im}).  However, given that short-distance
physics can only be isolated well away from sharp resonances $q^2 <
m_{J/\psi}^2$ and $q^2 > m_{\psi'}^2$ one might choose to use the
lattice results for low recoil observables and sum rule results for
large recoil observables.  If it is possible to obtain precise data in
the range $m_{J/\psi}^2 < q^2 < m_{\psi'}^2$, well separated from the
resonances, then a combination of lattice and sum rule results would
be well motivated.

In addition to providing results useful as inputs to Standard Model
(or BSM) predictions for observables, we can consider ratios of
form factors which test the accuracy of the heavy quark expansion.
To leading order in $\Lambda_{\subrm{QCD}}/m_b$, the one-loop improved
Isgur-Wise relations \cite{Isgur:1990kf} are
\begin{equation}
\frac{V}{T_1} \;=\; \kappa ~~~\mbox{and}~~~ \frac{A_1}{T_2} \;=\; \kappa\,,
\label{eq:isgurwise}
\end{equation}
where \cite{Grinstein:2004vb,Bobeth:2010wg}
\begin{equation}
\kappa \;=\; 1 \;-\; \frac{2\alpha_s}{3\pi}\,\log\frac{\mu}{m_b} \,.
\end{equation}
In our calculation, the tensor form factors are quoted with $\mu = m_b$.
The lattice data for these ratios are tabulated in Table~\ref{tab:ffratios2}.
We note that the $O(\Lambda_{\subrm{QCD}}/m_b)$ corrections to 
(\ref{eq:isgurwise}) are rather large for $V/T_1$ at about 25\%, while
significantly smaller for $A_1/T_2$, less than 10\%.

The calculations presented in this paper improve on what is known
about the $B \to V$ form factors, especially at large $q^2$, which
corresponds to the kinematic range accessible on the lattice.  Effects
of dynamical up, down, and strange quarks are included, allowing us to
move beyond the previous quenched determinations of the $B \to V$ form
factors.  These earlier studies also extrapolated results in heavy
quark mass from charm-scale simulations, while NRQCD permits us to
calculate form factors directly in the bottom quark sector with presently
accessible lattice spacings.
We performed a high-statistics study in order to combat the
poorer signal-to-noise effects present with vector meson correlation
functions.  In order to improve upon our calculation, the uncertainty
due to the current matching must be reduced, and calculations with lighter
quark masses and a finer lattice spacing should be done. 

The implications of the lattice QCD calculations presented here
are the subject for another paper \cite{Horgan:2013pva}.  There
we present results for $B\to K^*\mu^+\mu^-$ and $B_s\to\phi\mu^+\mu^-$
observables at low recoil, both for the Standard Model and beyond.  
It may be interesting to combine results from lattice and light cone
sum rules in order to determine the form factors over the whole
kinematic range.  Whether this would aid the search for BSM effects
in rare $b$ decays remains to be seen. 

\section*{ACKNOWLEDGMENTS}

This work was supported in part by an STFC Special Programme Grant
(PP/E006957/1).  RRH and MW are supported by an STFC Consolidated
Grant.  MW thanks the IPPP in Durham for an Associateship which
supported travel facilitating this work.  ZL is partially supported by
NSFC under the Project No.~11105153, the Youth Innovation Promotion
Association of CAS and the Scientific Research Foundation for ROCS,
SEM.  SM is supported by the U.S.~Department of Energy under
cooperative research agreement Contract Number
No.~DE-FG02-94ER40818. UKQCD and USQCD computational resources made
this work possible, including the DiRAC facility jointly funded by the
STFC, the Large Facilities Capital Fund of BIS and the Universities of
Cambridge and Glasgow.  We are grateful to our HPQCD Collaboration
colleagues for discussions and support.

\clearpage

\section*{APPENDIX: DATA TABLES AND $B_s \to K^*$ FORM FACTORS}
\label{sec:appendix}

\begin{table}[t]
\caption{\label{tab:va_ff_sl}Form factors parametrizing $B\to K^*$ vector and
axial-vector matrix elements.}
\begin{center}
\begin{tabular}{rcrrrr} \hline \hline
Ensemble & $|\VEC{n}|^2$ & \multicolumn{1}{c}{$V$} & \multicolumn{1}{c}{$A_0$}
 & \multicolumn{1}{c}{$A_1$} & \multicolumn{1}{c}{$A_{12}$} \\ \hline 
f0062 & 0 & \multicolumn{1}{r}{$\cdots$} & \multicolumn{1}{r}{$\cdots$} & 0.647(13) & \multicolumn{1}{r}{$\cdots$} \\
~ & 1 & 1.61(6) & 1.60(6) & 0.57(3) & 0.399(15) \\
~ & 2 & 1.40(5) & 1.44(5) & 0.53(4) & 0.37(2) \\
~ & 3 & 1.23(7) & 1.36(9) & \multicolumn{1}{r}{$\cdots$} & 0.37(2) \\
~ & 4 & 1.13(14) & 1.10(9) & 0.52(5) & 0.36(3) \\
c007 & 0 & \multicolumn{1}{r}{$\cdots$} & \multicolumn{1}{r}{$\cdots$} & 0.61(2) & \multicolumn{1}{r}{$\cdots$} \\
~ & 1 & 1.59(8) & 1.38(15) & 0.59(3) & 0.36(5) \\
~ & 2 & 1.55(12) & 1.48(12) & 0.57(3) & 0.45(6) \\
~ & 3 & 1.51(13) & 1.1(2) & \multicolumn{1}{r}{$\cdots$} & 0.31(17) \\
~ & 4 & 1.16(17) & 0.99(14) & 0.46(5) & 0.34(5) \\
c02 & 0 & \multicolumn{1}{r}{$\cdots$} & \multicolumn{1}{r}{$\cdots$} & 0.653(12) & \multicolumn{1}{r}{$\cdots$} \\
~ & 1 & 1.60(8) & 1.72(7) & 0.622(18) & 0.40(3) \\
~ & 2 & 1.39(13) & 1.36(7) & 0.58(3) & 0.36(2) \\
~ & 3 & 1.27(8) & 1.34(8) & \multicolumn{1}{r}{$\cdots$} & 0.36(4) \\
~ & 4 & 0.92(18) & 1.17(13) & 0.51(5) & 0.37(4) \\
\hline\hline
\end{tabular}
\end{center}
\end{table}

\begin{table}[t]
\caption{\label{tab:t_ff_sl}Form factors parametrizing $B\to K^*$ 
tensor matrix elements.}
\begin{center}
\begin{tabular}{rcrrr} \hline \hline
Ensemble & $|\VEC{n}|^2$ & \multicolumn{1}{c}{$T_1$} & \multicolumn{1}{c}{$T_2$}
 & \multicolumn{1}{c}{$T_{23}$} \\ \hline 
f0062 & 0 & \multicolumn{1}{r}{$\cdots$} & 0.648(14) & \multicolumn{1}{r}{$\cdots$} \\
~ & 1 & 1.29(5) & 0.58(2) & 1.09(3) \\
~ & 2 & 1.15(5) & 0.54(3) & 0.99(5) \\
~ & 3 & 1.02(5) & \multicolumn{1}{r}{$\cdots$} & 0.94(5) \\
~ & 4 & 1.00(9) & 0.51(5) & 0.96(8) \\
c007 & 0 & \multicolumn{1}{r}{$\cdots$} & 0.63(4) & \multicolumn{1}{r}{$\cdots$} \\
~ & 1 & 1.23(6) & 0.57(2) & 1.01(13) \\
~ & 2 & 1.16(9) & 0.55(2) & 1.21(19) \\
~ & 3 & 1.14(9) & \multicolumn{1}{r}{$\cdots$} & 0.7(4) \\
~ & 4 & 1.05(17) & 0.47(5) & 0.86(13) \\
c02 & 0 & \multicolumn{1}{r}{$\cdots$} & 0.624(14) & \multicolumn{1}{r}{$\cdots$} \\
~ & 1 & 1.29(6) & 0.596(15) & 1.03(2) \\
~ & 2 & 1.14(6) & 0.56(3) & 0.93(4) \\
~ & 3 & 1.01(6) & \multicolumn{1}{r}{$\cdots$} & 0.91(5) \\
~ & 4 & 0.82(13) & 0.49(4) & 0.96(7) \\
\hline \hline
\end{tabular}
\end{center}
\end{table}

\begin{table}[b]
\caption{\label{tab:kinematic_sl}Kinematic variables for $B\to K^*$
form factor calculations.}
\begin{center}
\begin{tabular}{rcrrr} \hline\hline
Ensemble & $|\VEC{n}|^2$ & $E_{K^*} (\mathrm{GeV})$ & $q^2 (\mathrm{GeV}^2)$ & 
$z(q^2, 12~\mathrm{GeV}^2)$ \\ \hline 
f0062 & 0 & 1.033(6) & 20.66(5) & $-0.0795(7)$ \\
 & 1 & 1.151(10) & 19.33(10) & $-0.0655(10)$ \\
 & 2 & 1.302(10) & 17.73(9) & $-0.0497(9)$ \\
 & 3 & 1.391(13) & 16.71(11) & $-0.0401(10)$ \\
 & 4 & 1.49(2) & 15.64(17) & $-0.0305(15)$ \\ \hline
c007 & 0 & 1.031(9) & 20.37(8) & $-0.0773(10)$ \\
 & 1 & 1.176(11) & 18.81(9) & $-0.0610(10)$ \\
 & 2 & 1.296(18) & 17.50(15) & $-0.0481(14)$ \\
 & 3 & 1.391(18) & 16.43(14) & $-0.0380(14)$ \\
 & 4 & 1.50(3) & 15.3(2) & $-0.0276(19)$ \\ \hline
c02 & 0 & 1.103(5) & 20.12(4) & $-0.0709(5)$ \\
 & 1 & 1.234(13) & 18.70(9) & $-0.0570(9)$ \\
 & 2 & 1.362(13) & 17.33(10) & $-0.0443(10)$ \\
 & 3 & 1.465(15) & 16.21(12) & $-0.0342(9)$ \\
 & 4 & 1.59(2) & 14.93(18) & $-0.0234(15)$ \\
\hline \hline
\end{tabular}
\end{center}
\end{table}

In Tables~\ref{tab:va_ff_sl} and \ref{tab:t_ff_sl} we give the $B \to
K^*$ form factors computed on each lattice ensemble.  For each
ensemble, as listed in Table~\ref{tab:gaugeparam}, we compute matrix
elements with the $K^*$ momentum equal to $\VEC{k} =
2\pi\VEC{n}/aN_x$, with $\VEC{n} = (0,0,0)$, $(1,0,0)$, $(1,1,0)$,
$(1,1,1)$, $(2,0,0)$ and their rotational equivalents.  There are some
blank entries in the tables: in our scheme for extracting the form
factors from Eq.~(\ref{eq:Vformula})--(\ref{eq:T23formula}), some form
factors cannot be extracted when $\VEC{k} = 0$, and $A_1$ and $T_2$
cannot be isolated without having some component of $\VEC{k}$ equal to
0.  For reference, in Table~\ref{tab:kinematic_sl} we also provide the
$K^*$ energies, $q^2$ values, and the numerical value for $z(q^2,
t_0=12~\mathrm{GeV}^2)$.  Tables~\ref{tab:va_ff_ss} and
\ref{tab:t_ff_ss} give the lattice results for the $B_s\to \phi$ form
factors, and Table~\ref{tab:kinematic_ss} give the kinematic values.

\begin{figure*}
\begin{center}
\includegraphics[width=0.45\textwidth]{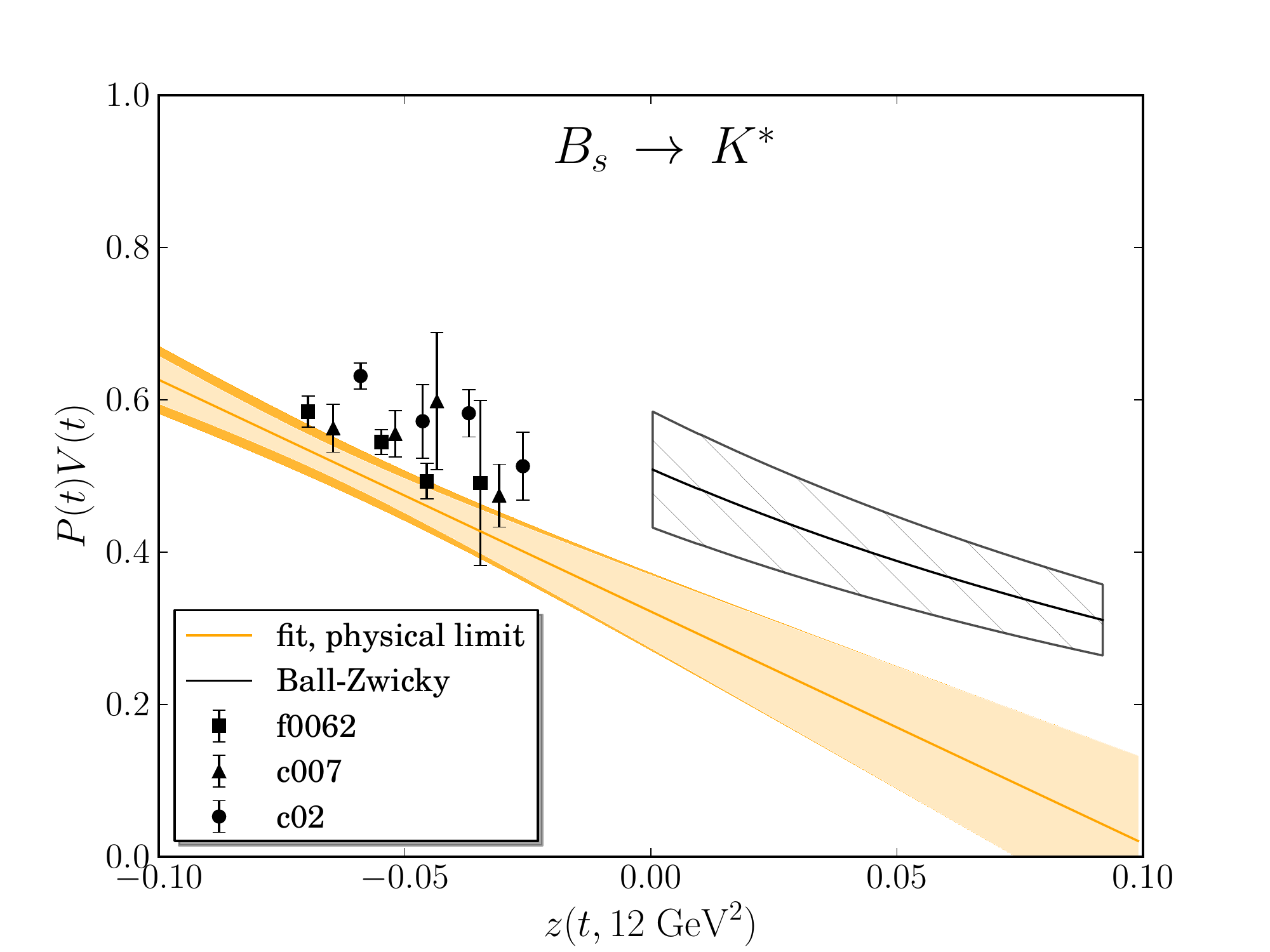}
\includegraphics[width=0.45\textwidth]{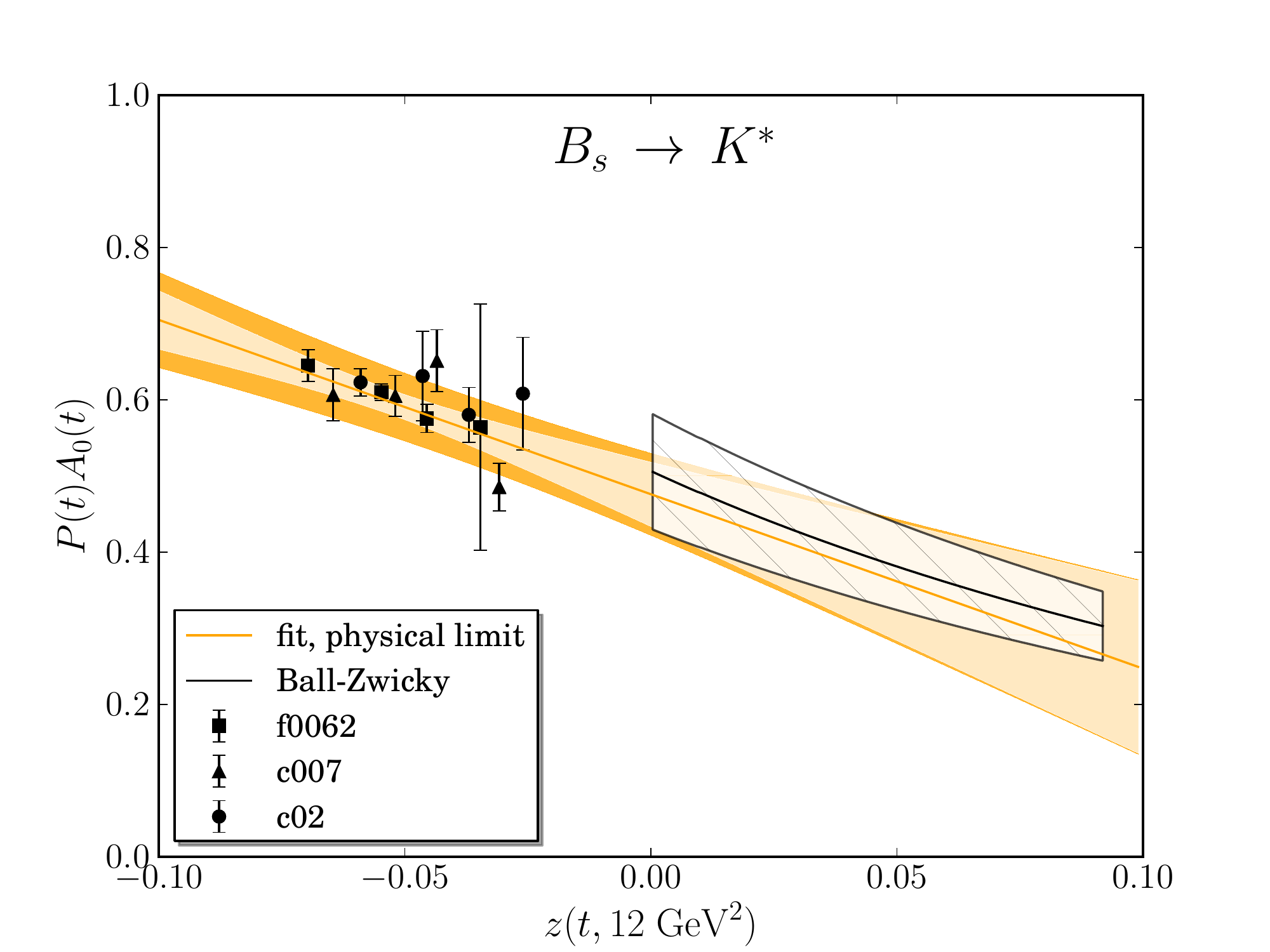}
\includegraphics[width=0.45\textwidth]{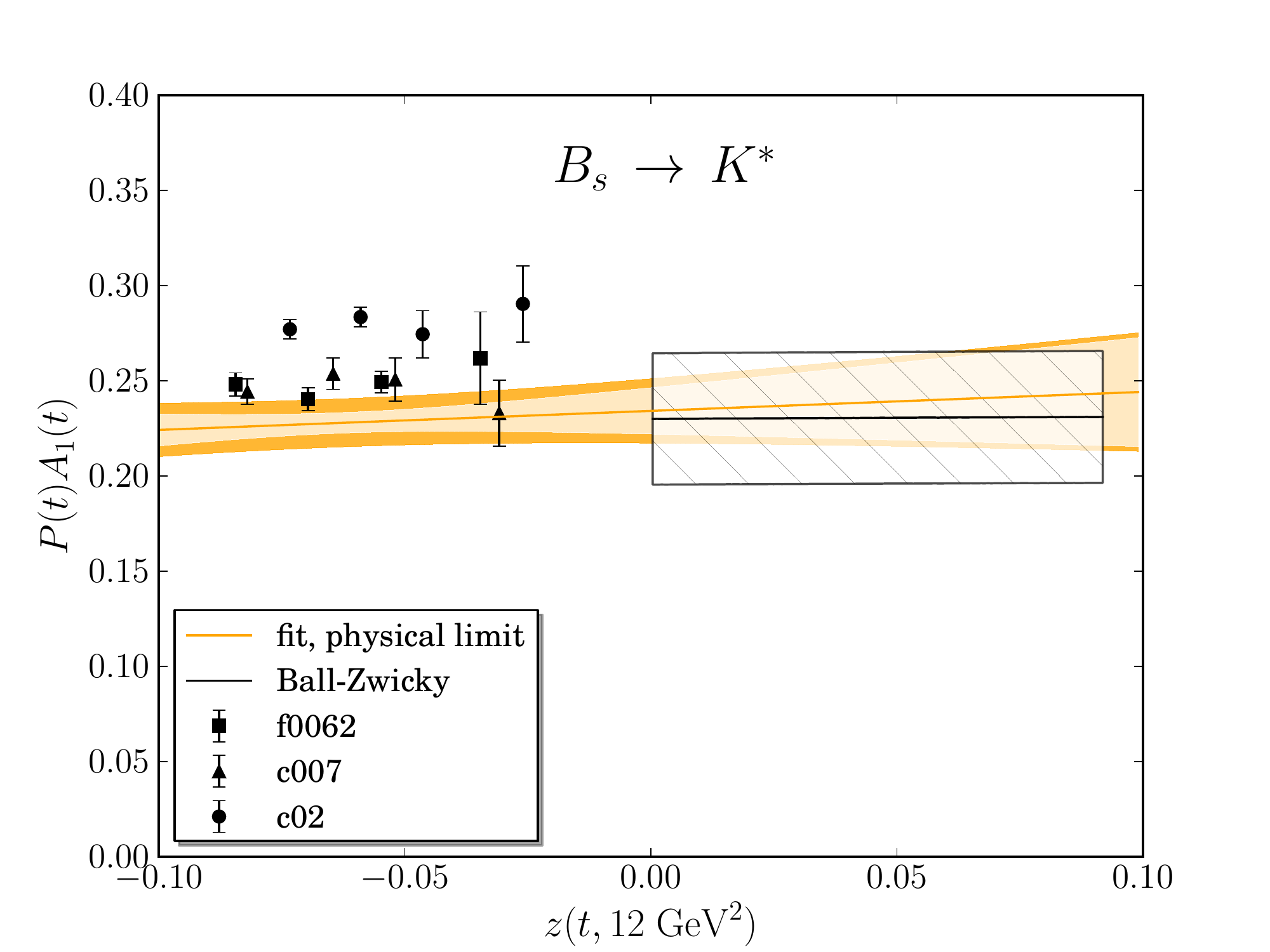}
\includegraphics[width=0.45\textwidth]{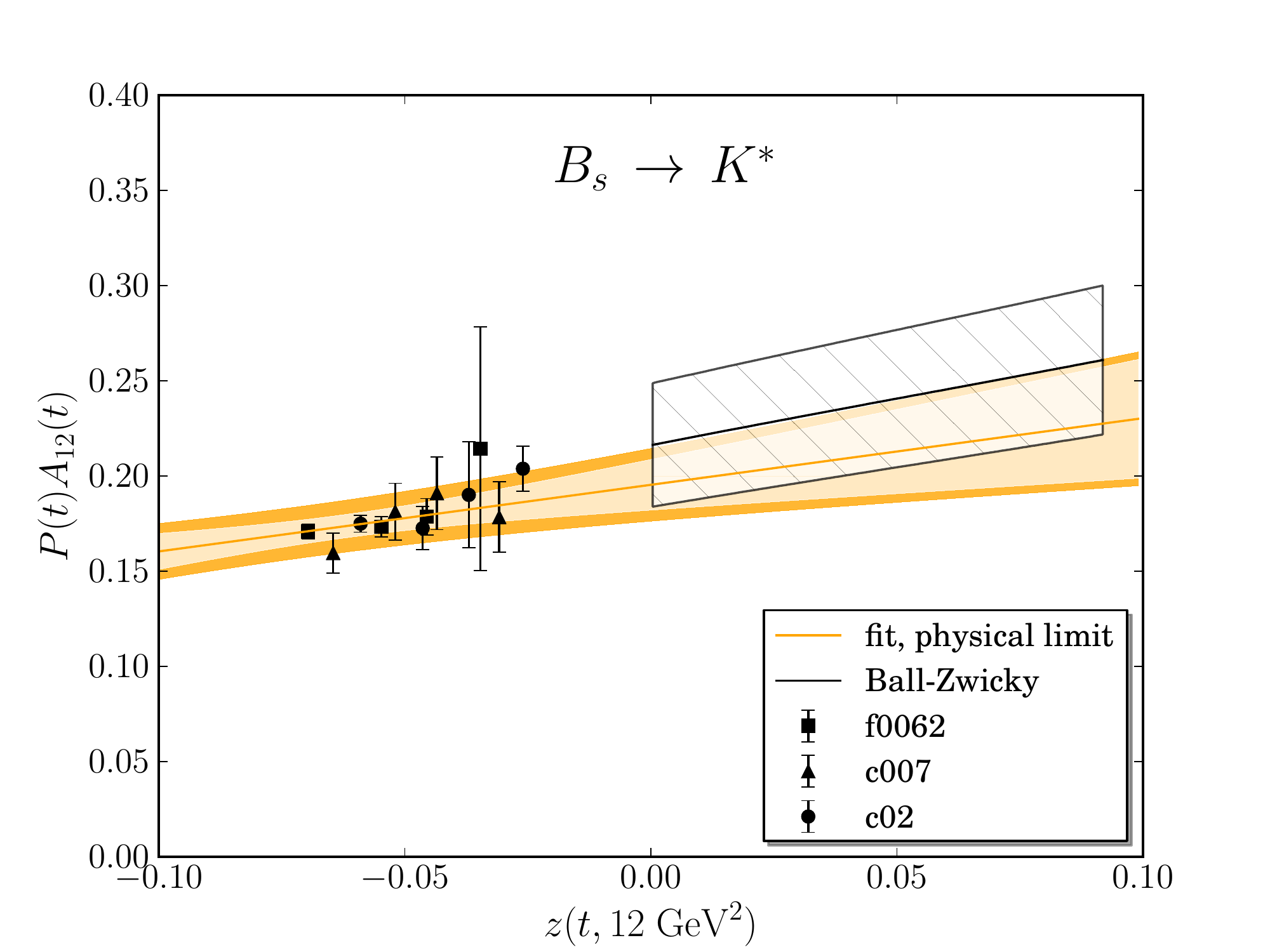}
\includegraphics[width=0.45\textwidth]{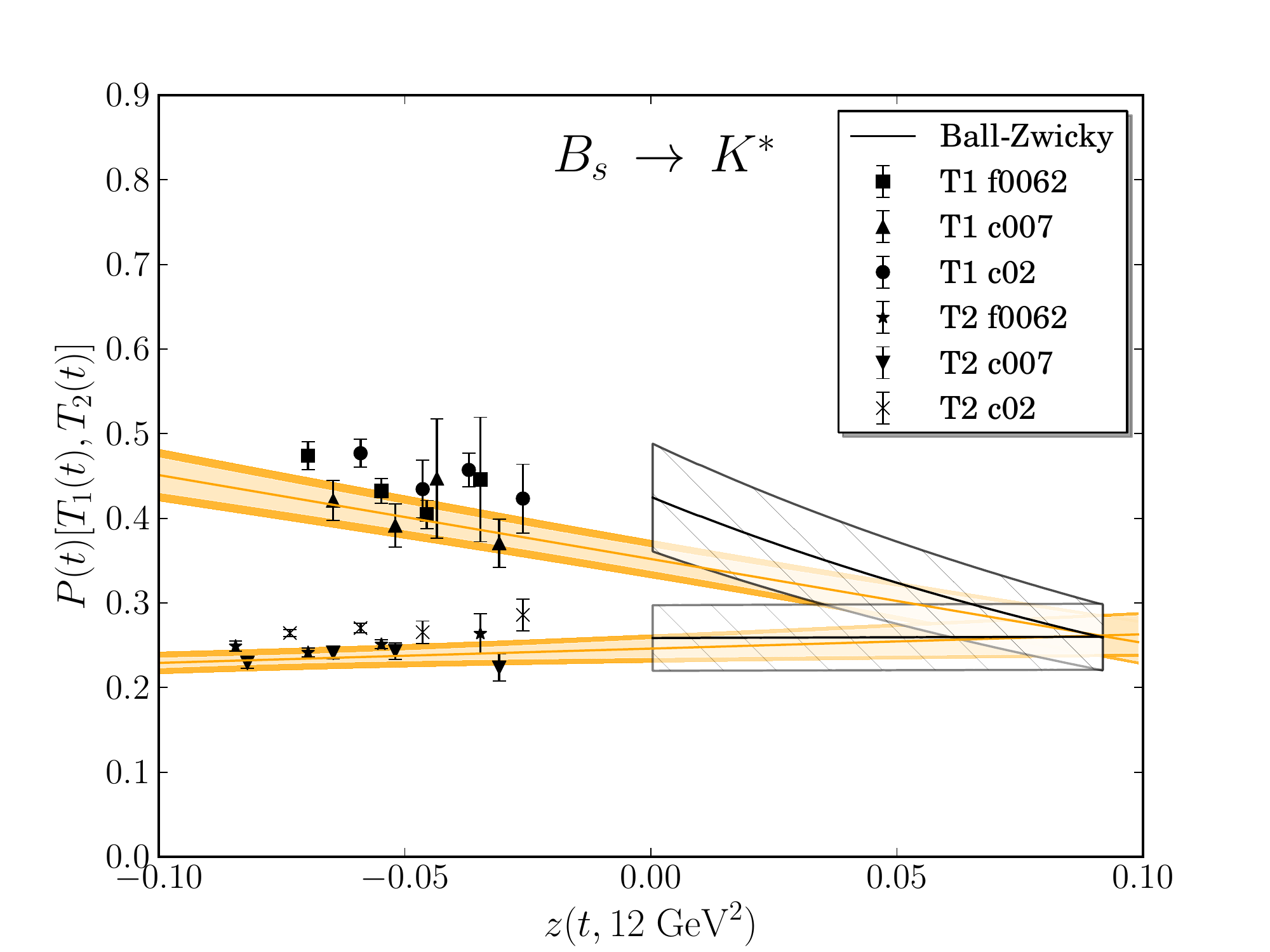}
\includegraphics[width=0.45\textwidth]{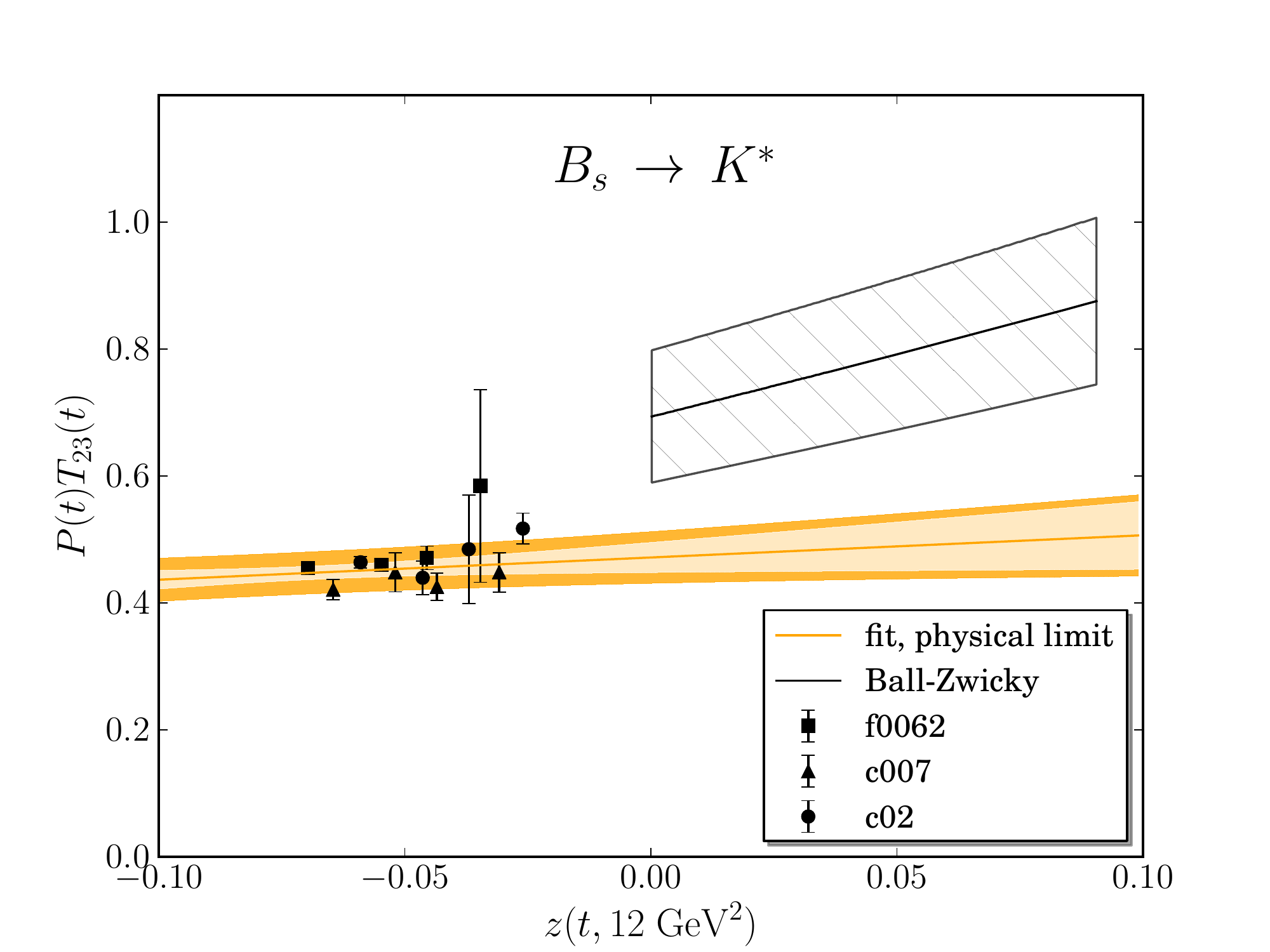}
\end{center}
\caption{\label{fig:pff_ls}$B_s \to K^*$ form factors, as in
Fig.~\ref{fig:pff_sl}.}
\end{figure*}

The decay $B_s \to K^* \ell\nu$ occurs at tree level in the Standard
Model, so its precise measurement and comparison to the Standard Model
is less likely to reveal physics beyond the Standard Model.  In fact
the $B_s\to K^*$ form factors may become useful in future analyses of
$b\to d$ decays using the mode $B_s \to \bar{K}^{*0}\ell^+\ell^-$.  We
have calculated the same set of form factors for $B_s \to K^*$ exactly
in the manner described for $B \to K^*$ and $B_s \to \phi$ in the main
body of the paper, simply by changing the quark masses appropriately.
In Table~\ref{tab:va_ff_ls} we give the data obtained for the $B_s \to
K^*$ form factors for vector and axial-vector matrix elements.  We
also give the tensor-current form factor data in
Table~\ref{tab:t_ff_ls}.  The data and corresponding fits are shown in
Fig.~\ref{fig:pff_ls}.  For reference, we also provide a table of
the kinematic variables used in the fits
(Table~\ref{tab:kinematic_ls}) and a table of the $B_s \to K^*$ form
factor ratio data (Table~\ref{tab:ffratios_ls}).  Tables
\ref{tab:sse_va_3par_ls} and \ref{tab:sse_t_6par_ls} give the final
fit parameters and correlation matrices for the seven $B_s \to K^*$ form
factors.

\begin{table}[t]
\caption{\label{tab:va_ff_ss}Form factors parametrizing $B_s \to \phi$ 
vector and axial-vector matrix elements.}
\begin{center}
\begin{tabular}{rcrrrr}\hline \hline
Ensemble & $|\VEC{n}|^2$ & \multicolumn{1}{c}{$V$} & \multicolumn{1}{c}{$A_0$}
 & \multicolumn{1}{c}{$A_1$} & \multicolumn{1}{c}{$A_{12}$} \\ \hline 
f0062 & 0 & \multicolumn{1}{r}{$\cdots$} & \multicolumn{1}{r}{$\cdots$} & 0.638(7) & \multicolumn{1}{r}{$\cdots$} \\
~ & 1 & 1.63(3) & 1.61(2) & 0.577(13) & 0.367(8) \\
~ & 2 & 1.38(4) & 1.411(19) & 0.555(10) & 0.353(9) \\
~ & 3 & 1.24(2) & 1.29(2) & \multicolumn{1}{r}{$\cdots$} & 0.36(2) \\
~ & 4 & 1.18(4) & 1.21(7) & 0.519(19) & 0.38(5) \\
c007 & 0 & \multicolumn{1}{r}{$\cdots$} & \multicolumn{1}{r}{$\cdots$} & 0.649(8) & \multicolumn{1}{r}{$\cdots$} \\
~ & 1 & 1.57(3) & 1.57(3) & 0.588(12) & 0.369(11) \\
~ & 2 & 1.41(7) & 1.40(3) & 0.567(18) & 0.332(18) \\
~ & 3 & 1.28(6) & 1.29(5) & \multicolumn{1}{r}{$\cdots$} & 0.34(3) \\
~ & 4 & 1.14(12) & 1.14(4) & 0.51(3) & 0.335(14) \\
c02 & 0 & \multicolumn{1}{r}{$\cdots$} & \multicolumn{1}{r}{$\cdots$} & 0.666(7) & \multicolumn{1}{r}{$\cdots$} \\
~ & 1 & 1.66(5) & 1.63(4) & 0.622(10) & 0.371(11) \\
~ & 2 & 1.54(9) & 1.45(10) & 0.578(19) & 0.35(2) \\
~ & 3 & 1.37(10) & 1.24(4) & \multicolumn{1}{r}{$\cdots$} & 0.337(16) \\
~ & 4 & 1.21(13) & 1.15(8) & 0.55(2) & 0.337(17) \\
\hline \hline
\end{tabular}
\end{center}
\end{table}

\begin{table}[t]
\caption{\label{tab:t_ff_ss}Form factors parametrizing $B_s\to \phi$ 
tensor matrix elements.}
\begin{center}
\begin{tabular}{rcrrr} \hline \hline
Ensemble & $|\VEC{n}|^2$ & \multicolumn{1}{c}{$T_1$} & \multicolumn{1}{c}{$T_2$}
 & \multicolumn{1}{c}{$T_{23}$} \\ \hline 
f0062 & 0 & \multicolumn{1}{r}{$\cdots$} & 0.642(8) & \multicolumn{1}{r}{$\cdots$} \\
~ & 1 & 1.30(3) & 0.584(12) & 1.032(8) \\
~ & 2 & 1.10(3) & 0.560(9) & 0.976(10) \\
~ & 3 & 1.01(2) & \multicolumn{1}{r}{$\cdots$} & 0.98(6) \\
~ & 4 & 0.95(4) & 0.526(17) & 1.06(12) \\
c007 & 0 & \multicolumn{1}{r}{$\cdots$} & 0.617(8) & \multicolumn{1}{r}{$\cdots$} \\
~ & 1 & 1.19(3) & 0.568(13) & 0.980(15) \\
~ & 2 & 1.07(5) & 0.544(13) & 0.922(20) \\
~ & 3 & 0.95(5) & \multicolumn{1}{r}{$\cdots$} & 0.87(2) \\
~ & 4 & 0.79(9) & 0.49(3) & 0.89(2) \\
c02 & 0 & \multicolumn{1}{r}{$\cdots$} & 0.638(6) & \multicolumn{1}{r}{$\cdots$} \\
~ & 1 & 1.26(5) & 0.595(8) & 1.007(11) \\
~ & 2 & 1.14(6) & 0.556(16) & 0.90(5) \\
~ & 3 & 1.03(5) & \multicolumn{1}{r}{$\cdots$} & 0.88(3) \\
~ & 4 & 0.94(8) & 0.534(19) & 0.90(3) \\
\hline \hline
\end{tabular}
\end{center}
\end{table}

\begin{table}[t]
\caption{\label{tab:kinematic_ss}Kinematic variables for $B_s \to \phi$
form factor calculations.}
\begin{center}
\begin{tabular}{rcrrr} \hline\hline
Ensemble & $|\VEC{n}|^2$ & $E_\phi (\mathrm{GeV})$ & $q^2 (\mathrm{GeV}^2)$ & 
$z(q^2, 12~\mathrm{GeV}^2)$ \\ \hline 
f0062 & 0 & 1.133(3) & 20.52(3) & $-0.0716(3)$ \\
 & 1 & 1.242(5) & 19.28(4) & $-0.0598(4)$ \\
 & 2 & 1.360(6) & 17.98(6) & $-0.0480(5)$ \\
 & 3 & 1.459(5) & 16.86(5) & $-0.0383(4)$ \\
 & 4 & 1.565(14) & 15.72(11) & $-0.0287(9)$ \\ \hline
c007 & 0 & 1.137(5) & 20.13(5) & $-0.0688(6)$ \\
 & 1 & 1.263(4) & 18.74(4) & $-0.0556(4)$ \\
 & 2 & 1.358(13) & 17.65(11) & $-0.0457(10)$ \\
 & 3 & 1.448(18) & 16.62(15) & $-0.0368(13)$ \\
 & 4 & 1.591(14) & 15.17(11) & $-0.0246(9)$ \\ \hline
c02 & 0 & 1.161(5) & 20.02(4) & $-0.0667(4)$ \\
 & 1 & 1.276(6) & 18.73(5) & $-0.0547(5)$ \\
 & 2 & 1.403(10) & 17.36(8) & $-0.0426(8)$ \\
 & 3 & 1.501(6) & 16.27(5) & $-0.0333(4)$ \\
 & 4 & 1.620(14) & 15.03(11) & $-0.0232(9)$ \\
\hline \hline
\end{tabular}
\end{center}
\end{table}

\begin{table}[t]
\caption{\label{tab:va_ff_ls}Form factors parametrizing $B_s \to K^*$ 
vector and axial-vector matrix elements.}
\begin{center}
\begin{tabular}{rcrrrr}
\hline \hline
Ensemble & $|\VEC{n}|^2$ & \multicolumn{1}{c}{$V$} & \multicolumn{1}{c}{$A_0$}
 & \multicolumn{1}{c}{$A_1$} & \multicolumn{1}{c}{$A_{12}$} \\ \hline 
f0062 & 0 & \multicolumn{1}{r}{$\cdots$} & \multicolumn{1}{r}{$\cdots$} & 0.609(14) & \multicolumn{1}{r}{$\cdots$} \\
~ & 1 & 1.59(6) & 1.80(6) & 0.538(12) & 0.382(8) \\
~ & 2 & 1.32(4) & 1.52(3) & 0.510(10) & 0.356(11) \\
~ & 3 & 1.10(5) & 1.32(5) & \multicolumn{1}{r}{$\cdots$} & 0.346(19) \\
~ & 4 & 1.0(2) & 1.2(3) & 0.48(4) & 0.39(12) \\
c007 & 0 & \multicolumn{1}{r}{$\cdots$} & \multicolumn{1}{r}{$\cdots$} & 0.598(16) & \multicolumn{1}{r}{$\cdots$} \\
~ & 1 & 1.50(8) & 1.66(9) & 0.558(16) & 0.35(2) \\
~ & 2 & 1.34(7) & 1.50(6) & 0.51(2) & 0.37(3) \\
~ & 3 & 1.33(19) & 1.48(10) & \multicolumn{1}{r}{$\cdots$} & 0.37(4) \\
~ & 4 & 0.99(8) & 1.03(6) & 0.43(3) & 0.33(3) \\
c02 & 0 & \multicolumn{1}{r}{$\cdots$} & \multicolumn{1}{r}{$\cdots$} & 0.649(12) & \multicolumn{1}{r}{$\cdots$} \\
~ & 1 & 1.64(4) & 1.66(5) & 0.605(11) & 0.378(10) \\
~ & 2 & 1.35(11) & 1.52(14) & 0.54(2) & 0.35(2) \\
~ & 3 & 1.24(6) & 1.26(7) & \multicolumn{1}{r}{$\cdots$} & 0.35(5) \\
~ & 4 & 1.01(9) & 1.22(15) & 0.51(3) & 0.359(20) \\
\hline \hline
\end{tabular}
\end{center}
\end{table}

\begin{table}[t]
\caption{\label{tab:t_ff_ls}Form factors parametrizing $B_s\to K^*$ 
tensor matrix elements.}
\begin{center}
\begin{tabular}{rcrrr} \hline \hline
Ensemble & $|\VEC{n}|^2$ & \multicolumn{1}{c}{$T_1$} & \multicolumn{1}{c}{$T_2$}
 & \multicolumn{1}{c}{$T_{23}$} \\ \hline 
f0062 & 0 & \multicolumn{1}{r}{$\cdots$} & 0.611(14) & \multicolumn{1}{r}{$\cdots$} \\
~ & 1 & 1.29(4) & 0.541(11) & 1.017(11) \\
~ & 2 & 1.05(4) & 0.514(11) & 0.946(14) \\
~ & 3 & 0.91(4) & \multicolumn{1}{r}{$\cdots$} & 0.91(3) \\
~ & 4 & 0.89(15) & 0.48(4) & 1.1(3) \\
c007 & 0 & \multicolumn{1}{r}{$\cdots$} & 0.560(15) & \multicolumn{1}{r}{$\cdots$} \\
~ & 1 & 1.12(7) & 0.531(15) & 0.93(3) \\
~ & 2 & 0.95(6) & 0.498(18) & 0.92(6) \\
~ & 3 & 0.99(15) & \multicolumn{1}{r}{$\cdots$} & 0.82(4) \\
~ & 4 & 0.77(6) & 0.41(3) & 0.82(5) \\
c02 & 0 & \multicolumn{1}{r}{$\cdots$} & 0.619(10) & \multicolumn{1}{r}{$\cdots$} \\
~ & 1 & 1.24(4) & 0.577(11) & 1.004(17) \\
~ & 2 & 1.03(7) & 0.52(3) & 0.89(5) \\
~ & 3 & 0.97(4) & \multicolumn{1}{r}{$\cdots$} & 0.90(16) \\
~ & 4 & 0.84(8) & 0.50(3) & 0.91(4) \\
\hline \hline
\end{tabular}
\end{center}
\end{table}

\begin{table}[t]
\caption{\label{tab:kinematic_ls}Kinematic variables for $B_s \to K^*$ form
factor calculations.}
\begin{center}
\begin{tabular}{rcrrr}
\hline\hline
Ensemble & $|\VEC{n}|^2$ & $E_{K^*}$ (GeV) & $q^2$ (GeV${}^2$) & $
z(q^2, 12 \mathrm{GeV}^2)$ \\ \hline 
f0062 & 0 & 1.034(4) & 21.44(4) & $-0.0844(5)$ \\
 & 1 & 1.161(8) & 20.00(7) & $-0.0696(7)$ \\
 & 2 & 1.302(9) & 18.47(7) & $-0.0548(7)$ \\
 & 3 & 1.385(14) & 17.47(12) & $-0.0455(11)$ \\
 & 4 & 1.50(3) & 16.3(3) & $-0.035(3)$ \\ \hline
c007 & 0 & 1.031(7) & 21.10(7) & $-0.0820(9)$ \\
 & 1 & 1.189(7) & 19.40(6) & $-0.0646(8)$ \\
 & 2 & 1.306(12) & 18.09(11) & $-0.0519(11)$ \\
 & 3 & 1.38(3) & 17.2(3) & $-0.043(3)$ \\
 & 4 & 1.518(14) & 15.77(12) & $-0.0308(11)$ \\ \hline
c02 & 0 & 1.106(5) & 20.52(5) & $-0.0733(6)$ \\
 & 1 & 1.241(5) & 19.04(5) & $-0.0590(5)$ \\
 & 2 & 1.367(8) & 17.66(7) & $-0.0464(6)$ \\
 & 3 & 1.461(11) & 16.60(8) & $-0.0370(7)$ \\
 & 4 & 1.59(2) & 15.31(16) & $-0.0260(14)$ \\
\hline\hline
\end{tabular}
\end{center}
\end{table}

\begin{table}
\caption{\label{tab:ffratios_ls}$B_s\to K^*$ form factor ratios.  
Statistical uncertainties were determined by bootstrap analysis.}
\begin{center}
\begin{tabular}{rcrrrr} \hline \hline
Ensemble & $|\VEC{n}|^2$ & \multicolumn{1}{c}{$V/A_1$} & 
\multicolumn{1}{c}{$A_{12}/A_1$} & \multicolumn{1}{c}{$T_1/T_2$} & 
\multicolumn{1}{c}{$T_{23}/T_2$} \\ \hline 
f0062 & 1 & 2.96(11) & 0.71(2) & 2.39(11) & 1.88(4) \\
~ & 2 & 2.60(9) & 0.70(3) & 2.04(8) & 1.84(4) \\
~ & 4 & 2.1(4) & 0.8(2) & 1.9(4) & 2.2(6) \\
c007 & 1 & 2.69(16) & 0.63(4) & 2.12(10) & 1.75(7) \\
~ & 2 & 2.62(16) & 0.73(7) & 1.90(13) & 1.85(14) \\
~ & 4 & 2.3(2) & 0.76(9) & 1.89(14) & 2.01(20) \\
c02 & 1 & 2.71(8) & 0.625(16) & 2.15(7) & 1.74(3) \\
~ & 2 & 2.49(20) & 0.64(5) & 1.96(15) & 1.69(11) \\
~ & 4 & 2.0(2) & 0.71(6) & 1.68(14) & 1.83(14) \\
\hline\hline
\end{tabular}
\end{center}
\end{table}

\begin{table}[t]
\caption{\label{tab:sse_va_3par_ls}Results and correlation matrices of
  fits to $B_s\to K^*$ form factors. }
\begin{center}
\begin{tabular}{c|c|rrr} \hline \hline
\multicolumn{5}{c}{$P(t; -42\mathrm{MeV}) V(t)$} \\ \hline
$p$ & Value  & \multicolumn{1}{c}{$C(p,a_0)$}  & \multicolumn{1}{c}{$C(p,a_1)$}  & \multicolumn{1}{c}{$C(p,c_{01})$} \\ \hline
$a_0$ & $0.322(48)$ \\
$a_1$ & $-3.04(67)$ & $0.91$ \\
$c_{01}$ & $4.82(1.66)$ & $-0.87$ & $-0.62$ \\
$c_{01s}$ & $-0.052(134)$ & $-0.04$ & $-0.00$ & $0.04$
\\ \hline
\multicolumn{5}{c}{$P(t; -87\mathrm{MeV}) A_0(t)$} \\ \hline
$p$ & Value  & \multicolumn{1}{c}{$C(p,a_0)$}  & \multicolumn{1}{c}{$C(p,a_1)$}  & \multicolumn{1}{c}{$C(p,c_{01})$} \\ \hline
$a_0$ & $0.476(42)$ \\
$a_1$ & $-2.29(74)$ & $0.92$ \\
$c_{01}$ & $0.46(68)$ & $-0.31$ & $0.03$ \\
$c_{01s}$ & $-0.157(144)$ & $-0.06$ & $0.01$ & $0.05$
\\ \hline
\multicolumn{5}{c}{$P(t; 350\mathrm{MeV}) A_1(t)$} \\ \hline
$p$ & Value  & \multicolumn{1}{c}{$C(p,a_0)$}  & \multicolumn{1}{c}{$C(p,a_1)$}  & \multicolumn{1}{c}{$C(p,c_{01})$} \\ \hline
$a_0$ & $0.2342(122)$ \\
$a_1$ & $0.100(174)$ & $0.90$ \\
$c_{01}$ & $2.38(48)$ & $-0.75$ & $-0.44$ \\
$c_{01s}$ & $-0.056(78)$ & $-0.06$ & $0.00$ & $0.05$
\\ \hline
\multicolumn{5}{c}{$P(t; 350\mathrm{MeV}) A_{12}(t)$} \\ \hline
$p$ & Value  & \multicolumn{1}{c}{$C(p,a_0)$}  & \multicolumn{1}{c}{$C(p,a_1)$}  & \multicolumn{1}{c}{$C(p,c_{01})$} \\ \hline
$a_0$ & $0.1954(133)$ \\
$a_1$ & $0.350(190)$ & $0.89$ \\
$c_{01}$ & $0.19(49)$ & $-0.53$ & $-0.12$ \\
$c_{01s}$ & $-0.590(137)$ & $-0.06$ & $0.03$ & $0.06$
\\ \hline
\multicolumn{5}{c}{$P(t; 350\mathrm{MeV}) T_{23}(t)$} \\ \hline
$p$ & Value  & \multicolumn{1}{c}{$C(p,a_0)$}  & \multicolumn{1}{c}{$C(p,a_1)$}  & \multicolumn{1}{c}{$C(p,c_{01})$} \\ \hline
$a_0$ & $0.472(24)$ \\
$a_1$ & $0.35(31)$ & $0.89$ \\
$c_{01}$ & $0.47(42)$ & $-0.61$ & $-0.23$ \\
$c_{01s}$ & $-0.306(102)$ & $-0.06$ & $0.03$ & $0.06$
\\ \hline\hline
\end{tabular}
\end{center}
\end{table}

\begin{table*}[t]
\caption{\label{tab:sse_t_6par_ls}Results and correlation matrix of the
 fit to $B_s\to K^*$ form factors 
$P(t; -42\mathrm{MeV}) T_1(t)$ and $P(t;440\mathrm{MeV}) T_2(t)$.  The fit
implements the constraint that $T_1(0) = T_2(0)$.}
\begin{center}
\begin{tabular}{c|c|rrrrrrr} \hline \hline
$p$ & Value  & \multicolumn{1}{c}{$C(p,a_0^{T_1})$} & 
\multicolumn{1}{c}{$C(p,a_1^{T_1})$}  & \multicolumn{1}{c}{$C(p,c_{01}^{T_1})$} & 
\multicolumn{1}{c}{$C(p,a_0^{T_2})$}  & \multicolumn{1}{c}{$C(p,a_1^{T_2})$} 
& \multicolumn{1}{c}{$C(p,c_{01}^{T_2})$} & 
\multicolumn{1}{c}{$C(p,c_{01s}^{T_1})$} \\ \hline
$a_0^{T_1}$& $0.3519(137)$ \\
$a_1^{T_1}$& $-0.992(168)$ & $0.09$ \\
$c_{01}^{T_1}$& $2.06(73)$ & $-0.74$ & $0.42$ \\
$a_0^{T_2}$& $0.2460(108)$ & $0.71$ & $0.72$ & $-0.20$ \\
$a_1^{T_2}$ & $0.169(124)$ & $0.65$ & $0.78$ & $-0.14$ & $0.88$ \\
$c_{01}^{T_2}$ & $1.46(37)$ & $-0.51$ & $-0.30$ & $0.27$ & $-0.71$ & $-0.34$ \\
$c_{01s}^{T_1}$ & $-0.248(85)$ & $-0.06$ & $0.05$ & $0.03$ & $-0.00$ & $-0.00$ & $0.00$ \\
$c_{01s}^{T_2}$ & $-0.093(47)$ & $-0.01$ & $-0.01$ & $0.01$ & $-0.03$ & $0.01$ & $0.03$ & $0.00$
\\ \hline\hline
\end{tabular}
\end{center}
\end{table*}

Finally, in order to facilitate comparison between our results and
others, we provide final results for the form factors at a few values
of $q^2$ in Table~\ref{tab:ff_vs_qsq}.  In this table we have combined
in quadrature the 5\% systematic uncertainty with the statistical and
fitting errors.

\begin{table*}
\caption{\label{tab:ff_vs_qsq}Values for form factors in the physical
  limit at several reference values of $q^2$ using the final results
  of our fits. Systematic uncertainties are included in the tabulated
  error estimates.}
\begin{center}
\begin{tabular}{cccccccc} \hline\hline
\multicolumn{8}{c}{$B\to K^*$} \\ 
$q^2$ (GeV${}^2$) & $V$ & $A_0$ & $A_1$ & $A_{12}$ & $T_1$ & $T_2$ & $T_{23}$ \\ \hline 
$q_{\mathrm{max}}^2$ & 1.93(15) & 1.93(15) & 0.62(4) & 0.44(4) & 1.55(10) & 0.63(4) & 1.20(9) \\
16 & 1.28(11) & 1.25(10) & 0.53(4) & 0.38(3) & 1.05(7) & 0.53(3) & 0.98(7) \\
12 & 0.84(12) & 0.80(11) & 0.44(4) & 0.33(4) & 0.71(5) & 0.44(4) & 0.80(8) \\
0 & 0.30(15) & 0.27(14) & 0.30(5) & 0.25(6) & 0.29(4) & 0.29(4) & 0.52(10) \\
\hline
\multicolumn{8}{c}{$B_s\to \phi$} \\ 
$q^2$ (GeV${}^2$) & $V$ & $A_0$ & $A_1$ & $A_{12}$ & $T_1$ & $T_2$ & $T_{23}$ \\ \hline 
$q_{\mathrm{max}}^2$ & 1.74(10) & 1.85(10) & 0.62(3) & 0.41(2) & 1.36(8) & 0.62(3) & 1.10(6) \\
16 & 1.19(7) & 1.32(7) & 0.52(3) & 0.37(2) & 0.99(5) & 0.53(3) & 0.95(5) \\
12 & 0.77(6) & 0.90(6) & 0.44(3) & 0.33(2) & 0.69(4) & 0.45(3) & 0.81(5) \\
0 & 0.24(7) & 0.38(6) & 0.29(3) & 0.25(3) & 0.31(2) & 0.31(2) & 0.56(5) \\
\hline
\multicolumn{8}{c}{$B_s\to K^*$} \\ 
$q^2$ (GeV${}^2$) & $V$ & $A_0$ & $A_1$ & $A_{12}$ & $T_1$ & $T_2$ & $T_{23}$ \\ \hline 
$q_{\mathrm{max}}^2$ & 1.99(13) & 2.38(16) & 0.58(3) & 0.43(3) & 1.48(10) & 0.60(3) & 1.14(6) \\
16 & 1.02(8) & 1.33(8) & 0.45(3) & 0.36(2) & 0.90(6) & 0.47(3) & 0.90(5) \\
12 & 0.56(9) & 0.84(9) & 0.37(3) & 0.31(3) & 0.61(4) & 0.39(3) & 0.75(5) \\
0 & 0.04(11) & 0.27(11) & 0.24(3) & 0.23(3) & 0.26(3) & 0.26(3) & 0.50(6) \\
\hline \hline
\end{tabular}
\end{center}
\end{table*}


\bibliographystyle{h-physrev5}
\bibliography{mbw}

\end{document}